\documentclass[lettersize,journal]{IEEEtran}
\usepackage{amsmath,amsfonts}
\usepackage{algorithm}
\usepackage{array}

\usepackage{textcomp}
\usepackage{stfloats}
\usepackage{url}
\usepackage{verbatim}
\usepackage{graphicx}
\usepackage{cite}

\usepackage{threeparttable}
\usepackage{xcolor}
\usepackage{tablefootnote}
\usepackage{multirow}
\usepackage{epstopdf}

\usepackage{amssymb} 

\usepackage{algorithmicx}
\usepackage{algpseudocode} 

\usepackage{subfigure}

\begin{document}

\title{Towards the Development of A Three-Dimensional SBP-SAT FDTD Method: Theory and Validation}

\author{Yu Cheng*, Hanhong Liu*, Xinsong Wang, Guangzhi Chen, Xiang-Hua Wang, Xingqi Zhang, \\ Shunchuan Yang, ~\IEEEmembership{Member,~IEEE}, and Zhizhang Chen, ~\IEEEmembership{Fellow,~IEEE}
        % <-this % stops a space
\thanks{Manuscript received xxx; revised xxx.}
\thanks{
%	This work was supported in part by the National Natural Science Foundation
%	of China through Grant 61801010, 62071125, 61671257, in part by Pre-Research Project
%	through Grant J2019-VIII-0009-0170. 
	This work was supported in part by the National Natural Science Foundation of China under Grant 62141405, 62101020, 62071125, in part by Defense Industrial Technology Development Program under Grant JCKY2019601C005, in part by Pre-Research Project under Grant J2019-VIII-0009-0170.
	{\it{(Authors marked by * are equally contributed to this paper. Corresponding author: Shunchuan Yang.)}}
	
	Y. Cheng, H. H. Liu and X. S. Wang are with the School of Electronic and Information Engineering, Beihang University, Beijing, 100083, China (e-mail: yucheng@buaa.edu.cn, liu759753745@buaa.edu.cn, wxs20@buaa.edu.cn).
	
	X. Q. Zhang is with School of Electrical and Electronic Engineering, University College Dublin, Dublin, Ireland (e-mail:xingqi.zhang@ucd.ie).
	
	X. H. Wang is with School of Science, Tianjin University of Technology and Education, Tianjin, China (e-mail: xhwang199@outlook.com).	
	
	G. Z. Chen and S. C. Yang is with the Research Institute for Frontier Science and School of Electronic and Information Engineering, Beihang University, Beijing, 100083, China (e-mail: dazhihaha@buaa.edu.cn, scyang@buaa.edu.cn).
	
	Z. Z. Chen is currently with the College of Physics and Information Engineering, Fuzhou University, Fuzhou, Fujian. P. R. China, on leave from the Department of Electrical and Computer Engineering, Dalhousie University, Halifax, Nova Scotia, Canada B3H 4R2 (email: zz.chen@ieee.org).}

}

\markboth{Journal of \LaTeX\ Class Files,~Vol.~xx, No.~x, August~xxxx}%
{Shell \MakeLowercase{\textit{et al.}}: A Sample Article Using IEEEtran.cls for IEEE Journals}


\maketitle

\begin{abstract}
%To enhance scalability and performance of the traditional finite-difference time-domain (FDTD) methods, a three-dimensional summation-by-parts simultaneous approximation terms (SBP-SAT) FDTD method is developed to solve complex electromagnetic problems. It is theoretically stable and can be further used for multiple mesh blocks with different mesh cells. This paper mainly focuses on the fundamental theoretical aspects upon its three-dimensional implementation, the SAT for various boundary conditions, its numerical dispersion properties, and the comparison with the FDTD method. The proposed SBP-FDTD method inherits all the merits of the FDTD method, which has only local operators without matrix inversion, easy to implement, and has the same level of accuracy with a negligible overhead of runtime (0.13\%) and memory usage (1.2\%). four numerical examples are carried out to indicate the effectiveness of the proposed method. 
To enhance the scalability and performance of the traditional finite-difference time-domain (FDTD) methods, a three-dimensional summation-by-parts simultaneous approximation term (SBP-SAT) FDTD method is developed to solve complex electromagnetic problems. It is theoretically stable and can be further used for multiple mesh blocks with different mesh sizes. This paper mainly focuses on the fundamental theoretical aspects upon its three-dimensional implementation, the SAT for various boundary conditions, and the numerical dispersion properties and the comparison with the FDTD method. The proposed SBP-SAT FDTD method inherits all the merits of the FDTD method, which is matrix-free, easy to implement, and has the same level of accuracy with a negligible overhead of runtime (0.13\%) and memory usage (1.2\%). Four numerical examples are carried out to validate the effectiveness of the proposed method. 
\end{abstract}


\vspace{+0.3cm}
\section{Introduction}
\vspace{+0.3cm}

\IEEEPARstart{T}{he} finite-difference time-domain (FDTD) method has been widely used in scattering analysis \cite{1radar}, designs of waveguides \cite{2design}, antennas \cite{3antenna}, and biomedicine \cite{4biomedicine} due to its simplicity, high parallel efficiency, and strong capability to handle complex media. However, it suffers from accuracy issues due to staircase errors when multiscale or complex structures are involved.

The subgridding technique is one of the approaches to effectively decrease staircase errors through local refinement meshes in regions including geometrically fine structures. Many efforts have been made in the last few decades \cite{SubgriddingConsistentTweiland, SubgriddingFlexibleTweiland, SubgriddingMichal, SubgriddingRobert, SubgriddingLowReflection, SubgriddingHangingMichal, 25wave, 5Huygens}. Especially, a theoretically stable subgridding technique through filtering out unstable modes was proposed in \cite{8filter}. However, it may be computationally expensive for large-scale problems to calculate those stable modes in simulations. In \cite{26MOR}\cite{27MOR}, a reduced-order model (MOR) was used in the subgridding scheme to extend the Courant-Friedrichs-Lewy (CFL) condition for improving the efficiency. Another high-order smoothing technique was proposed to interpolate fields on the interfaces in the non-standard (NS)-FDTD method \cite{28NSFDTD}.  An asymmetric FDTD subgridding technique was proposed in \cite{30SPD}, which can be used for any mesh refinement ratio. It's well-known that the long-time stability of those subgridding algorithms can not be always guaranteed since the theoretical proofs can hardly be given through making interpolation operators meet the reciprocity principle \cite{ReciprocitySubgridding} or the dissipation theory \cite{DissipationSubgridding}. 

Recently, the summation by parts simultaneous approximation term (SBP-SAT) technique provides the possibility of implementing long-time stable subgridding techniques. The finite-difference methods with the SBP-SAT techniques were originally proposed to solve the Euler and Navier-Stokes equations in \cite{9N-Sequation}\cite{10N-Sequation}, and other applications are carried out in \cite{11other}\cite{12other}. Then, it has been introduced to solve the two-dimensional Maxwell's equations in \cite{13two}\cite{14two}, in which electric and magnetic field components are collocated at field nodes. Since staggered grids can decrease numerical dispersion errors \cite{15disper}, efforts have been done to extend the SBP-SAT techniques to solving acoustic scattering problems with staggered grids \cite{16acoustic}\cite{17acoustic}. In \cite{24SBPSATChengyu}\cite{24SBPSATYuHui}, the SBP-SAT FDTD method is developed to solve the two-dimensional Maxwell's equations on staggered grids.

However, to the best of the authors’ knowledge, there are no reports or implementations based on the three-dimensional SBP-SAT FDTD method to solve the Maxwell's equations. Based on our previous experience, extensions of the two-dimensional time-domain methods into their three-dimensional counterparts are nontrivial, and instability may occur. In this article, a three-dimensional theoretically stable FDTD method based on the SBP-SAT technique is proposed, which can be used for multiple mesh blocks with different cell sizes. It will be discussed in a follow-up article. The paper mainly focuses on the fundamental theoretical aspects of the proposed method and its validation. In the proposed method, additional field nodes are added on the boundaries of computational domain to satisfy the SBP property, and nodes inside the computational domain are the same as those of the traditional FDTD method. Then, by using the properties of the SBP operator, the energy of the whole computational domain is fully determined by fields on the boundaries. The perfectly electrical conducting (PEC), perfectly magnetic conducting (PMC) boundary conditions, and the periodic boundary condition (PBC) are weakly enforced through the SAT technique. In addition, the numerical errors are comprehensively investigated. These theoretical proofs guarantee the long-time stability of the proposed three-dimensional SBP-SAT FDTD method, and it can be the spurious-free alternative for the FDTD method. The main contributions of this article are divided into four aspects.

\begin{enumerate}
	\item A three-dimensional SBP-SAT FDTD method is proposed in this article. To make the discrete operators satisfy the SBP properties, additional electric and magnetic nodes are sampled on the boundaries of the computational domain. Its time-marching formulations are comprehensively derived based on modified grids, and the matrix-free implementations are also presented in detail. Although several matrices are involved in our derivation, the time-marching formulations can be decomposed into the elemental manner as that in the FDTD method. Therefore, it is as efficient as the FDTD method.  
	
	\item The PEC, PMC, and PBC are derived through the SAT technique to guarantee their stability. Unlike implementations in the FDTD method, those boundary conditions are weakly enforced through the SAT technique. It can be used to develop theoretically stable subgridding methods for multiple mesh blocks with different mesh sizes. Since the SATs only exist on the boundaries of computational domain, a negligible overhead of memory and runtime is imposed.
	
	\item The numerical dispersion of the proposed method is comprehensively investigated and compared with that of the FDTD method. It is found that the proposed SBP-SAT FDTD method has the same level of accuracy and anisotropy as the FDTD method. 	
	
	\item Four practical numerical examples are carried out to validate its stability, accuracy, and efficiency. Numerical results show that the proposed SBP-SAT FDTD method shares the same merits as the FDTD method. It is simple, easy to implement, matrix-free and has strong capability of handling complex media.  
\end{enumerate}

This paper is organized as follows. In Section II, grids used in the proposed three-dimensional SBP-SAT FDTD method are first presented in detial. Then, the time-marching formulations based on the modified grids are shown. In Section III, treatments of the PEC, PMC boundary conditions and PBC by the SAT technique are rigorously derived to guarantee the long-time stability. In Section IV, its numerical dispersion error based on PBC is comprehensively investigated and compared with that of the FDTD method. Then, the practical implementation and its efficiency comparison are carried out in Section V. In Section VI, four numerical examples are carried out to domesticate the effectiveness of the proposed method. Finally, conclusions are drawn in Section VII.

\section{The Grids and Formulations in the SBP-SAT FDTD Method}

\subsection{Fields Nodes Distribution on Grids}
\begin{figure*}[h]
	\subfigure[]{
		\includegraphics[scale=0.25]{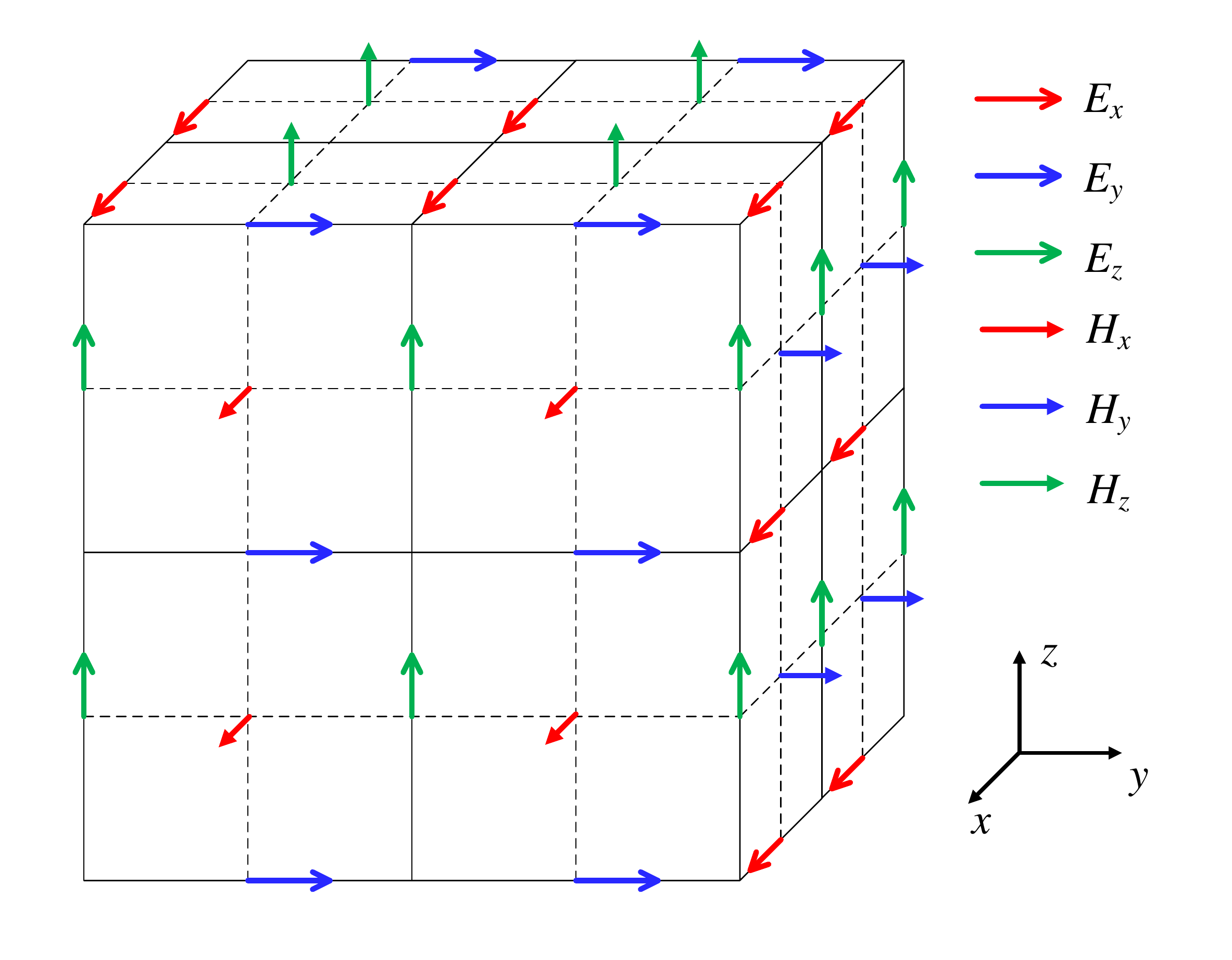}\label{F_2_3}
	}
	\subfigure[]{
		\includegraphics[scale=0.25]{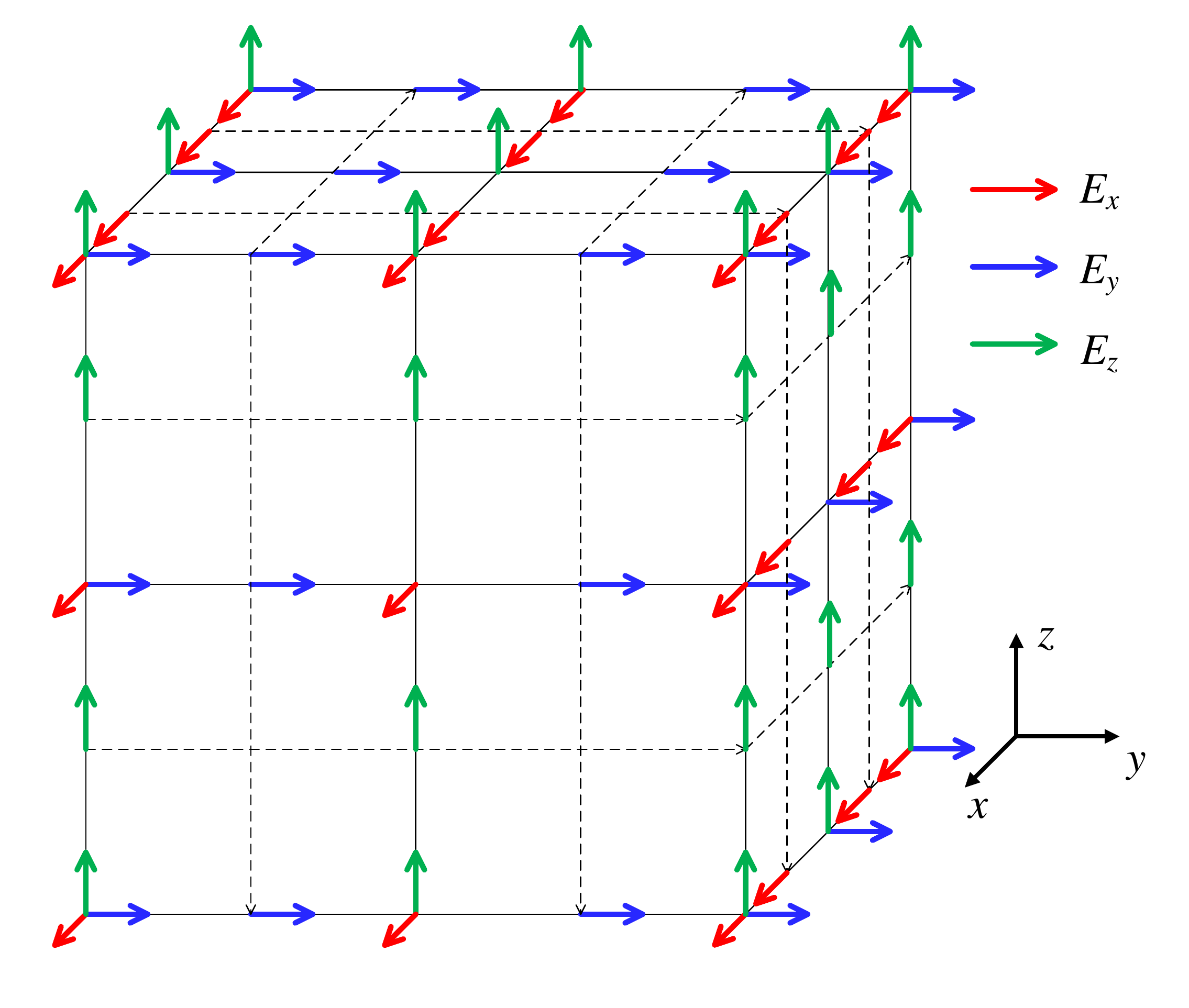}\label{F_2_1}
	}
	\vspace{-0.1cm}
	\subfigure[]{
		\includegraphics[scale=0.25]{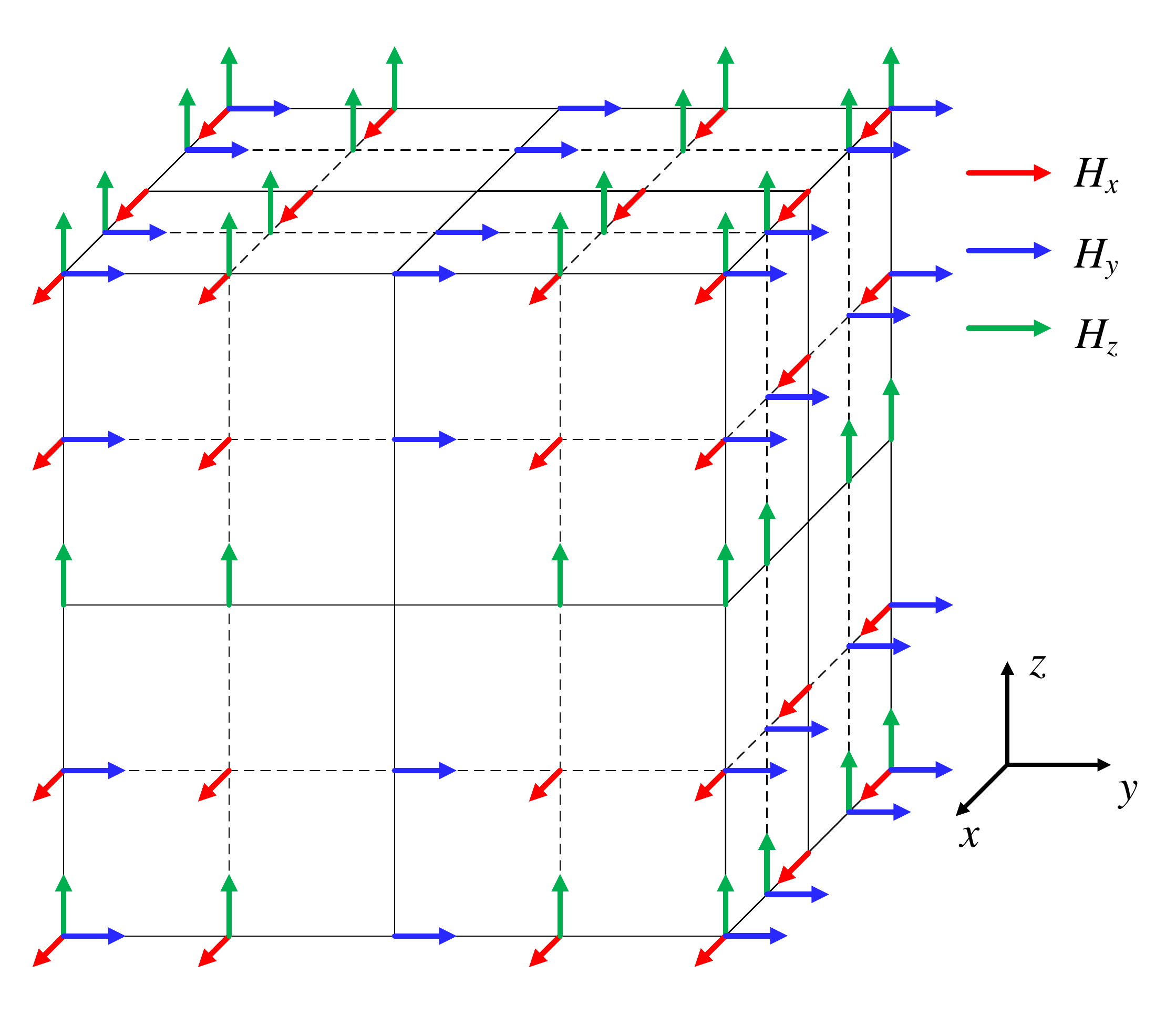}\label{F_2_2}
	}
	\caption{The distribution of fields in the SBP-SAT FDTD method: (a) $\bf E$- and $\bf H$-nodes in the FDTD method, (b) $\bf E$-nodes in the SBP-SAT FDTD method, (c) $\bf H$-nodes in the SBP-SAT FDTD method.}
	\label{F_2}
\end{figure*}

Without loss of generality, a lossless, homogenous and isotropic medium is considered. The three-dimensional Maxwell's equations [31] are given by 
\begin{subequations}
	\begin{align}
		\nabla \times & \boldsymbol{\rm{H}} = \varepsilon \frac{\partial \boldsymbol{\rm{E}}}{\partial t}, \label{E1_1} \\
		\nabla \times & \boldsymbol{\rm{E}} = - \mu \frac{\partial \boldsymbol{\rm{H}}}{\partial t}, \label{E1_2}
	\end{align}
\end{subequations}
where $\varepsilon$ and $\mu$ are the permittivity and the permeability of the medium, respectively. In order to solve (\ref{E1_1}) and (\ref{E1_2}), the FDTD method uses Yee's grids to sample electromagnetic fields in the spatial domain, as shown in Fig. \ref{F_2_3}. Electric field nodes ({\textbf{E}}-nodes) are located in the middle of each cell edge, and magnetic field nodes ({\textbf{H}}-nodes) are located at the center of each cell face. {\textbf{E}}- and {\textbf{H}}-nodes are interlaced with each other on Yee's grids.

The electromagnetic fields in the proposed three-dimensional SBP-SAT FDTD method are similar to those in the FDTD method. The SBP-SAT FDTD method and the FDTD method have exactly the same field node distributions inside the computational domain. However, to meet the SBP properties, both {\textbf{E}-} and {\textbf{H}}-node distributions in the SBP-SAT FDTD method have to be modified on the boundaries of computational domain.

To clearly demonstrate grids used in the proposed method, two kinds of one-dimensional grids, ${{x}_ + } = {\left[ {{x_0},{x_1}, \ldots,{x_i},\ldots,{x_n}} \right]^T}$ and ${x_ - } = {\left[ {{x_0},{x_{1/2}}, \ldots,{x_{i-1/2}},\ldots, {x_{n - 1/2}},{x_n}} \right]^T}$, where subscripts denote field node locations and $x_i = ih$ with $h$ as the interval, are used to sample electromagnetic fields in the one-dimensional spatial domain. In our implementation, three-dimensional grids are extended from ${{x}_ + }$ and ${{x}_ - }$ with appropriate combinations, which implies that they are decomposed into three one-dimensional grids in the $x$, $y$, and $z$ directions, respectively. Table \ref{T1} lists how six three-dimensional grids for each fields are decomposed into corresponding three one-dimensional grids in the $x$, $y$, and $z$ directions, respectively. 
\begin{table}[h]
	\renewcommand\arraystretch{1.5}
	\centering
	\caption{The Three-Dimensional Grids in the x, y, z Directions}
	\label{T1}
%	\resizebox{8cm}{!}
%	{
	\setlength{\tabcolsep}{6mm}
	{
		\begin{threeparttable}[b]
			\begin{tabular}{ c| c| c| c| c}
				\hline
				\hline
				\multicolumn{2}{c|}{\textbf{Field Nodes}}  &\multicolumn{3}{c}{\textbf{Axis}}\\
				\hline
				\multirow{3}*{\textbf{E-nodes}} &$E_x$ &$x_-$  &$y_+$ &$z_+$ \\
					\cline{2-5}
				 									   &$E_y$ &$x_+$  &$y_-$ &$z_+$ \\
				 	\cline{2-5}
												 	   &$E_z$ &$x_+$  &$y_+$ &$z_-$ \\
					\hline
				\multirow{3}*{\textbf{H-nodes}} &$H_x$ &$x_+$  &$y_-$ &$z_-$ \\
				\cline{2-5}
											 	 	   &$H_y$ &$x_-$  &$y_+$ &$z_-$ \\
				\cline{2-5}
											 	 	   &$H_z$ &$x_-$  &$y_-$ &$z_+$ \\
%				\cline{3-6}
				\hline
				\hline

			\end{tabular}
			%\tablefootnote{Ratio is defined as the ratio of time cost used in the LOD-FDTD method with fine grid to that in the correspond method.}
		\end{threeparttable}
	}
%	}
\end{table}

{\textbf{E}}- and {\textbf{H}}-nodes on the boundaries of computational domain in the SBP-SAT FDTD method are shown in Fig. \ref{F_2_1} and \ref{F_2_2}, respectively. It can be found that additional magnetic fields need to be added in middle of boundary edges in their vertical direction, and four corner nodes of their vertical boundary edges. To clearly demonstrate field distribution on the boundaries, we take ${E_x}$ and ${H_x}$ as examples.

Two ${E_x}$ nodes are added at each edge along the $x$ direction. Therefore, two additional electric field nodes are added at the end and beginning of each edge along the corresponding directions. Additional ${H_x}$ nodes are added in the middle of boundary edges along the $y$ and $z$  direction, respectively, and intersections of two edges.  

As it is stated above, node distributions are exactly the same as those in Yee's grids inside the computational domain. Only additional field nodes are required to be added on the boundaries to meet the SBP properties. Therefore, only a small amount of memory is required to store those additional nodes.  

\subsection{The SBP Operators in the One-Dimensional Space}
In order to discrete (\ref{E1_1}) and (\ref{E1_2}) with the SBP grids, several operators, which are similar to those in \cite{14two}, are first  defined in the one-dimensional space. Two discrete finite-difference matrices ${\mathbb{D_+}}$ and ${\mathbb{D_-}}$ are defined on ${\bf{x}}_+$ and ${\bf{x}}_-$ \cite{24SBPSATChengyu}. ${\mathbb{D_+}}$ and ${\mathbb{D_-}}$ should satisfy the following accuracy relationship
\begin{equation}\label{E2}
		{\mathbb{D}_ + }{\bf{x}}_ - ^k = k{\bf{x}}_ + ^{k - 1}{\rm{, }}{\mathbb{D}_ - }{\bf{x}}_ + ^k = k{\bf{x}}_ - ^{k - 1}{\rm{, }}{\kern 1pt} k = 0,{\rm{ }}1,{\rm{ }}
\end{equation}
where the dimensions of ${\mathbb{D_+}}$ and ${\mathbb{D_-}}$ are ${N_ + } \times {N_ - }$ and $ {N_ - } \times {N_ + }$, respectively. When $k = 0$, ${\bf{x}}_ - ^{ - 1} = 0$ and ${\bf{x}}_ + ^{ - 1} = 0$. ${\mathbb{D_+}}$ and ${\mathbb{D_-}}$ can be further  expressed as
\begin{equation}\label{E3}
		{\mathbb{D}_ + }={\mathbb{P}^{-1}_ + }{\mathbb{Q}_ + }, {\mathbb{D}_ - }={\mathbb{P}^{-1}_ - }{\mathbb{Q}_ - }, 
\end{equation}
where $\mathbb{P}_ +$ and $\mathbb{P}_ -$ are positive definite matrices, and their entities denote the Gaussian weights associated with corresponding field nodes. $\mathbb{Q}_ +$ and $\mathbb{Q}_ -$ satisfy 

\begin{equation} \label{E_QB}
{\mathbb{Q}_ + } + \mathbb{Q}_ - ^T = \mathbb{B}
\end{equation}
where 
\begin{equation}\label{E4}
	\mathbb{B} = \left[
	 {\begin{array}{*{20}{c}}
				{ - 1}&0& \cdots &0&0\\
				0&0&{}&{}&0\\
				\vdots &{}& \ddots &{}& \vdots \\
				0&{}&{}&0&0\\
				0&0& \cdots &0&1
		\end{array}} \right]_{{N_ + } \times {N_ - }}.
\end{equation}
When difference operators satisfy (\ref{E3}) and (\ref{E_QB}), they can be regarded as the SBP operators. With the definition of $\mathbb{B}$, we get
%\begin{equation}\label{E5}
%	{\bf{x}}{^2} = {{\bf{x}}^T}\mathbb{P}{\bf{x}} \approx \int {{x^2}} dl,
%\end{equation}
\begin{equation}\label{E6}
	{{\bf{x}}_+^T}\mathbb{B}{\bf{x}_-} = {x_n}{x_n} - {x_0}{x_0},
\end{equation}
where ${\left\| \cdot \right\|^2}$ is the two-norm of a column vector, which will be used to estimate the discrete energy. From (\ref{E6}), it can be noted that the energy in computational domain only depends on the boundary nodes. 

Entities of ${\mathbb{D_+}}$, ${\mathbb{D_-}}$, ${\mathbb{P_+}}$, ${\mathbb{P_-}}$, ${\mathbb{Q_+}}$ and ${\mathbb{Q_-}}$ can be found in \cite{24SBPSATChengyu}. For readers' convenience, they are also listed in the Appendix.

\subsection{The Semi-Discrete Formulations of the SBP-SAT FDTD Method without the Boundary Conditions}
Several discrete finite-difference matrices are used to approximate the partial differential operators in (\ref{E1_1})-(\ref{E1_2}). By using these discrete matrices, the semi-discrete Maxwell's equations can be written as 
\begin{subequations}
	\begin{align}
		&\frac{{d{{\bf{E}}_x}}}{{dt}} = {\mathbb{D}_{{H_z}}^y}{{\bf{H}}_z} - {\mathbb{D}_{{H_y}}^z}{{\bf{H}}_y}, \label{E7}\\
		&\frac{{d{{\bf{E}}_y}}}{{dt}} = {\mathbb{D}_{{H_x}}^z}{{\bf{H}}_x} - {\mathbb{D}_{{H_z}}^x}{{\bf{H}}_z}, \label{E8}\\
		&\frac{{d{{\bf{E}}_z}}}{{dt}} = {\mathbb{D}_{{H_y}}^x}{{\bf{H}}_y} - {\mathbb{D}_{{H_x}}^y}{{\bf{H}}_x},\label{E9}\\
		&\frac{{d{{\bf{H}}_x}}}{{dt}} = {\mathbb{D}_{{E_y}}^z}{{\bf{E}}_y} - {\mathbb{D}_{{E_z}}^y}{{\bf{E}}_z},\label{E10}\\
		&\frac{{d{{\bf{H}}_y}}}{{dt}} = {\mathbb{D}_{{E_z}}^x}{{\bf{E}}_z} - {\mathbb{D}_{{E_x}}^z}{{\bf{E}}_x},\label{E11}\\
		&\frac{{d{{\bf{H}}_z}}}{{dt}} = {\mathbb{D}_{{E_x}}^y}{{\bf{E}}_x} - {\mathbb{D}_{{E_y}}^x}{{\bf{E}}_y},\label{E12}
	\end{align}
\end{subequations}
where ${{\bf{E}}_x}$ is a column vector collecting all ${\bf{E}}_x$ nodes in the computational domain in the $x$, $y$, $z$ directions, and it is similar for ${{\bf{E}}_y}$, ${{\bf{E}}_z}$, ${{\bf{H}}_x}$, ${{\bf{H}}_y}$, and ${{\bf{H}}_z}$. The discrete finite-difference matrices are defined as 
\begin{subequations}
	\begin{align}
		{\mathbb{D}_{{i}}^x} =  {{\mathbb{I}_{z}} \otimes {\mathbb{I}_{y}} \otimes {\mathbb{D}_{x}}},\label{ED_1}\\
		{\mathbb{D}_{{j}}^y} =  {{\mathbb{I}_{z}} \otimes {\mathbb{D}_{y}} \otimes {\mathbb{I}_{x}}},\label{ED_2}\\
		{\mathbb{D}_{{k}}^z} =  {{\mathbb{D}_{z}} \otimes {\mathbb{I}_{y}} \otimes {\mathbb{I}_{x}}},\label{ED_3}
	\end{align}
\end{subequations}
where $\mathbb{I}$ are identity matrices with the corresponding dimensions. $\otimes$ is the Kronecker product operator, which can be regarded as an extension of the low-dimensional operator into its high-dimensional counterpart. The subscripts of these matrices denote finite-difference operators defined for the corresponding components. Since ${{\bf{E}}_x}$, ${{\bf{E}}_y}$, ${{\bf{E}}_z}$, ${{\bf{H}}_x}$, ${{\bf{H}}_y}$, and ${{\bf{H}}_z}$ are arranged in the order of $x$, $y$, and $z$, respectively, the Kronecker product operation should be performed in the corresponding order.

%Four types of finite-difference matrices for (\ref{ED_1}) - (\ref{ED_3}) are used in the following derivation. For (\ref{ED_1}), we have (1)  $i = {H}_z$, $x = x+$, $y = y-$, and $z = z+$; (2) $i = {H}_y$, $x = x+$, $y = y+$, and $z = z-$; (3) $i = {E}_z$, $x = x-$, $y = y+$, and $z = z-$; (4) $i = {E}_y$, $x = x-$, $y = y-$, and $z = z+$. For (\ref{ED_2}), we have (1) $j = {H}_z$, $x = x-$, $y = y+$, and $z = z+$; (2) $j = {H}_x$, $x = x+$, $y = y+$, and $z = z-$; (3) $j = {E}_z$, $x = x+$, $y = y-$, and $z = z-$; (4) $j = {E}_x$, $x = x-$, $y = y-$, and $z = z+$. For (\ref{ED_3}), we have (1) $k = {H}_y$, $x = x-$, $y = y+$, and $z = z+$; (2) $k = {H}_x$, $x = x+$, $y = y-$, and $z = z+$; (3) $k = {E}_y$, $x = x+$, $y = y-$, and $z = z-$; (4) $k = {E}_x$, $x = x-$, $y = y+$, and $z = z-$. 
Four types of finite-difference matrices for (\ref{ED_1})-(\ref{ED_3}) are used in the following derivation. For (\ref{ED_1}), we have (1)  $i = {H}_z$; (2) $i = {H}_y$; (3) $i = {E}_z$; (4) $i = {E}_y$. For (\ref{ED_2}), we have (1) $j = {H}_z$; (2) $j = {H}_x$; (3) $j = {E}_z$; (4) $j = {E}_x$. For (\ref{ED_3}), we have (1) $k = {H}_y$; (2) $k = {H}_x$; (3) $k = {E}_y$; (4) $k = {E}_x$. The subscripts $x$, $y$ and $z$ correspond to field nodes $i$ in Table \ref{T1}.

By using (\ref{E7})-(\ref{E12}), electromagnetic fields can be discretized in the temporal domain. Then it is necessary to add boundary conditions by using the SAT technique.

\section{Boundary Implementation in the Proposed SBP-SAT FDTD Method}

%\begin{subequations}
%	\begin{align}
%		{\mathbb{P}_{{H_x}}} = {{\mathbb{P}_{z - }} \otimes {\mathbb{P}_{y - }} \otimes {\mathbb{P}_{x + }}}, \label{EP_1}\\
%		{\mathbb{P}_{{H_y}}} = {{\mathbb{P}_{z - }} \otimes {\mathbb{P}_{y + }} \otimes {\mathbb{P}_{x - }}}, \label{EP_2}\\
%		{\mathbb{P}_{{H_z}}} = {{\mathbb{P}_{z + }} \otimes {\mathbb{P}_{y - }} \otimes {\mathbb{P}_{x - }}}, \label{EP_3}\\
%		{\mathbb{P}_{{E_x}}} = {{\mathbb{P}_{z + }} \otimes {\mathbb{P}_{y + }} \otimes {\mathbb{P}_{x - }}}, \label{EP_4}\\
%		{\mathbb{P}_{{E_y}}} = {{\mathbb{P}_{z + }} \otimes {\mathbb{P}_{y - }} \otimes {\mathbb{P}_{x + }}}, \label{EP_5}\\
%		{\mathbb{P}_{{E_z}}} = {{\mathbb{P}_{z - }} \otimes {\mathbb{P}_{y + }} \otimes {\mathbb{P}_{x + }}}. \label{EP_6}
%	\end{align}
%\end{subequations}

\subsection{The Proposed SBP-SAT FDTD Method with the PEC Boundary Condition}
In the SBP-SAT FDTD method, the boundary conditions are weakly enforced by the penalty terms using the SAT technique. The SAT for the PEC boundary condition is added in (\ref{E10})-(\ref{E12}) to weakly enforce that tangential electric fields vanish. Then, (\ref{E10})-(\ref{E12}) can be expressed as
%\begin{equation}
	\begin{align}\label{E13}
			&\frac{{d{{\bf{H}}_x}}}{{dt}} - {\mathbb{D}_{{E_y}}^z}{{\bf{E}}_y} + {\mathbb{D}_{{E_z}}^y}{{\bf{E}}_z}  \\ 
			&={\sigma _{{f_1}}}{\left( {\mathbb{P}_{{H_x}}^{\rm{'''}}} \right)^{ - 1}}\mathbb{R}_{{H_{{x_r}}}}^T{\mathbb{P}_{zx}^{\rm{''}}}{{\bf{E}}_{{z_r}}} 
			+ {\sigma _{{b_1}}}{\left( {\mathbb{P}_{{H_x}}^{\rm{'''}}} \right)^{ - 1}}\mathbb{R}_{{H_{{x_l}}}}^T{\mathbb{P}_{zx}^{\rm{''}}}{{\bf{E}}_{{z_l}}}  \notag \\
			&+ {\sigma _{{t_1}}}{\left( {\mathbb{P}_{{H_x}}^{\rm{'''}}} \right)^{ - 1}}\mathbb{R}_{{H_{{x_t}}}}^T{\mathbb{P}_{yx}^{\rm{''}}}{{\bf{E}}_{{y_t}}}
			+ {\sigma _{{d_1}}}{\left( {\mathbb{P}_{{H_x}}^{\rm{'''}}} \right)^{ - 1}}\mathbb{R}_{{H_{{x_d}}}}^T{\mathbb{P}_{yx}^{\rm{''}}}{{\bf{E}}_{{y_d}}},\notag
	\end{align}													
%\end{equation}

%\begin{equation}
	\begin{align}\label{E14}
		&\frac{{d{{\bf{H}}_y}}}{{dt}} - {\mathbb{D}_{{E_z}}^x}{{\bf{E}}_z} + {\mathbb{D}_{{E_x}}^z}{{\bf{E}}_x} \\
		&={\sigma _{{l_1}}}{\left( {\mathbb{P}_{{H_y}}^{\rm{'''}}} \right)^{ - 1}}\mathbb{R}_{{H_{{y_b}}}}^T{\mathbb{P}_{zy}^{\rm{''}}}{{\bf{E}}_{{z_b}}}
		+ {\sigma _{{r_1}}}{\left( {\mathbb{P}_{{H_y}}^{\rm{'''}}} \right)^{ - 1}}\mathbb{R}_{{H_{{y_f}}}}^T{\mathbb{P}_{zy}^{\rm{''}}}{{\bf{E}}_{{z_f}}} \notag \\
		&+ {\sigma _{{t_2}}}{\left( {\mathbb{P}_{{H_y}}^{\rm{'''}}} \right)^{ - 1}}\mathbb{R}_{{H_{{y_t}}}}^T{\mathbb{P}_{xy}^{\rm{''}}}{{\bf{E}}_{{x_t}}}
		+ {\sigma _{{d_2}}}{\left( {\mathbb{P}_{{H_y}}^{\rm{'''}}} \right)^{ - 1}}\mathbb{R}_{{H_{{y_d}}}}^T{\mathbb{P}_{xy}^{\rm{''}}}{{\bf{E}}_{{x_d}}},\notag
	\end{align}	
%\end{equation}

%\begin{equation}
	\begin{align}\label{E15}
		&\frac{{d{{\bf{H}}_z}}}{{dt}} - {\mathbb{D}_{{E_x}}^y}{{\bf{E}}_x} + {\mathbb{D}_{{E_y}}^x}{{\bf{E}}_y}\\  
		&={\sigma _{{f_2}}}{\left( {\mathbb{P}_{{H_z}}^{\rm{'''}}} \right)^{ - 1}}\mathbb{R}_{{H_{{z_r}}}}^T{\mathbb{P}_{xz}^{\rm{''}}}{{\bf{E}}_{{x_r}}}
		+ {\sigma _{{b_2}}}{\left( {\mathbb{P}_{{H_z}}^{\rm{'''}}} \right)^{ - 1}}\mathbb{R}_{{H_{{z_l}}}}^T{\mathbb{P}_{xz}^{\rm{''}}}{{\bf{E}}_{{x_l}}} \notag \\
		&+ {\sigma _{{l_2}}}{\left( {\mathbb{P}_{{H_z}}^{\rm{'''}}} \right)^{ - 1}}\mathbb{R}_{{H_{{z_b}}}}^T{\mathbb{P}_{yz}^{\rm{''}}}{{\bf{E}}_{{y_b}}}
		+ {\sigma _{{r_2}}}{\left( {\mathbb{P}_{{H_z}}^{\rm{'''}}} \right)^{ - 1}}\mathbb{R}_{{H_{{z_f}}}}^T{\mathbb{P}_{yz}^{\rm{''}}}{{\bf{E}}_{{y_f}}}.	\notag
	\end{align}
%\end{equation}
${{\bf{E}}_{{x_r}}}$, ${{\bf{E}}_{{x_l}}}$, ${{\bf{E}}_{{x_t}}}$, ${{\bf{E}}_{{x_d}}}$, ${{\bf{E}}_{{y_t}}}$, ${{\bf{E}}_{{y_d}}}$, ${{\bf{E}}_{{y_b}}}$, ${{\bf{E}}_{{y_f}}}$, ${{\bf{E}}_{{z_r}}}$, ${{\bf{E}}_{{z_l}}}$, ${{\bf{E}}_{{z_b}}}$, ${{\bf{E}}_{{z_f}}}$ are column vectors collecting components nodes on the boundaries, where subscripts denote different electric components on boundary faces of the computational domain. Take ${{\bf{E}}_{{y_t}}}$ as an example, it can be calculated by ${{\bf{E}}_{{y_t}}} = \mathbb{R}_{{E_{{y_t}}}}^T{{\bf{E}}_y}$, where $\mathbb{R}_{{E_{{y_t}}}}^T =  {{{{\bf{e}}_{N + }}} \otimes  {{\mathbb{I}_{x + }} \otimes {\mathbb{I}_{y - }}} } $ with ${{{\bf{e}}_{N + }}} = {\left[ {{1},{0}, \ldots,{0}} \right]^T}$. It selects ${\bf E}_y$ nodes on the top boundary face. $\sigma$ and $\chi$ with subscripts represent free parameters to meet the stability of semi-discrete system in the SBP-SAT FDTD method. The normal matrices are defined as 
\begin{subequations}
	\begin{align}
		{\mathbb{P}_{j}^{\rm{''}}} &= {\mathbb{P}_{m}} \otimes {\mathbb{P}_{n}}. \label{EP_2} \\
		{\mathbb{P}_{i}^{\rm{'''}}}  &= {{\mathbb{P}_{z}} \otimes {\mathbb{P}_{y}} \otimes {\mathbb{P}_{x}}}, \label{EP_1}
	\end{align}
\end{subequations}
For (\ref{EP_2}), six scenarios need to be considered as follows. (1) $j=xy$, $m=y_+$ and $n=x_-$; (2) $j=yx$, $m=y_-$ and $n=x_+$; (3) $j=xz$, $m=z_+$ and $n=x_-$; (4) $j=zx$, $m=z_-$ and $n=x_+$; (5) $j=yz$, $m=z_+$ and $n=y_-$; (6) $j=zy$, $m=z_-$ and $n=y_+$. As for (\ref{EP_1}), six scenarios will be used in the our derivation, which can be expressed as (1) $i=E_x$; (2) $i=E_y$; (3) $i=E_z$; (4) $i=H_x$; (5) $i=H_y$; (6) $i=H_z$, and subscripts $x$, $y$ and $z$ also correspond to field nodes $i$ in Table \ref{T1}. 

The electromagnetic energy in the whole computational domain can be calculated as follows
\begin{equation}\label{E16}
	\begin{aligned}
		{\cal E} &= {\frac{1}{2}}{\bf{E}}_x^H{\mathbb{P}_{{E_x}}^{\rm{'''}}}{{\bf{E}}_x} 
		+ {\frac{1}{2}}{\bf{E}}_y^H{\mathbb{P}_{{E_y}}^{\rm{'''}}}{{\bf{E}}_y}
		+ {\frac{1}{2}}{\bf{E}}_z^H{\mathbb{P}_{{E_z}}^{\rm{'''}}}{{\bf{E}}_z} \\
		&+ {\frac{1}{2}}{\bf{H}}_x^H{\mathbb{P}_{{H_x}}^{\rm{'''}}}{{\bf{H}}_x}
		+ {\frac{1}{2}}{\bf{H}}_y^H{\mathbb{P}_{{H_y}}^{\rm{'''}}}{{\bf{H}}_y} 
		+{\frac{1}{2}}{\bf{H}}_z^H{\mathbb{P}_{{H_z}}^{\rm{'''}}}{{\bf{H}}_z},
	\end{aligned}
\end{equation}
where ${\left(  \cdot  \right)^H}$ is conjugate transpose operator. After taking the partial derivative with respect to time, we can obtain ${d\cal E }/{dt}$ by substituting (\ref{E7})-(\ref{E9}) and (\ref{E13})-(\ref{E15}) into (\ref{E16}).

\begin{equation}\label{E17}
	\begin{aligned}
		\frac{{d\cal E }}{{dt}} &= \left( { {\sigma _{{t_2}}} - 1} \right){\bf{E}}_{{x_t}}^T{\mathbb{P}_{xy}^{\rm{''}}}{{\bf{H}}_{{y_t}}} 
		+ \left( { {\sigma _{{d_2}}} + 1} \right){\bf{E}}_{{x_d}}^T{\mathbb{P}_{xy}^{\rm{''}}}{{\bf{H}}_{{y_d}}}   \\
		&+ \left( { {\sigma _{{r_2}}} + 1} \right){\bf{E}}_{{x_r}}^T{\mathbb{P}_{xz}^{\rm{''}}}{{\bf{H}}_{{z_r}}} 
		+ \left( { {\sigma _{{l_2}}} - 1} \right){\bf{E}}_{{x_l}}^T{\mathbb{P}_{xz}^{\rm{''}}}{{\bf{H}}_{{z_l}}}  \\
		&+ \left( { {\sigma _{{b_2}}} + 1} \right){\bf{E}}_{{y_b}}^T{\mathbb{P}_{yz}^{\rm{''}}}{{\bf{H}}_{{z_b}}} 
		+ \left( { {\sigma _{{f_2}}} - 1} \right){\bf{E}}_{{y_f}}^T{\mathbb{P}_{yz}^{\rm{''}}}{{\bf{H}}_{{z_f}}} \\
		&+ \left( { {\sigma _{{t_1}}} + 1} \right){\bf{E}}_{{y_t}}^T{\mathbb{P}_{yx}^{\rm{''}}}{{\bf{H}}_{{x_t}}} 
		+ \left( { {\sigma _{{d_1}}} - 1} \right){\bf{E}}_{{y_d}}^T{\mathbb{P}_{yx}^{\rm{''}}}{{\bf{H}}_{{x_d}}}\\
		&+ \left( { {\sigma _{{b_1}}} - 1} \right){\bf{E}}_{{z_b}}^T{\mathbb{P}_{zy}^{\rm{''}}}{{\bf{H}}_{{y_b}}} 
		+ \left( { {\sigma _{{f_1}}} + 1} \right){\bf{E}}_{{z_f}}^T{\mathbb{P}_{zy}^{\rm{''}}}{{\bf{H}}_{{y_f}}} \\
		&+ \left( { {\sigma _{{r_1}}} - 1} \right){\bf{E}}_{{z_r}}^T{\mathbb{P}_{zx}^{\rm{''}}}{{\bf{H}}_{{x_r}}} 
		+ \left( { {\sigma _{{l_1}}} + 1} \right){\bf{E}}_{{z_l}}^T{\mathbb{P}_{zx}^{\rm{''}}}{{\bf{H}}_{{x_l}}}.
	\end{aligned}
\end{equation}
To guarantee the stability of the system, which implies that there is no dissipation of the energy, $d{\cal E} /dt = 0$ should be satisfied. From (\ref{E17}), we can find that the stability of the proposed method can be analytically proved with appropriate free parameters. One option is
\begin{equation}\label{E18}
	\begin{aligned}
			&{\sigma _{{t_2}}} = {\sigma _{{l_2}}} = {\sigma _{{f_2}}} = {\sigma _{{d_1}}} = {\sigma _{{b_1}}} = {\sigma _{{r_1}}} = 1,\\
			&{\sigma _{{r_2}}} = {\sigma _{{d_2}}} = {\sigma _{{b_2}}} = {\sigma _{{t_1}}} = {\sigma _{{f_1}}} = {\sigma _{{l_1}}} =  - 1.
	\end{aligned}
\end{equation}

\subsection{The Proposed SBP-SAT FDTD Method with the PMC Boundary Condition}
Similarly, the penalty terms from the SAT technique are added in (\ref{E10})-(\ref{E12}) to weakly enforce the PMC boundary conditions, which can be expressed as
%\begin{equation}
	\begin{align}\label{E19}
			&\frac{{d{{\bf{E}}_x}}}{{dt}} - {\mathbb{D}_{{H_z}}^y}{{\bf{H}}_z} + {\mathbb{D}_{{H_y}}^z}{{\bf{H}}_y}  \\
			&={\chi _{{t_1}}}{\left( {\mathbb{P}_{{E_x}}^{\rm{'''}}} \right)^{ - 1}}\mathbb{R}_{{E_{{x_t}}}}^T{\mathbb{P}_{xy}^{\rm{''}}}{{\bf{H}}_{{y_t}}}
			+ {\chi _{{d_1}}}{\left( {\mathbb{P}_{{E_x}}^{\rm{'''}}} \right)^{ - 1}}\mathbb{R}_{{E_{{x_d}}}}^T{\mathbb{P}_{xy}^{\rm{''}}}{{\bf{H}}_{{y_d}}} \notag \\
			&+ {\chi _{{r_1}}}{\left( {\mathbb{P}_{{E_x}}^{\rm{'''}}} \right)^{ - 1}}\mathbb{R}_{{E_{{x_r}}}}^T{\mathbb{P}_{xz}^{\rm{''}}}{{\bf{H}}_{{z_r}}}
			+ {\chi _{{l_1}}}{\left( {\mathbb{P}_{{E_x}}^{\rm{'''}}} \right)^{ - 1}}\mathbb{R}_{{E_{{x_l}}}}^T{\mathbb{P}_{xz}^{\rm{''}}}{{\bf{H}}_{{z_l}}},\notag
	\end{align}
%\end{equation}
%\begin{equation}
	\begin{align}\label{E20}
	&\frac{{d{{\bf{E}}_y}}}{{dt}} -{\mathbb{D}_{{H_x}}^z}{{\bf{H}}_x} + {\mathbb{D}_{{H_z}}^x}{{\bf{H}}_z}  \\
	&={\chi _{{b_1}}}{\left( {\mathbb{P}_{{E_y}}^{\rm{'''}}} \right)^{ - 1}}\mathbb{R}_{{E_{{y_b}}}}^T{\mathbb{P}_{yz}^{\rm{''}}}{{\bf{H}}_{{z_b}}}
	+ {\chi _{{f_1}}}{\left( {\mathbb{P}_{{E_y}}^{\rm{'''}}} \right)^{ - 1}}\mathbb{R}_{{E_{{y_f}}}}^T{\mathbb{P}_{yz}^{\rm{''}}}{{\bf{H}}_{{z_f}}} \notag \\
	&+ {\chi _{{t_2}}}{\left( {\mathbb{P}_{{E_y}}^{\rm{'''}}} \right)^{ - 1}}\mathbb{R}_{{E_{{y_t}}}}^T{\mathbb{P}_{yx}^{\rm{''}}}{{\bf{H}}_{{x_t}}}
	+ {\chi _{{d_2}}}{\left( {\mathbb{P}_{{E_y}}^{\rm{'''}}} \right)^{ - 1}}\mathbb{R}_{{E_{{y_d}}}}^T{\mathbb{P}_{yx}^{\rm{''}}}{{\bf{H}}_{{x_d}}},\notag
	\end{align}
%\end{equation}

%\begin{equation}
	\begin{align}\label{E21}
		&\frac{{d{{\bf{E}}_z}}}{{dt}} - {\mathbb{D}_{{H_y}}^x}{{\bf{H}}_y} + {\mathbb{D}_{{H_x}}^y}{{\bf{H}}_x} \\
		&= {\chi _{{b_2}}}{\left( {\mathbb{P}_{{E_z}}^{\rm{'''}}} \right)^{ - 1}}\mathbb{R}_{{E_{{z_b}}}}^T{\mathbb{P}_{zy}^{\rm{''}}}{{\bf{H}}_{{y_b}}}
		+ {\chi _{{f_2}}}{\left( {\mathbb{P}_{{E_z}}^{\rm{'''}}} \right)^{ - 1}}\mathbb{R}_{{E_{{z_f}}}}^T{\mathbb{P}_{zy}^{\rm{''}}}{{\bf{H}}_{{y_f}}}\notag \\
		&+ {\chi _{{r_2}}}{\left( {\mathbb{P}_{{E_z}}^{\rm{'''}}} \right)^{ - 1}}\mathbb{R}_{{E_{{z_r}}}}^T{\mathbb{P}_{zx}^{\rm{''}}}{{\bf{H}}_{{x_r}}}
		+ {\chi _{{l_2}}}{\left( {\mathbb{P}_{{E_z}}^{\rm{'''}}} \right)^{ - 1}}\mathbb{R}_{{E_{{z_l}}}}^T{\mathbb{P}_{zx}^{\rm{''}}}{{\bf{H}}_{{x_l}}}.\notag
	\end{align}
%\end{equation}
For the PMC boundary condition, the proof of the stability is similar to that with the PEC boundary condition. After taking the derivative of the electromagnetic energy in the computational domain with respect to time, ${{d\cal E }}/{{dt}}$ in computational domain can be obtained by substituting (\ref{E10})-(\ref{E12}) and (\ref{E19})-(\ref{E21}) into (\ref{E16}).
\begin{equation}\label{E22}
	\begin{aligned}
			\frac{{d\cal E }}{{dt}} 
			&=  \left( {  {\chi _{{t_1}}} - 1} \right){\bf{E}}_{{x_t}}^T{\mathbb{P}_{xy}^{\rm{''}}}{{\bf{H}}_{{y_t}}} 
			+ \left( {  {\chi _{{d_1}}} + 1} \right){\bf{E}}_{{x_d}}^T{\mathbb{P}_{xy}^{\rm{''}}}{{\bf{H}}_{{y_d}}} \\
			&+ \left( {{\chi _{{r_1}}} + 1} \right){\bf{E}}_{{x_r}}^T{\mathbb{P}_{xz}^{\rm{''}}}{{\bf{H}}_{{z_r}}} 
			+ \left( { {\chi _{{l_1}}} - 1} \right){\bf{E}}_{{x_l}}^T{\mathbb{P}_{xz}^{\rm{''}}}{{\bf{H}}_{{z_l}}}\\
			&+ \left( {{\chi _{{b_1}}} + 1} \right){\bf{E}}_{{y_b}}^T{\mathbb{P}_{yz}^{\rm{''}}}{{\bf{H}}_{{z_b}}} 
			+ \left( { {\chi _{{f_1}}} - 1} \right){\bf{E}}_{{y_f}}^T{\mathbb{P}_{yz}^{\rm{''}}}{{\bf{H}}_{{z_f}}}\\
			&+ \left( { {\chi _{{t_2}}} + 1} \right){\bf{E}}_{{y_t}}^T{\mathbb{P}_{yx}^{\rm{''}}}{{\bf{H}}_{{x_t}}} 
			+ \left( {  {\chi _{{d_2}}} - 1} \right){\bf{E}}_{{y_d}}^T{\mathbb{P}_{yx}^{\rm{''}}}{{\bf{H}}_{{x_d}}}\\
			&+ \left( {  {\chi _{{b_2}}} - 1} \right){\bf{E}}_{{z_b}}^T{\mathbb{P}_{zy}^{\rm{''}}}{{\bf{H}}_{{y_b}}} 
			+ \left( { {\chi _{{f_2}}} + 1} \right){\bf{E}}_{{z_f}}^T{\mathbb{P}_{zy}^{\rm{''}}}{{\bf{H}}_{{y_f}}}\\
			&+ \left( {  {\chi _{{r_2}}} - 1} \right){\bf{E}}_{{z_r}}^T{\mathbb{P}_{zx}^{\rm{''}}}{{\bf{H}}_{{x_r}}} 
			+ \left( { {\chi _{{l_2}}} + 1} \right){\bf{E}}_{{z_l}}^T{\mathbb{P}_{zx}^{\rm{''}}}{{\bf{H}}_{{x_l}}}.
	\end{aligned}
\end{equation} 
To ensure ${{d\cal E }}/{{dt}} = 0$, the free parameters can be chosen as follows
\begin{equation}\label{E23}
	\begin{aligned}
			&{\chi _{{t_1}}} = {\chi _{{l_1}}} = {\chi _{{f_1}}} = {\chi _{{d_2}}} = {\chi _{{b_2}}} = {\chi _{{r_2}}} = 1,\\
			&{\chi _{{r_1}}} = {\chi _{{d_1}}} = {\chi _{{b_1}}} = {\chi _{{t_2}}} = {\chi _{{f_2}}} = {\chi _{{l_2}}} =  - 1.
	\end{aligned}
\end{equation}
%\subsection{The Proposed SBP-SAT FDTD Method with the PBC Boundary Conditions}
\subsection{The Proposed SBP-SAT FDTD Method with the PBC}

Since both electric and magnetic fields exist on the boundaries in the proposed SBP-SAT FDTD method, the additional penalty terms will be added in (\ref{E7})-(\ref{E12}) to weakly enforce the PBC. The phase shift $e^{-j{\alpha}_x},e^{-j{\alpha}_y},e^{-j{\alpha}_z}$ are considered in the $x$, $y$, and $z$ directions, respectively. ${\alpha}_x,{\alpha}_y,{\alpha}_z$ are defined as ${\alpha}_i = k_{0i}h_i$ where ${\bf{k_0}}=k_{0x}\widehat{\bf{x}}+k_{0y}\widehat{\bf{y}}+k_{0z}\widehat{\bf{z}}$ is the wave vector and $\widehat{\bf{x}},\widehat{\bf{y}},\widehat{\bf{z}}$ are the unit vector in the corresponding directions. $h_i$ denotes sizes of the computational domain in the $x$, $y$ and $z$ directions. (\ref{E7})-(\ref{E12}) with PBC can be written as
%\begin{equation}\label{EPBC1}
%	\begin{aligned}
\begin{align}	\label{EPBC1}
	\frac{{d{{\bf{H}}_x}}}{{dt}} &- {\mathbb{D}_{{E_y}}^z}{{\bf{E}}_y} + {\mathbb{D}_{{E_z}}^y}{{\bf{E}}_z}  \notag\\ 
	&={\sigma _{{f_1}}}{\left( {\mathbb{P}_{{H_x}}^{\rm{'''}}} \right)^{ - 1}}\mathbb{R}_{{H_{{x_r}}}}^T{\mathbb{P}_{zx}^{\rm{''}}} {\left({{\bf{E}}_{{z_r}}} - e^{-j{\alpha}_y} {{\bf{E}}_{{z_l}}}\right)}\notag\\ 
	&+ {\sigma _{{b_1}}}{\left( {\mathbb{P}_{{H_x}}^{\rm{'''}}} \right)^{ - 1}}\mathbb{R}_{{H_{{x_l}}}}^T{\mathbb{P}_{zx}^{\rm{''}}}{\left({{\bf{E}}_{{z_l}}} - e^{j{\alpha}_y} {{\bf{E}}_{{z_r}}}\right)}\\
	&+ {\sigma _{{t_1}}}{\left( {\mathbb{P}_{{H_x}}^{\rm{'''}}} \right)^{ - 1}}\mathbb{R}_{{H_{{x_t}}}}^T{\mathbb{P}_{yx}^{\rm{''}}}{\left({{\bf{E}}_{{y_t}}} - e^{-j{\alpha}_z} {{\bf{E}}_{{y_d}}}\right)}\notag\\
	&+ {\sigma _{{d_1}}}{\left( {\mathbb{P}_{{H_x}}^{\rm{'''}}} \right)^{ - 1}}\mathbb{R}_{{H_{{x_d}}}}^T{\mathbb{P}_{yx}^{\rm{''}}}{\left({{\bf{E}}_{{y_d}}} - e^{j{\alpha}_z} {{\bf{E}}_{{y_t}}}\right)},\notag
\end{align}
%	\end{aligned}													
%\end{equation}

%\begin{equation}\label{EPBC2}
%	\begin{aligned}
	\begin{align}\label{EPBC2}
	\frac{{d{{\bf{H}}_y}}}{{dt}} &- {\mathbb{D}_{{E_z}}^x}{{\bf{E}}_z} + {\mathbb{D}_{{E_x}}^z}{{\bf{E}}_x} \notag\\
	&={\sigma _{{l_1}}}{\left( {\mathbb{P}_{{H_y}}^{\rm{'''}}} \right)^{ - 1}}\mathbb{R}_{{H_{{y_b}}}}^T{\mathbb{P}_{zy}^{\rm{''}}}{\left({{\bf{E}}_{{z_b}}} - e^{-j{\alpha}_x} {{\bf{E}}_{{z_f}}}\right)}\notag\\
	&+ {\sigma _{{r_1}}}{\left( {\mathbb{P}_{{H_y}}^{\rm{'''}}} \right)^{ - 1}}\mathbb{R}_{{H_{{y_f}}}}^T{\mathbb{P}_{zy}^{\rm{''}}}{\left({{\bf{E}}_{{z_f}}} - e^{j{\alpha}_x} {{\bf{E}}_{{z_b}}}\right)}\\
	&+ {\sigma _{{t_2}}}{\left( {\mathbb{P}_{{H_y}}^{\rm{'''}}} \right)^{ - 1}}\mathbb{R}_{{H_{{y_a}}}}^T{\mathbb{P}_{xy}^{\rm{''}}}{\left({{\bf{E}}_{{x_t}}} - e^{-j{\alpha}_z}{{\bf{E}}_{{x_d}}}\right)}\notag\\
	&+ {\sigma _{{d_2}}}{\left( {\mathbb{P}_{{H_y}}^{\rm{'''}}} \right)^{ - 1}}\mathbb{R}_{{H_{{y_d}}}}^T{\mathbb{P}_{xy}^{\rm{''}}}{\left({{\bf{E}}_{{x_d}}} - e^{j{\alpha}_z}{{\bf{E}}_{{x_t}}}\right)},\notag
\end{align}
%	\end{aligned}													
%\end{equation}

%\begin{equation}\label{EPBC3}
%	\begin{aligned}
	\begin{align}\label{EPBC3}
	\frac{{d{{\bf{H}}_z}}}{{dt}} &- {\mathbb{D}_{{E_x}}^y}{{\bf{E}}_x} + {\mathbb{D}_{{E_y}}^x}{{\bf{E}}_y}\notag\\  
	&={\sigma _{{f_2}}}{\left( {\mathbb{P}_{{H_z}}^{\rm{'''}}} \right)^{ - 1}}\mathbb{R}_{{H_{{z_r}}}}^T{\mathbb{P}_{xz}^{\rm{''}}}{\left({{\bf{E}}_{{x_r}}} - e^{-j{\alpha}_y} {{\bf{E}}_{{x_l}}}\right)}\notag\\
	&+ {\sigma _{{b_2}}}{\left( {\mathbb{P}_{{H_z}}^{\rm{'''}}} \right)^{ - 1}}\mathbb{R}_{{H_{{z_l}}}}^T{\mathbb{P}_{xz}^{\rm{''}}}{\left({{\bf{E}}_{{x_l}}} - e^{j{\alpha}_y} {{\bf{E}}_{{x_r}}}\right)}\\
	&+ {\sigma _{{l_2}}}{\left( {\mathbb{P}_{{H_z}}^{\rm{'''}}} \right)^{ - 1}}\mathbb{R}_{{H_{{z_b}}}}^T{\mathbb{P}_{yz}^{\rm{''}}}{\left({{\bf{E}}_{{y_b}}} - e^{-j{\alpha}_x} {{\bf{E}}_{{y_f}}}\right)}\notag\\
	&+ {\sigma _{{r_2}}}{\left( {\mathbb{P}_{{H_z}}^{\rm{'''}}} \right)^{ - 1}}\mathbb{R}_{{H_{{z_f}}}}^T{\mathbb{P}_{yz}^{\rm{''}}}{\left({{\bf{E}}_{{y_f}}} -e^{j{\alpha}_x} {{\bf{E}}_{{y_b}}}\right)},\notag
\end{align}
%	\end{aligned}													
%\end{equation}

%\begin{equation}\label{EPBC4}
%	\begin{aligned}
	\begin{align}\label{EPBC4}
		\frac{{d{{\bf{E}}_x}}}{{dt}} &- {\mathbb{D}_{{H_z}}^y}{{\bf{H}}_z} + {\mathbb{D}_{{H_y}}^z}{{\bf{H}}_y}  \notag\\
		&={\chi _{{t_1}}}{\left( {\mathbb{P}_{{E_x}}^{\rm{'''}}} \right)^{ - 1}}\mathbb{R}_{{E_{{x_a}}}}^T{\mathbb{P}_{xy}^{\rm{''}}}{\left({{\bf{H}}_{{y_t}}} - e^{-j{\alpha}_z}{{\bf{H}}_{{y_d}}}\right)}\notag\\
		&+ {\chi _{{d_1}}}{\left( {\mathbb{P}_{{E_x}}^{\rm{'''}}} \right)^{ - 1}}\mathbb{R}_{{E_{{x_d}}}}^T{\mathbb{P}_{xy}^{\rm{''}}}{\left({{\bf{H}}_{{y_d}}} - e^{j{\alpha}_z}{{\bf{H}}_{{y_t}}}\right)} \\
		&+ {\chi _{{r_1}}}{\left( {\mathbb{P}_{{E_x}}^{\rm{'''}}} \right)^{ - 1}}\mathbb{R}_{{E_{{x_r}}}}^T{\mathbb{P}_{xz}^{\rm{''}}}{\left({{\bf{H}}_{{z_r}}} -e^{-j{\alpha}_y} {{\bf{H}}_{{z_l}}}\right)}\notag\\
		&+ {\chi _{{l_1}}}{\left( {\mathbb{P}_{{E_x}}^{\rm{'''}}} \right)^{ - 1}}\mathbb{R}_{{E_{{x_l}}}}^T{\mathbb{P}_{xz}^{\rm{''}}}{\left({{\bf{H}}_{{z_l}}} -e^{j{\alpha}_y} {{\bf{H}}_{{z_r}}}\right)},\notag
	\end{align}
%	\end{aligned}													
%\end{equation}
%\begin{equation}\label{EPBC5}
%	\begin{aligned}
	\begin{align}\label{EPBC5}
		\frac{{d{{\bf{E}}_y}}}{{dt}} &-{\mathbb{D}_{{H_x}}^z}{{\bf{H}}_x} + {\mathbb{D}_{{H_z}}^x}{{\bf{H}}_z} \notag \\
		&={\chi _{{b_1}}}{\left( {\mathbb{P}_{{E_y}}^{\rm{'''}}} \right)^{ - 1}}\mathbb{R}_{{E_{{y_b}}}}^T{\mathbb{P}_{yz}^{\rm{''}}}{\left({{\bf{H}}_{{z_b}}} -e^{-j{\alpha}_x} {{\bf{H}}_{{z_f}}}\right)}\notag\\
		&+ {\chi _{{f_1}}}{\left( {\mathbb{P}_{{E_y}}^{\rm{'''}}} \right)^{ - 1}}\mathbb{R}_{{E_{{y_f}}}}^T{\mathbb{P}_{yz}^{\rm{''}}}{\left({{\bf{H}}_{{z_f}}} -e^{j{\alpha}_x} {{\bf{H}}_{{z_b}}}\right)}\\
		&+ {\chi _{{t_2}}}{\left( {\mathbb{P}_{{E_y}}^{\rm{'''}}} \right)^{ - 1}}\mathbb{R}_{{E_{{y_a}}}}^T{\mathbb{P}_{yx}^{\rm{''}}}{\left({{\bf{H}}_{{x_t}}} - e^{-j{\alpha}_z}{{\bf{H}}_{{x_d}}}\right)}\notag\\
		&+ {\chi _{{d_2}}}{\left( {\mathbb{P}_{{E_y}}^{\rm{'''}}} \right)^{ - 1}}\mathbb{R}_{{E_{{y_d}}}}^T{\mathbb{P}_{yx}^{\rm{''}}}{\left({{\bf{H}}_{{x_d}}} - e^{j{\alpha}_z}{{\bf{H}}_{{x_t}}}\right)},\notag
	\end{align}
%	\end{aligned}													
%\end{equation}

%\begin{equation}\label{EPBC6}
%	\begin{aligned}
	\begin{align}\label{EPBC6}
		\frac{{d{{\bf{E}}_z}}}{{dt}} &- {\mathbb{D}_{{H_y}}^x}{{\bf{H}}_y} + {\mathbb{D}_{{H_x}}^y}{{\bf{H}}_x} \notag\\
		&= {\chi _{{b_2}}}{\left( {\mathbb{P}_{{E_z}}^{\rm{'''}}} \right)^{ - 1}}\mathbb{R}_{{E_{{z_b}}}}^T{\mathbb{P}_{zy}^{\rm{''}}}{\left({{\bf{H}}_{{y_b}}} - e^{-j{\alpha}_z}{{\bf{H}}_{{y_f}}}\right)}\notag\\
		&+ {\chi _{{f_2}}}{\left( {\mathbb{P}_{{E_z}}^{\rm{'''}}} \right)^{ - 1}}\mathbb{R}_{{E_{{z_f}}}}^T{\mathbb{P}_{zy}^{\rm{''}}}{\left({{\bf{H}}_{{y_f}}} - e^{j{\alpha}_z}{{\bf{H}}_{{y_b}}}\right)}\\
		&+ {\chi _{{r_2}}}{\left( {\mathbb{P}_{{E_z}}^{\rm{'''}}} \right)^{ - 1}}\mathbb{R}_{{E_{{z_r}}}}^T{\mathbb{P}_{zx}^{\rm{''}}}{\left({{\bf{H}}_{{x_r}}} - e^{-j{\alpha}_z}{{\bf{H}}_{{x_l}}}\right)}\notag\\
		&+ {\chi _{{l_2}}}{\left( {\mathbb{P}_{{E_z}}^{\rm{'''}}} \right)^{ - 1}}\mathbb{R}_{{E_{{z_l}}}}^T{\mathbb{P}_{zx}^{\rm{''}}}{\left({{\bf{H}}_{{x_l}}} - e^{j{\alpha}_z}{{\bf{H}}_{{x_r}}}\right)}.\notag
	\end{align}
%	\end{aligned}													
%\end{equation}
By substituting (\ref{EPBC1})-(\ref{EPBC6}) into (\ref{E16}) and taking the partial derivative with respect to time, ${d\cal E }/{dt}$ with the PBC can be expressed as

%\begin{equation}\label{EPBCE}
%	\begin{aligned}
	\begin{align}\label{EPBCE}
		&\frac{{d\cal E }}{{dt}} 
		= \left( { {\sigma _{{t_2}}} + {\chi _{{t_1}}} - 1} \right){\bf{E}}_{{x_t}}^T{\mathbb{P}_{xy}^{\rm{''}}}{{\bf{H}}_{{y_t}}} \notag\\
		&+ \left( {{\sigma _{{d_2}}} + {\chi _{{d_1}}} + 1} \right){\bf{E}}_{{x_d}}^T{\mathbb{P}_{xy}^{\rm{''}}}{{\bf{H}}_{{y_d}}} \\
		&+ \left( {{\sigma _{{r_2}}} + {\chi _{{r_1}}} + 1} \right){\bf{E}}_{{x_r}}^T{\mathbb{P}_{xz}^{\rm{''}}}{{\bf{H}}_{{z_r}}} 
		+ \left( {{\sigma _{{l_2}}} + {\chi _{{l_1}}} - 1} \right){\bf{E}}_{{x_l}}^T{\mathbb{P}_{xz}^{\rm{''}}}{{\bf{H}}_{{z_l}}} \notag\\
		&+ \left( {{\sigma _{{b_2}}} + {\chi _{{b_1}}} + 1} \right){\bf{E}}_{{y_b}}^T{\mathbb{P}_{yz}^{\rm{''}}}{{\bf{H}}_{{z_b}}}  
		+ \left( {{\sigma _{{f_2}}} + {\chi _{{f_1}}} - 1} \right){\bf{E}}_{{y_f}}^T{\mathbb{P}_{yz}^{\rm{''}}}{{\bf{H}}_{{z_f}}} \notag\\
		&+ \left( {{\sigma _{{t_1}}} + {\chi _{{t_2}}} + 1} \right){\bf{E}}_{{y_t}}^T{\mathbb{P}_{yx}^{\rm{''}}}{{\bf{H}}_{{x_t}}}  
		+ \left( {{\sigma _{{d_1}}} + {\chi _{{d_2}}} - 1} \right){\bf{E}}_{{y_d}}^T{\mathbb{P}_{yx}^{\rm{''}}}{{\bf{H}}_{{x_d}}} \notag\\
		&+ \left( {{\sigma _{{b_1}}} + {\chi _{{b_2}}} - 1} \right){\bf{E}}_{{z_b}}^T{\mathbb{P}_{zy}^{\rm{''}}}{{\bf{H}}_{{y_b}}}  
		+ \left( {{\sigma _{{f_1}}} + {\chi _{{f_2}}} + 1} \right){\bf{E}}_{{z_f}}^T{\mathbb{P}_{zy}^{\rm{''}}}{{\bf{H}}_{{y_f}}}\notag\\
		&+ \left( {{\sigma _{{r_1}}} + {\chi _{{r_2}}} - 1} \right){\bf{E}}_{{z_r}}^T{\mathbb{P}_{zx}^{\rm{''}}}{{\bf{H}}_{{x_r}}}  
		+ \left( {{\sigma _{{l_1}}} + {\chi _{{l_2}}} + 1} \right){\bf{E}}_{{z_l}}^T{\mathbb{P}_{zx}^{\rm{''}}}{{\bf{H}}_{{x_l}}}.\notag
	\end{align}
%	\end{aligned}													
%\end{equation}

In order to ensure the stability, $\sigma$ and $\chi$ can be
 \begin{equation}\label{EPBCP}
 	\begin{aligned}
 		&{\chi _{{t_1}}} = {\chi _{{l_1}}} = {\chi _{{f_1}}} = {\chi _{{d_2}}} = {\chi _{{b_2}}} = {\chi _{{r_2}}} = \frac{1}{2},\\
 		&{\chi _{{r_1}}} = {\chi _{{d_1}}} = {\chi _{{b_1}}} = {\chi _{{t_2}}} = {\chi _{{f_2}}} = {\chi _{{l_2}}} =  - \frac{1}{2},\\
 		&{\sigma _{{t_2}}} = {\sigma _{{l_2}}} = {\sigma _{{f_2}}} = {\sigma _{{d_1}}} = {\sigma _{{b_1}}} = {\sigma _{{r_1}}} = \frac{1}{2},\\
 		&{\sigma _{{r_2}}} = {\sigma _{{d_2}}} = {\sigma _{{b_2}}} = {\sigma _{{t_1}}} = {\sigma _{{f_1}}} = {\sigma _{{l_1}}} =  - \frac{1}{2}.
 	\end{aligned}
 \end{equation}

\section{Dispersion Analysis}
In order to illustrate the numerical dispersion error of the proposed SBP-SAT FDTD method in free space, a cubic computation domain with PBC is considered. The whole computational domain is filled with air. The analytic wavenumber is $k_0=w{\sqrt{{{\varepsilon}_0}{\mu}_0}}$. The numerical wavenumber $\tilde{k}_0=\tilde{k}_{real}+ j \tilde{k}_{imag}$ can be calculated by the SBP-SAT FDTD method and the FDTD method. The dispersion error, dissipation error and global error are defined similar to \cite{dispersionDGTD} as 
\begin{align}
	&{\text{Dispersion~Error:}} \left| e^{-jk_0{\lambda}}-e^{-j\tilde{k}_{real}{\lambda}} \right|, \label{E_DISPER} \\
	&{\text{Dissipation~Error:}} \left| 1-e^{-j{\tilde{k}_{imag}}{\lambda}} \right|, \label{E_DISSIP}\\
	&{\text{Global~Error:}} \left| e^{-jk_0{\lambda}}-e^{-j\tilde{k}_{0}{\lambda}} \right|, \label{E_GLOB}
\end{align}
where $\lambda$ is the wave length and satisfies $\lambda=2\pi/k_0$.

$\tilde{k}_0$ can be calculated by eigenvalues of the amplification matrix. A vector ${\bf U}^n = \left[{\bf{E}}_x^n,\,{\bf{E}}_y^n,\,{\bf{E}}_z^n,\,{\bf{H}}_x^{n-\frac{1}{2}},\,{\bf{H}}_y^{n-\frac{1}{2}},\,{\bf{H}}_z^{n-\frac{1}{2}}\right]^T$, where ${\bf{E}}_x^{n}$, ${\bf{E}}_y^{n}$, ${\bf{E}}_z^{n}$, ${\bf{H}}_x^{n-\frac{1}{2}}$, ${\bf{H}}_y^{n-\frac{1}{2}}$, and ${\bf{H}}_z^{n-\frac{1}{2}}$ containing all field nodes in the $x$, $y$, and $z$ directions, is defined. Take ${\bf{E}}_x^n$ as an example, ${{\bf{E}}_x^n} = \left[{ E}_x|^n_{1,1,1}, \,{\ E}_x|_{2,1,1}^{n},\,\dots,\,{ E}_x|_{m,p+1,q+1}^{n}\right]^T$, where the subscripts $m$, $p$, and $q$ denote field components' indices in $x$, $y$ and $z$ directions, respectively. By using ${\bf U}^n$, the time-marching formulations in the FDTD method or the SBP-SAT FDTD method in the whole computational domain can be written as 
\begin{equation}\label{EDISPLAMBDA}
	\begin{aligned}
		e^{j\omega \Delta t}{\bf{U}}^{n}
		=
		{\bf{\Lambda}} {\bf{U}}^{n}.
	\end{aligned}
\end{equation}
After eigenvalues of $	{\bf{\Lambda}}$ are solved,  $\tilde{k}_0$ can be calculated as 
\begin{equation}\label{E_NUMWN}
	\begin{aligned}
		\tilde{k}_0^m = \frac{{\rm{ln}}\left(\lambda^m \right)}{jc{\Delta}t}, 
	\end{aligned}
\end{equation}
where $c = w/k_0$ is the speed of light in vacuum, the superscript $m$ denote the $m$th eigenvalue of matrix ${\bf{\Lambda}}$. Since the eigenvalues of matrix ${\bf{\Lambda}}$ are different, we choose the $\tilde{k}_0^m$ nearest to the analytic wavenumber $k_0$ to calculate the numerical error of the SBP-SAT FDTD method and the FDTD method.

\begin{figure*}[htbp]
	\subfigure[]{
		\includegraphics[width=0.325\linewidth]{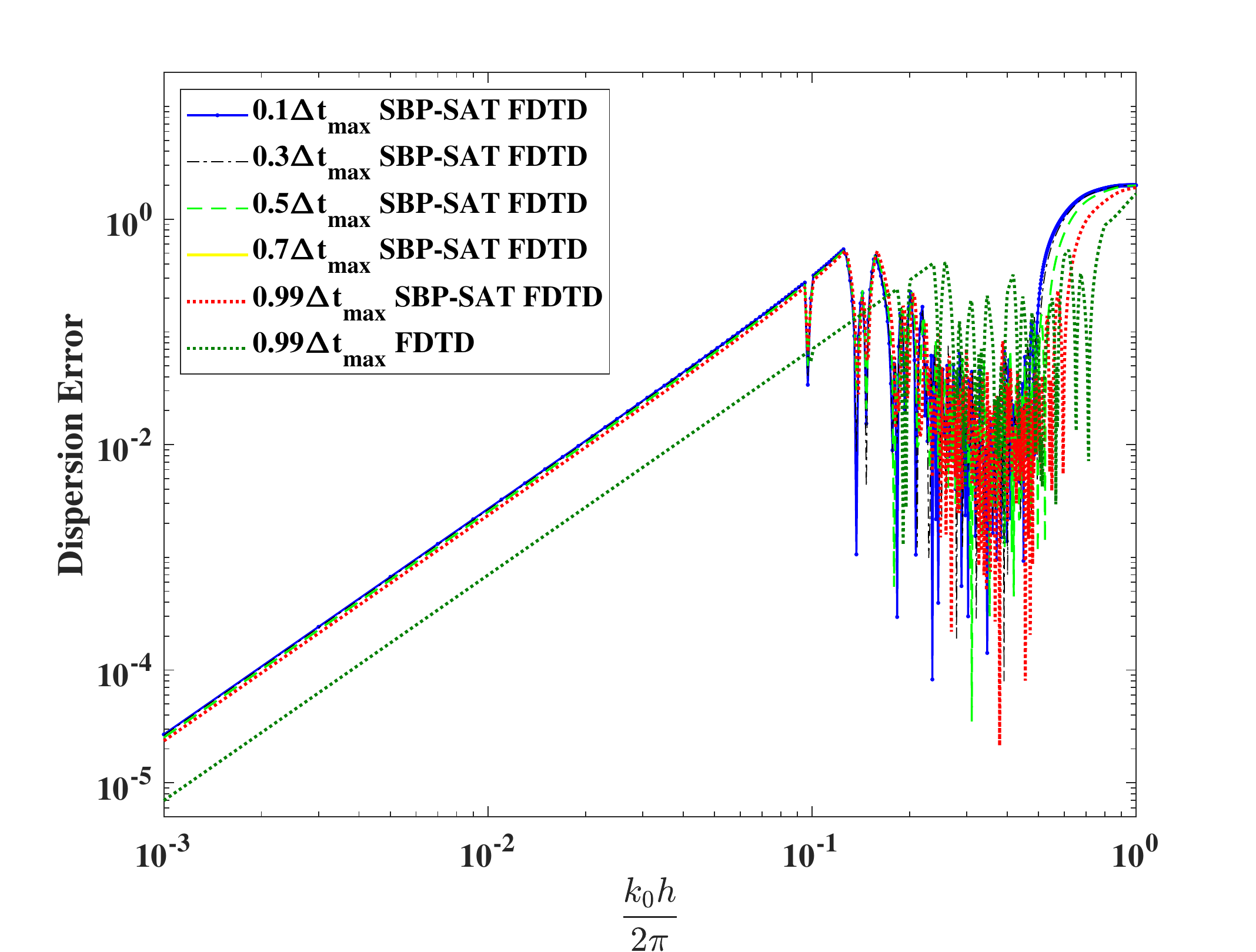}
	}
	\subfigure[]{
		\includegraphics[width=0.325\linewidth]{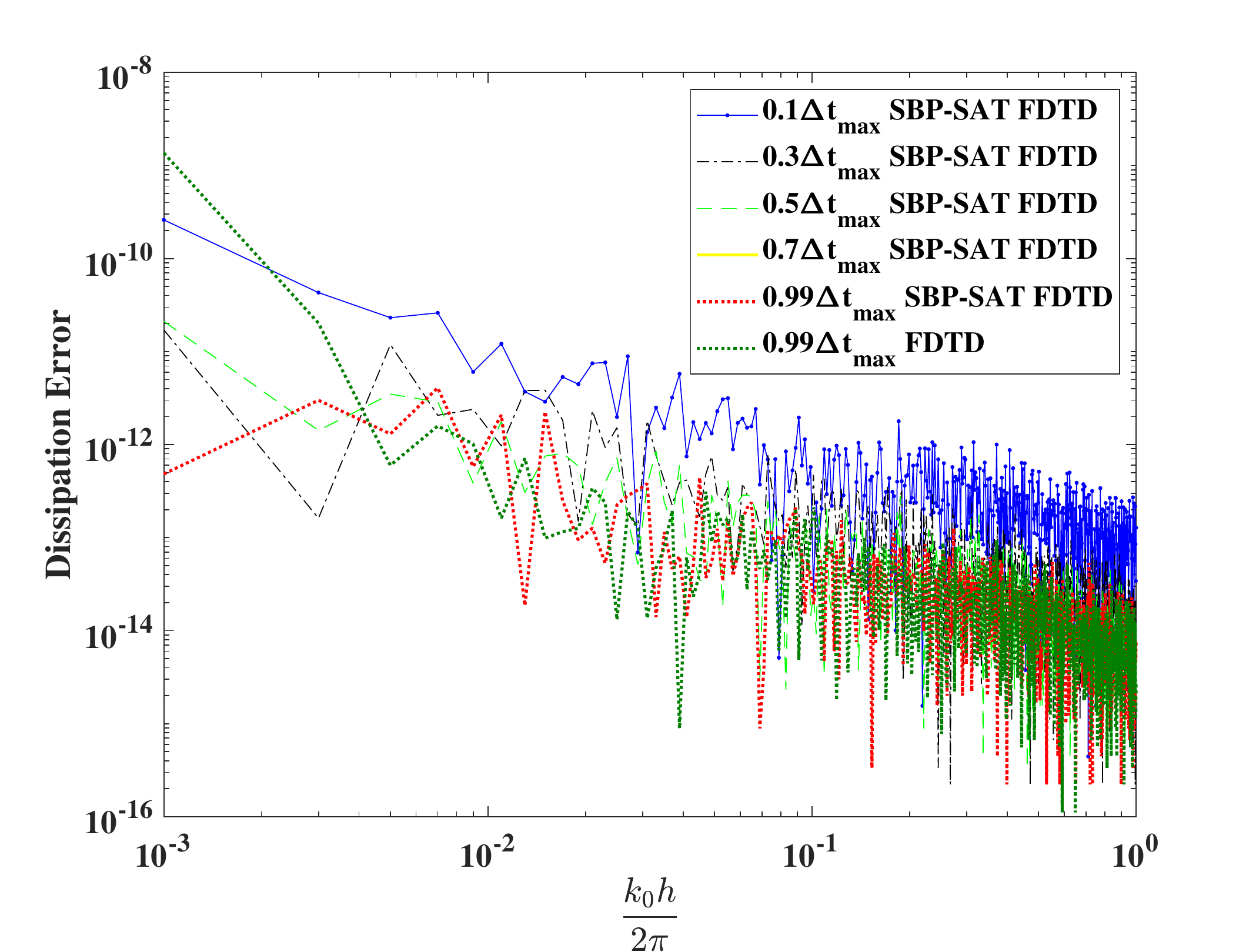}
	}
	\subfigure[]{
		\includegraphics[width=0.325\linewidth]{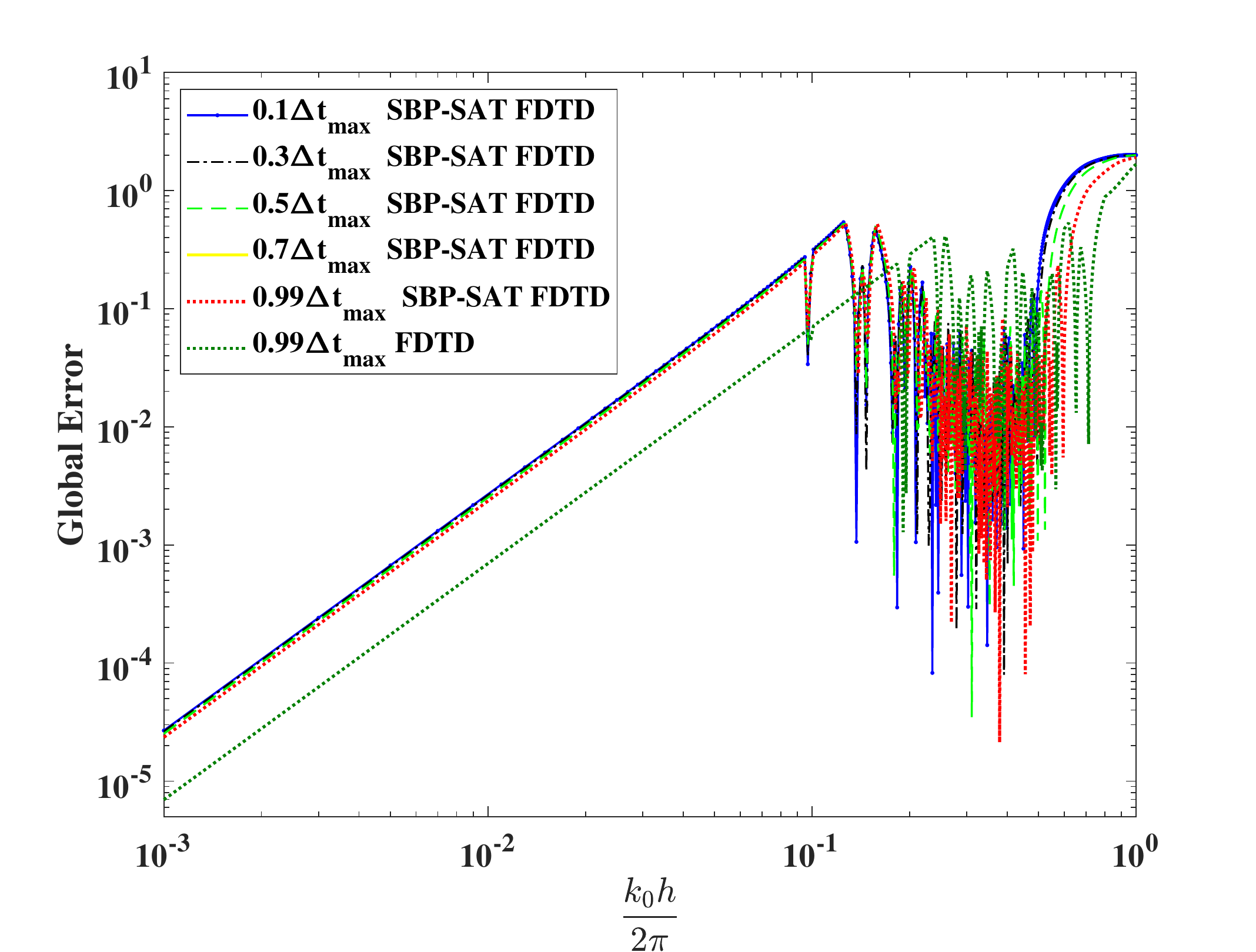}
	}
	\caption{The numerical error of the SBP-SAT FDTD method and the FDTD method when normal incidence: (a) the dispersion error, (b) the dissipation error, (c) the global error.}
	\label{F_ERR_z}
\end{figure*}

We set the phase shift of PBC as $e^{-j{\alpha}_z}$ in the $z$ direction, where ${\alpha}_z = k_{0z}h_z$, and zero in the $x$ and $y$ directions to simulate a plane wave propagating along the $z$ direction. The numerical dispersion error of the SBP-SAT FDTD method and the FDTD method with different time steps are shown in Fig. \ref{F_ERR_z}. It can be noted that both the FDTD method and the proposed SBP-SAT FDTD method have relatively large numerical dispersion error when $k_0h/2\pi \ge 1/10$. Comparing with the FDTD method, the SBP-SAT FDTD method has larger dispersion error and less dissipation error. However, the dissipation errors in two methods are much less than dispersion errors, which can be negligible. Therefore, the numerical dispersion errors in two methods mainly depend on the dispersion errors.

\begin{figure*}[htbp]
	\subfigure[]{
		\includegraphics[width=0.23\linewidth]{Picture/NumError/fdtd_sesan.pdf}
	}
	\subfigure[]{
		\includegraphics[width=0.23\linewidth]{Picture/NumError/fdtd_zong.pdf}
	}
	\subfigure[]{
		\includegraphics[width=0.23\linewidth]{Picture/NumError/sbp_sesan.pdf}
	}
	\subfigure[]{
		\includegraphics[width=0.24\linewidth]{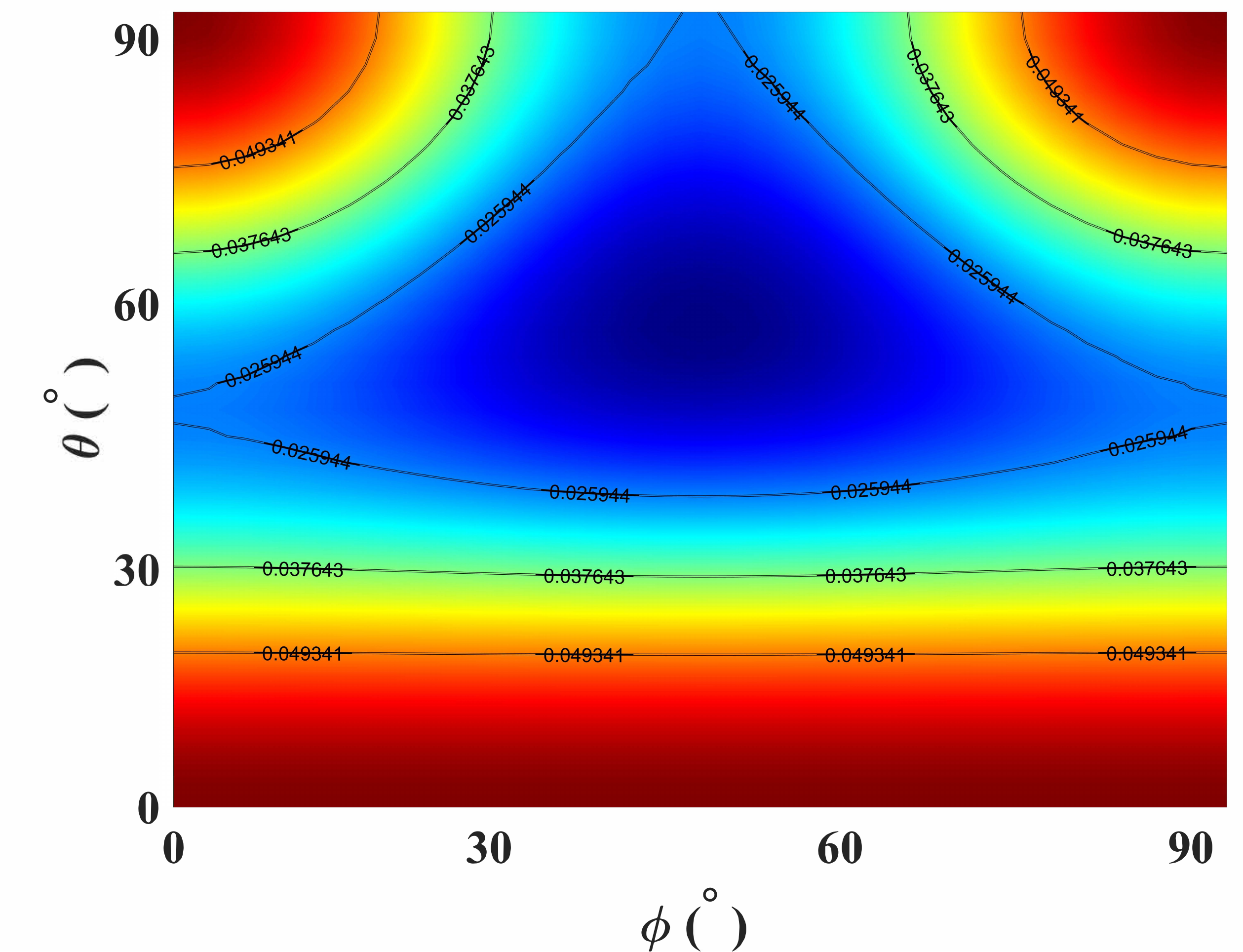}
	}
	\caption{The numerical dispersion error of the SBP-SAT FDTD method and the FDTD method verse ${\theta}$ and ${\phi}$: (a) the dispersion error of the FDTD method, (b) the global error of the FDTD method, (c) the dispersion error of the SBP-SAT FDTD method, (d) the global error of the SBP-SAT FDTD method.}
	\label{F_ERR_FC}
\end{figure*}

When the plane wave obliquely incidents, the phase shift  $e^{-j{\alpha}_x},e^{-j{\alpha}_y}$, and $e^{-j{\alpha}_z}$ in the $x$, $y$, and $z$ directions must be considered. $k_{0x}, k_{0y}, k_{0z}$ can be calculated as $k_{0x} = k_{0}sin{\theta}cos{\phi}$, $k_{0y} = k_{0}sin{\theta}sin{\phi}$, $k_{0z} = k_{0}cos{\theta}$, where ${\theta}$ and ${\phi}$ are the azimuth and zenith angles. ${\Delta}t=0.99{\Delta}t_{max}$ and the mesh size of $\lambda/20$ are used. The numerical errors obtained from the FDTD method and the SBP-SAT FDTD method verse ${\theta}$ and ${\phi}$ are shown in Fig. \ref{F_ERR_FC}. It can be found that the numerical error in two methods is the largest when the plane wave normally propagate along the $x$, $y$, and $z$ directions. Although the numerical dispersion error of the proposed SBP-SAT FDTD
method is slightly larger than that in the FDTD method, they have the same level of accuracy in the practical simulations as shown in our numerical results.

\section{Practical Implementation and Efficiency  Comparison}
After the central finite-difference scheme is used in the temporal domain, the leapfrog time-marching formulations can be obtained.
As shown in Section III, several matrices are used in the semi-discrete formulations.  It should be noted those matrices do not exist in the practical implementations. The time-marching procedure can be done in a matrix-free manner, which is similar to that in the FDTD mehtod. To make this point clear, ${E_x}$ and ${H_x}$ are used as examples to demonstrate the efficiency and the practical implementations of the SBP-SAT FDTD method.

Let's consider a rectangular cavity with PEC walls. It is discretized as ${N_x} \times {N_y} \times {N_z}$ cells, where ${N_x}$, ${N_y}$, and ${N_z}$ are the overall cell numbers in the $x$, $y$, and $z$ directions, respectively. In the SBP-SAT FDTD method,  ${E_x}$ is stored using an array with dimension of $\left( {{N_x} + 2} \right) \times \left( {{N_y} + 1} \right) \times \left( {{N_z} + 1} \right)$, and an array with dimension of $\left( {{N_x} + 1} \right) \times \left( {{N_y} + 2} \right) \times \left( {{N_z} + 2} \right)$ for ${H_x}$ is required. In the FDTD method, ${E_x}$ and ${H_x}$ are stored in the arrays with dimension of ${N_x} \times \left( {{N_y} + 1} \right) \times \left( {{N_z} + 1} \right)$, $\left( {{N_x} + 1} \right) \times {N_y} \times {N_z}$. It is obvious that the overhead of memory usage is $\left( {{N_y} + 1} \right) \times \left( {{N_z} + 1} \right)$ for ${E_x}$, and $2\left( {{N_x} + 1} \right) \times \left( {N_y} + {{N_z}+ 2} \right) $ for ${H_x}$. Compared with the overall memory usage to store the three-dimensional electromagnetic fields, those overheads to store extra nodes on the boundaries of computational domain are negligible. Therefore, memory usage of the SBP-SAT FDTD method is almost the same as that of the FDTD method. 

Since the second-order central finite-difference scheme is considered in our implementation, the first and last two rows of ${\mathbb{D}_{{H_y}}^z}$ have only three non-zero entities as shown in (\ref{Dplus}), and other rows have two non-zero values. Therefore, the elemental time-marching formulation for $E_x$ in the $z$ direction can be expressed as
%\begin{equation}\label{E24}
%	\begin{aligned}
%		\left. {{{E}_x}} \right|_{i,j,k}^{n + 1} = \left. {{{E}_x}} \right|_{i,j,k}^n - \sum\limits_{m = 1}^{Nz + 1} {{d_{km}}} \left. {{{H}_y}} \right|_{i,j,m}^n,
%	\end{aligned}
%\end{equation}
\begin{equation}\label{E24}
	\begin{aligned}
		\left. {{{\bf{E}}_x}} \right|_{i,j,k}^{n + 1} = \left\{ {\begin{array}{*{20}{c}}
				\begin{array}{l}
					\left. {{{\bf{E}}_x}} \right|_{i,j,k}^n - \Delta t\sum\limits_{m = 1}^3 {{d_{km}}} \left. {{{\bf{H}}_y}} \right|_{i,j,m}^n,\\
					k = 1,2,
				\end{array}\\
				\begin{array}{l}
					\left. {{{\bf{E}}_x}} \right|_{i,j,k}^n - \Delta t\sum\limits_{m = k}^{k + 1} {{d_{km}}} \left. {{{\bf{H}}_y}} \right|_{i,j,m}^n,\\
					k = 3, \cdots ,{N_z} - 1,
				\end{array}\\
				\begin{array}{l}
					\left. {{{\bf{E}}_x}} \right|_{i,j,k}^n - \Delta t\sum\limits_{m = {N_z}}^{{N_z} + 2} {{d_{km}}} \left. {{{\bf{H}}_y}} \right|_{i,j,m}^n,\\
					k = {N_z},{N_z} + 1,
				\end{array}
		\end{array}} \right.
	\end{aligned}
\end{equation}
%\begin{equation}\label{E25}
%	\begin{aligned}
%		\left. {{{E}_x}} \right|_{i,j,k}^{n + 1} = \left. {{{E}_x}} \right|_{i,j,k}^n - \sum\limits_{m = k}^{k+1} {{d_{km}}} \left. {{{H}_y}} \right|_{i,j,m}^n, k=others
%	\end{aligned}
%\end{equation}
%\begin{equation}\label{E26}
%	\begin{aligned}
%		\left. {{{E}_x}} \right|_{i,j,k}^{n + 1} = \left. {{{E}_x}} \right|_{i,j,k}^n - \sum\limits_{m = N_z}^{N_z+2} {{d_{km}}} \left. {{{H}_y}} \right|_{i,j,m}^n, k=N_z,N_z+1 
%	\end{aligned}
%\end{equation}
where ${d_{km}}$ is the entity in the $k$th row and the $m$th column of ${\mathbb{D}_{{H_y}}^z}$.

\begin{algorithm}[h]
	\caption{Calculate ${E_x}$ in the $z$ direction} %算法的名字
	\hspace*{0.02in} {\bf Input:} %算法的输入， \hspace*{0.02in}用来控制位置，同时利用 \\ 进行换行
	${E_x}$, ${H_y}$\\
	\hspace*{0.02in} {\bf Output:} %算法的结果输出
	${E_x}$
	\begin{algorithmic}[1]
%		\State some description % \State 后写一般语句
		\For{$i=1:N_x+2$} % For 语句，需要和EndFor对应
			\For{$j=1:N_y+1$}
				\State $E_x|_{i,j,1}$ = Boundary-difference($H_y$)
				\State $E_x|_{i,j,2}$ = Boundary-difference($H_y$)
				\For{$k=3:N_z-1$}
					\State $E_x|_{i,j,k}$ = Finite-difference($H_y$)
				\EndFor
				\State $E_x|_{i,j,Nz}$ = Boundary-difference($H_y$)
				\State $E_x|_{i,j,Nz+1}$ = Boundary-difference($H_y$)
			\EndFor
		\EndFor
%		\State ...
%		\If{condition} % If 语句，需要和EndIf对应
%		\State ...
%		\Else
%		\State ...
%		\EndIf
%		\EndFor
%		\While{condition} % While语句，需要和EndWhile对应
%		\State ...
%		\EndWhile
%		\State 
	\end{algorithmic}
\end{algorithm}
To further demonstrate the time-marching procedure, $E_x$ in the $z$ direction is calculated through {\textbf{Algorithm 1}}. It can be found that the SBP-SAT FDTD method needs to add two additional $E_x$ components in the $x$ direction. Therefore, it needs to carry out two more surface component calculations and to perform special treatment in the time-marching procedure on the boundaries. However, the overall  count only slightly increase compared with that of the FDTD method. Therefore, the efficiency of the proposed SBP-SAT FDTD method is almost the same as that of the FDTD method. 

\begin{algorithm}[h]
	\caption{Calculate ${H_x}$ in the $z$ direction} %算法的名字
	\hspace*{0.02in} {\bf Input:} %算法的输入， \hspace*{0.02in}用来控制位置，同时利用 \\ 进行换行
	${H_x}$, ${E_y}$\\
	\hspace*{0.02in} {\bf Output:} %算法的结果输出
	${H_x}$
	\begin{algorithmic}[1]
		%		\State some description % \State 后写一般语句
		\For{$i=1:N_x+1$} % For 语句，需要和EndFor对应
		\For{$j=1:N_y+2$}
		\State $H_x|_{i,j,1}$ = Boundary-difference($E_y$)
		\State $H_x|_{i,j,2}$ = Boundary-difference($E_y$)
		\State $H_x|_{i,j,3}$ = Boundary-difference($E_y$)
		\For{$k=4:N_z-1$}
		\State $H_x|_{i,j,k}$ = Finite-difference($E_y$)
		\EndFor
		\State $H_x|_{i,j,Nz}$ = Boundary-difference($E_y$)
		\State $H_x|_{i,j,Nz+1}$ = Boundary-difference($E_y$)
		\State $H_x|_{i,j,Nz+2}$ = Boundary-difference($E_y$)
		\EndFor
		\EndFor
		\For{$i=1:N_x+1$} % For 语句，需要和EndFor对应
		\For{$j=1:N_y+2$}
			\State $H_x|_{i,j,1}$ = SAT-boundary($E_y|_{i,j,1}$)
			\State $H_x|_{i,j,Nz+2}$ = SAT-boundary($E_y|_{i,j,Nz+1}$)
		\EndFor
		\EndFor
	\end{algorithmic}
\end{algorithm}

Similarly, ${H_x}$ in the $z$ direction is calculated through {\textbf{Algorithm 2}}. Since the SATs are added on magnetic fields for PEC boundary conditions, extra operations are required to handle them on the boundaries of computational domain. When $H_x$ is calculated in the $z$ direction, it should be corrected by electric fields on the two $xoy$ planes additionally to satisfy the PEC boundary conditions. Since the overall count of nodes on the boundaries is relatively small compared with the overall number of spatial components, the overhead of runtime is negligible, especially for the large-scale problems.

Although the SATs impose a small overhead on runtime and memory usage, they can provide extra flexibility and some attractive properties to the proposed method. It provides many possibilities for the FDTD methods, such as the theoretically stable subgridding FDTD method, the {\it{hp}}-refinement techniques, and the energy stable hybrid time-domain method. Another follow-up article will report results upon the theoretically stable subgridding method based on the proposed three-dimensional SBP-SAT FDTD method.

%\section{Dispersion Analysis}

\section{Numerical Examples}
In this section, four numerical examples are carried out to validate the effectiveness of the proposed three-dimensional SAT-SBP FDTD method, which include a cavity with PEC boundary conditions, a dielectric rod (DR) resonator, an iris filter and the specific absorption rate (SAR) calculation of a human head model. The in-house solvers based on the proposed SBP-SAT FDTD method and the FDTD method were developed in C++. All examples in this section were run through a single thread for fair comparison, and was completed on a workstation with an Intel i7-7700 3.6 GHz CPU and 256 G memory. 
\subsection{A Cavity with PEC Boundary Condition}
A three-dimensional cavity with PEC walls is first used to verify the long-time stability and the accuracy of the proposed method. The cavity is filled with air and its dimension is $1~m\times1~m\times1~m$. A Gaussian pulse with the bandwidth of 2 GHz at the center of the cavity is used as the excitation source. Uniform meshes with $\Delta  = 4 \times {10^{ - 2}}~m$ are used to discretize the cavity. The total physical time is $1 \times {10^{ - 4}}~s$, and ${\Delta}t=76.98~ps$ is the maximum time step under the CFL condition, which is exactly the same as that of the FDTD method. The probe used to record electric fields is placed at (0.4,0.4,0.4)[$m$].

\begin{figure}[h]
	\centering
	\subfigure[]{
		\includegraphics[scale=0.17]{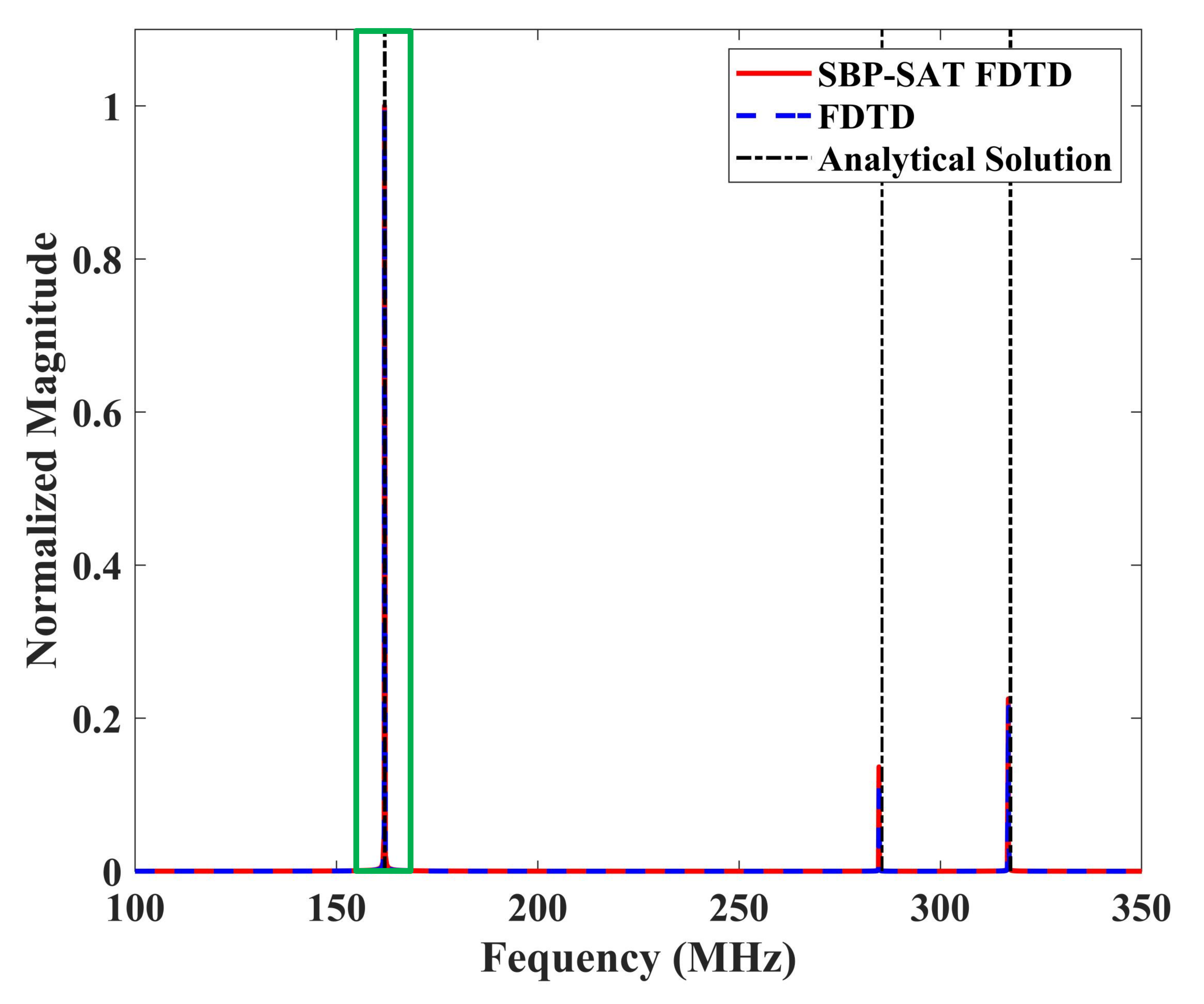}
	}
	\\
	\vspace{-0.1cm}
	\subfigure[]{
		\includegraphics[scale=0.45]{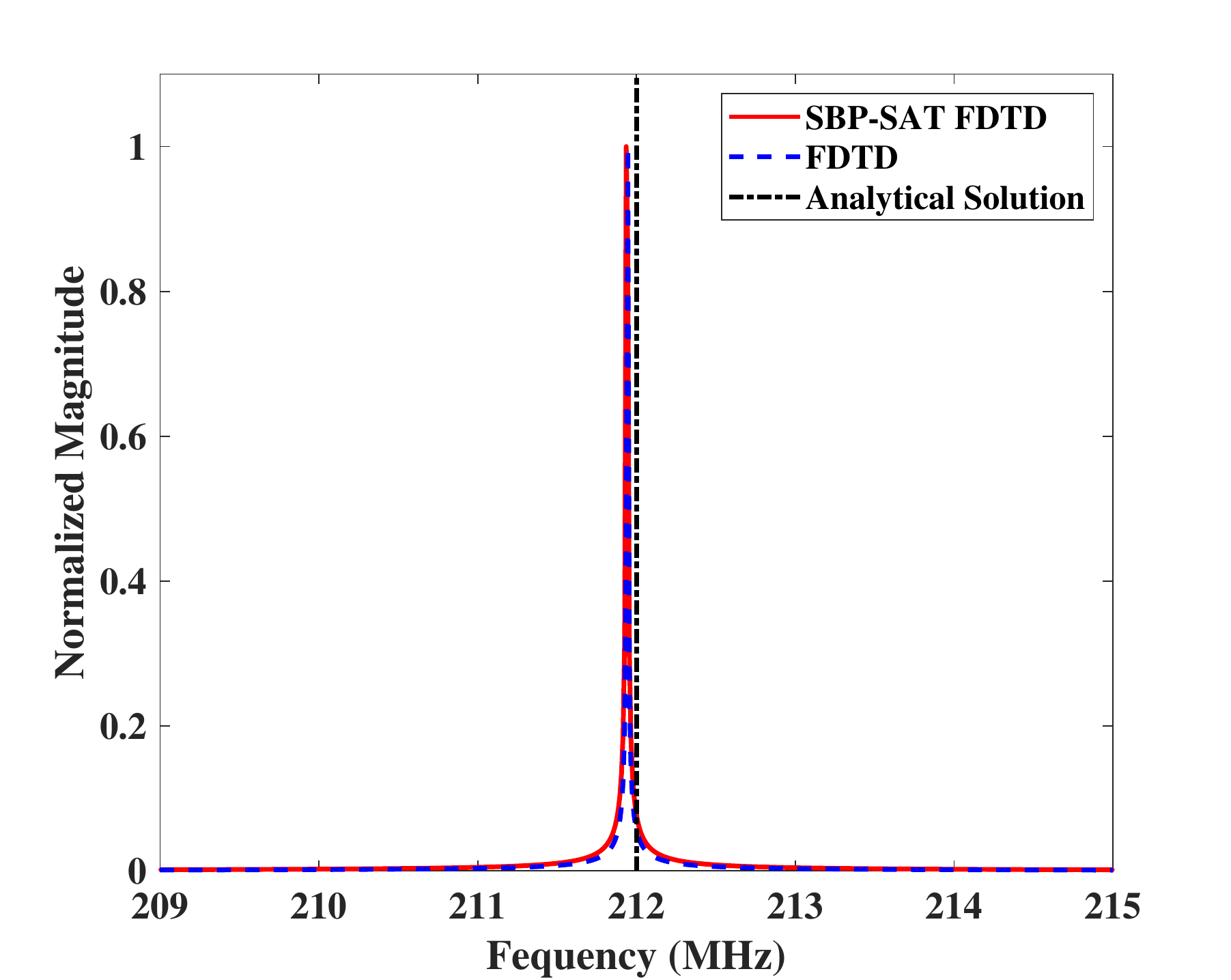}
	}
	\caption{Resonant frequencies calculated from the SBP-SAT FDTD method, the FDTD method and the analytical solution: (a) results in the frequency range 200 MHz to 400 MHz, (b) the zoom-in view of (a) in the range 209 MHz to 215 MHz.
	}
	\label{F_3}
\end{figure}

The resonant frequencies calculated by the SBP-SAT FDTD method and the FDTD method are shown in Fig. \ref{F_3}, and are compared with analytical solutions. It can be found that the resonant frequencies obtained from the FDTD method agree well with the analytical solutions in the whole frequency range. As for the proposed SBP-SAT FDTD method, results also show excellent agreement with the analytical solutions and those obtained from the FDTD method, as shown in Fig. \ref{F_3}, which confirms our previous analysis that the proposed SBP-SAT FDTD method has the same level of accuracy of the FDTD method. 

\begin{figure}[h]
	\centering
	\includegraphics[scale=0.45]{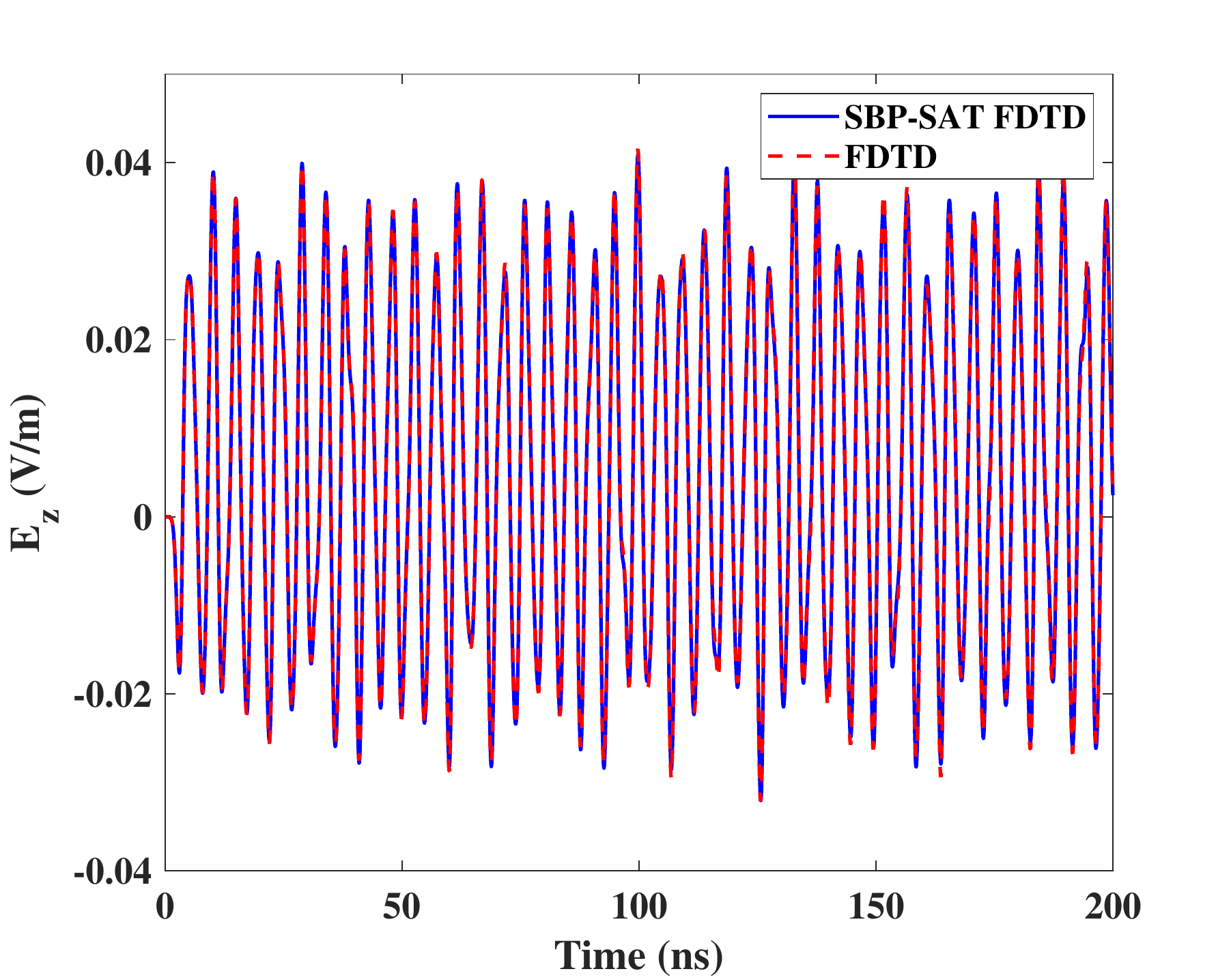}
	\caption{$E_z$ verse time steps obtained from the SBP-SAT FDTD method and the FDTD method.}
	\label{F_TIME}
\end{figure}

\begin{figure}[h]
	\centering
	\includegraphics[scale=0.13]{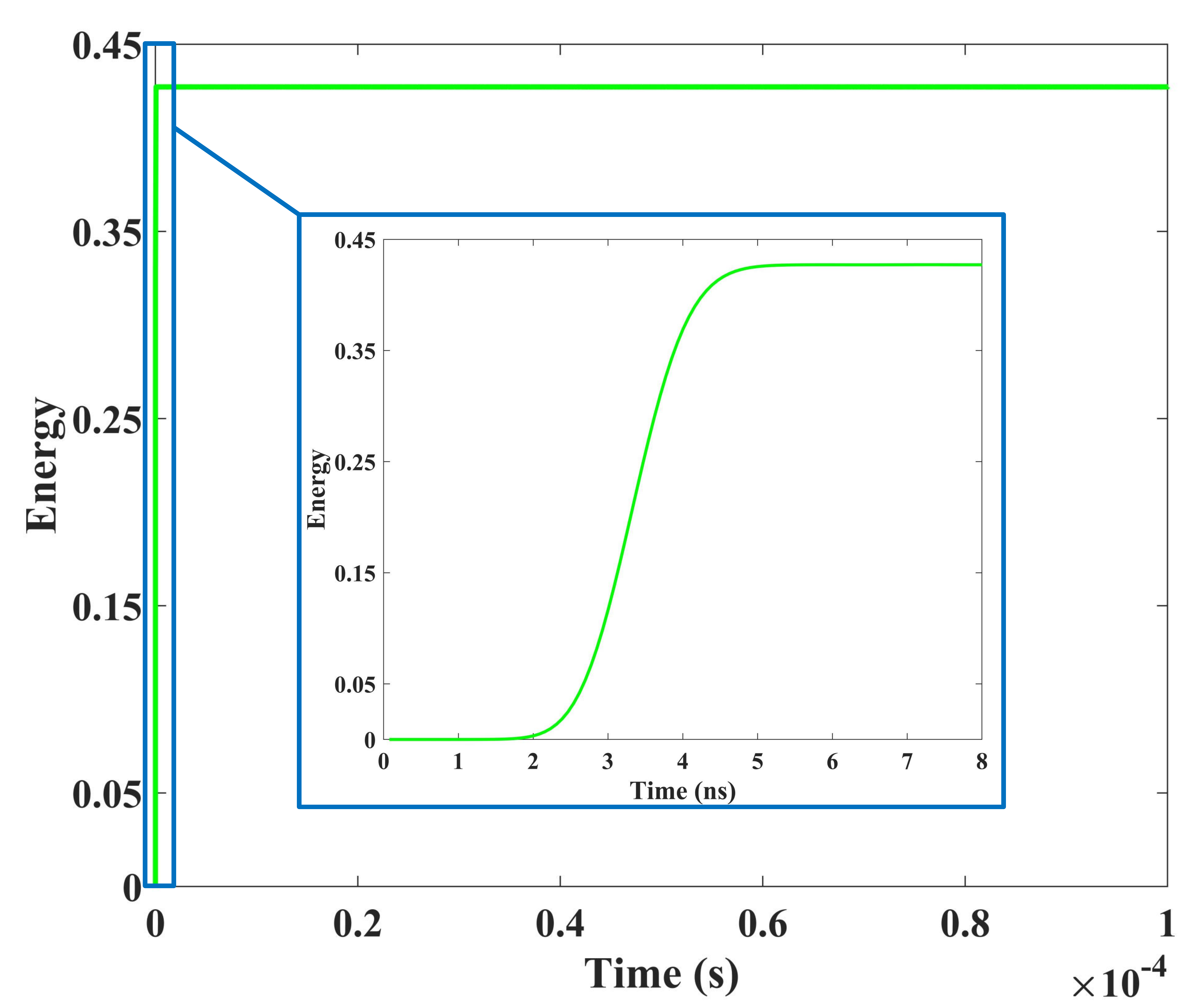}
	\caption{The energy verse time obtained from the SBP-SAT FDTD method.}
	\label{F_4}
\end{figure}

 Fig. \ref{F_TIME} shows $E_z$ at (0.4,0.4,0.4)[$m$] obtained from the proposed FDTD method and the proposed SBP-SAT FDTD method. It can be found that those results agree well with each other, and no signs of instability for the proposed FDTD method occurs. In addition, the energy in the computational domain is also calculated to further investigate the stability. As shown in Fig. \ref{F_4}, after one million time steps, the energy of the cavity is not divergent, which indicates that the proposed SBP-SAT FDTD method is long-time stable in the three-dimensional space.

\subsection{A DR Resonator}
\begin{figure}[h]
	\centering
	\includegraphics[scale=0.3]{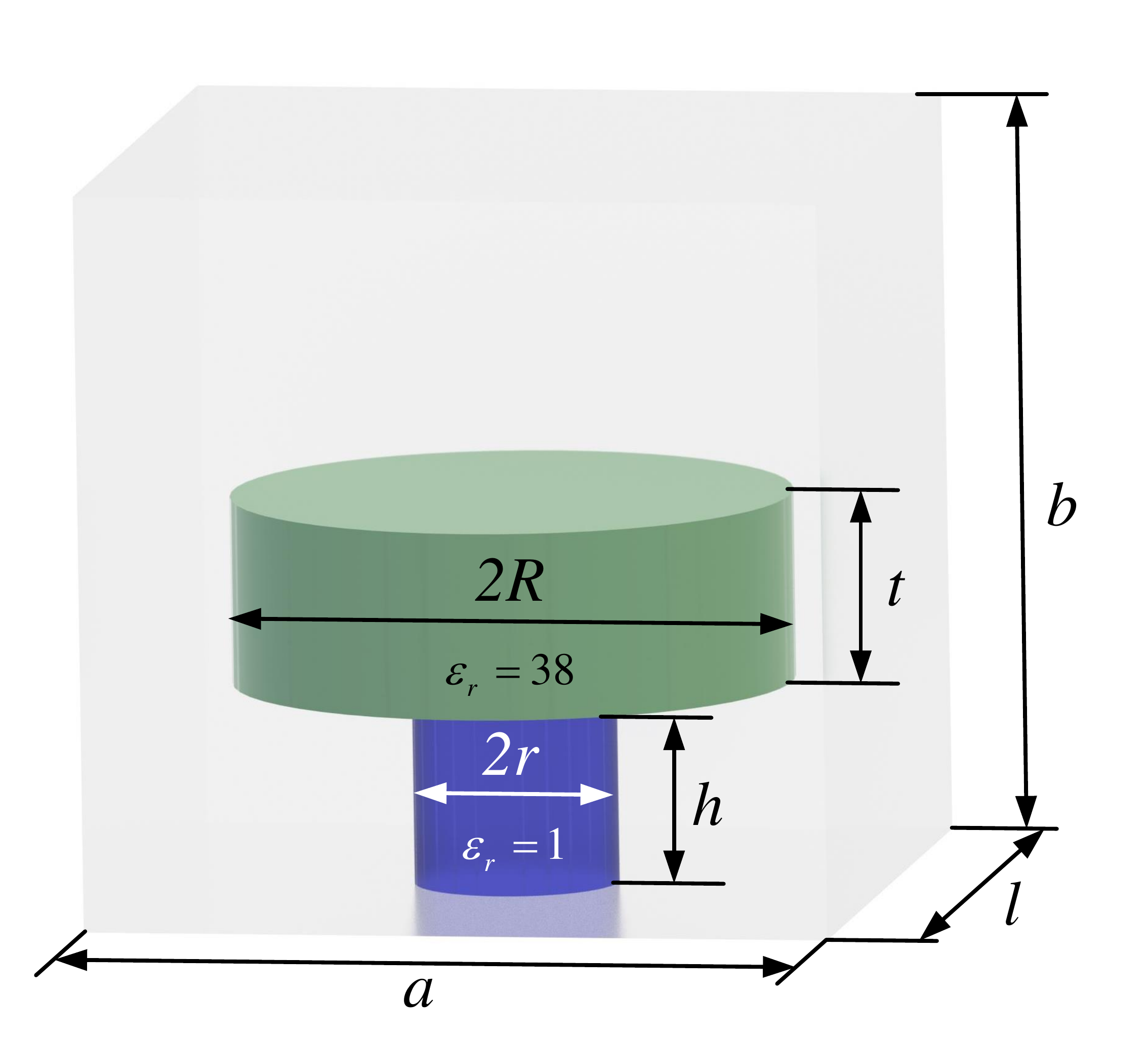}
	\caption{The geometrical configuration and the constant parameters of the DR resonator.}
	\label{F_5}
\end{figure}

A DR resonator is then considered to verify the accuracy of the proposed method. The geometrical configuration of the DR resonator is shown in Fig. \ref{F_5}, which can also be found in \cite{20DR} \cite{21DR}. The dimension of the cavity is $a = b = 2.5362~cm$, $l = 2.5718~cm$. Two dielectric cylinders are placed in the cavity. Their dimensions are $2R = 1.7551~cm$ with $t = 0.5893~cm$ and $2R = 1.9228~cm$ with $t = 0.6426~cm$. The height of the small cylinder is $h = 0.6985~cm$. The constant parameter of the large cylinder is ${\varepsilon _r} = 38$, and the other is ${\varepsilon _r} = 1$.  

Two different grids are used to complete the simulation to verify the accuracy of the proposed method, which are $26 \times 26 \times 26$ and $20 \times 20 \times 20$, respectively. The total count of time steps in our simulation is 35,000. The Gaussian pulse $f\left( t \right) = {e^{ - 4\pi {{\left( {t - {t_0}} \right)}^2}/{t_w^2}}}$, where ${t_w} = 0.35~ns$ and ${t_0} = 0.28~ns$, is used as the excitation source. We take the resonant frequencies calculated by the FDTD method and measurement results from \cite{20DR} as reference. 
\begin{table}[h]
	\renewcommand\arraystretch{1.5}
	\centering
	\caption{Comparision of Resonant Frequencies Obtained from the FDTD Method and the Proposed SBP-SAT FDTD Method with Measured Results}
	\label{T2}
	\resizebox{9cm}{!}
	{
		\begin{threeparttable}[b]
			\begin{tabular}{ c| c| c| c| c}
				\hline
				\hline
				\textbf{Method}	&\textbf{Meshes}	&\textbf{Simulation [GHz]} &\textbf{Measured [GHz]} &\textbf{Relative	error}\\
				\hline
				FDTD &\multirow{2}*{$26\times26\times26$} &$4.121$ &\multirow{2}*{$4.136$}  &$0.36\%$\\
				\cline{1-1} \cline{3-3} \cline{5-5}

				PROPOSED  & 			 &$4.121$ &	&$0.36\%$\\		
				\hline
				FDTD &\multirow{2}*{$20\times20\times20$} &$3.675$ &\multirow{2}*{$3.760$}  &$2.26\%$\\
				\cline{1-1} \cline{3-3} \cline{5-5}
				
				PROPOSED  & 			 &$3.675$ &	&$2.26\%$\\
				
%				\multicolumn{2}{c|}{\multirow{2}*{\textbf{Method}}}  &\multicolumn{3}{c}{\textbf{Method}}\\
%				\cline{3-5}
%				& 	  &x &y &z\\
%				\hline
%				\multirow{3}*{\textbf{{\it E} nodes}} &$E_x$ &$x_-$  &$y_+$ &$z_+$ \\
%				\cline{3-5}
%				&$E_y$ &$x_+$  &$y_-$ &$z_+$ \\
%				\cline{3-5}
%				&$E_z$ &$x_+$  &$y_+$ &$z_-$ \\
%				\hline
%				\multirow{3}*{\textbf{{\it H} nodes}} &$H_x$ &$x_+$  &$y_-$ &$z_-$ \\
%				\cline{3-5}
%				&$H_y$ &$x_-$  &$y_+$ &$z_-$ \\
%				\cline{3-5}
%				&$H_z$ &$x_-$  &$y_-$ &$z_+$ \\
				%				\cline{3-6}
				\hline
				\hline
				
			\end{tabular}
			%\tablefootnote{Ratio is defined as the ratio of time cost used in the LOD-FDTD method with fine grid to that in the correspond method.}
		\end{threeparttable}
	}
\end{table}

Table \ref{T2} lists the resonant frequencies calculated by the FDTD method, the proposed method and measured results. For $26\times26\times26$ meshes, the resonant frequency calculated by the FDTD method and the proposed method are both 4.121 GHz. To quantitatively measure the accuracy, the relative error (RE) is defined as $|{f_c}-{f_m}|/{f_m}$, where $f_c$ is the calculated resonant frequency, and $f_m$ is the measured value. Compared with the measured result 4.136 GHz, the RE of the FDTD method and the proposed method are both 0.36\%. The two methods show excellent performance in terms of accuracy. As for $20\times20\times20$ meshes, the resonant frequency calculated by two methods is 3.675 GHz, and the measured result is 3.760 GHz. Therefore, the RE is 2.26\%. Since relatively coarse meshes are used, which leads to large staircase errors, a slightly large RE is obtained compared with the first case. 

To sum up, the proposed SBP-SAT FDTD method can effectively calculate the resonant frequencies of the DR resonator, and show the same level of accuracy as that of the FDTD method.

\subsection{A 5-Pole H-Plane Iris Filter}
\begin{figure}[h]
	\centering
	\subfigure[]{
		\includegraphics[scale=0.5]{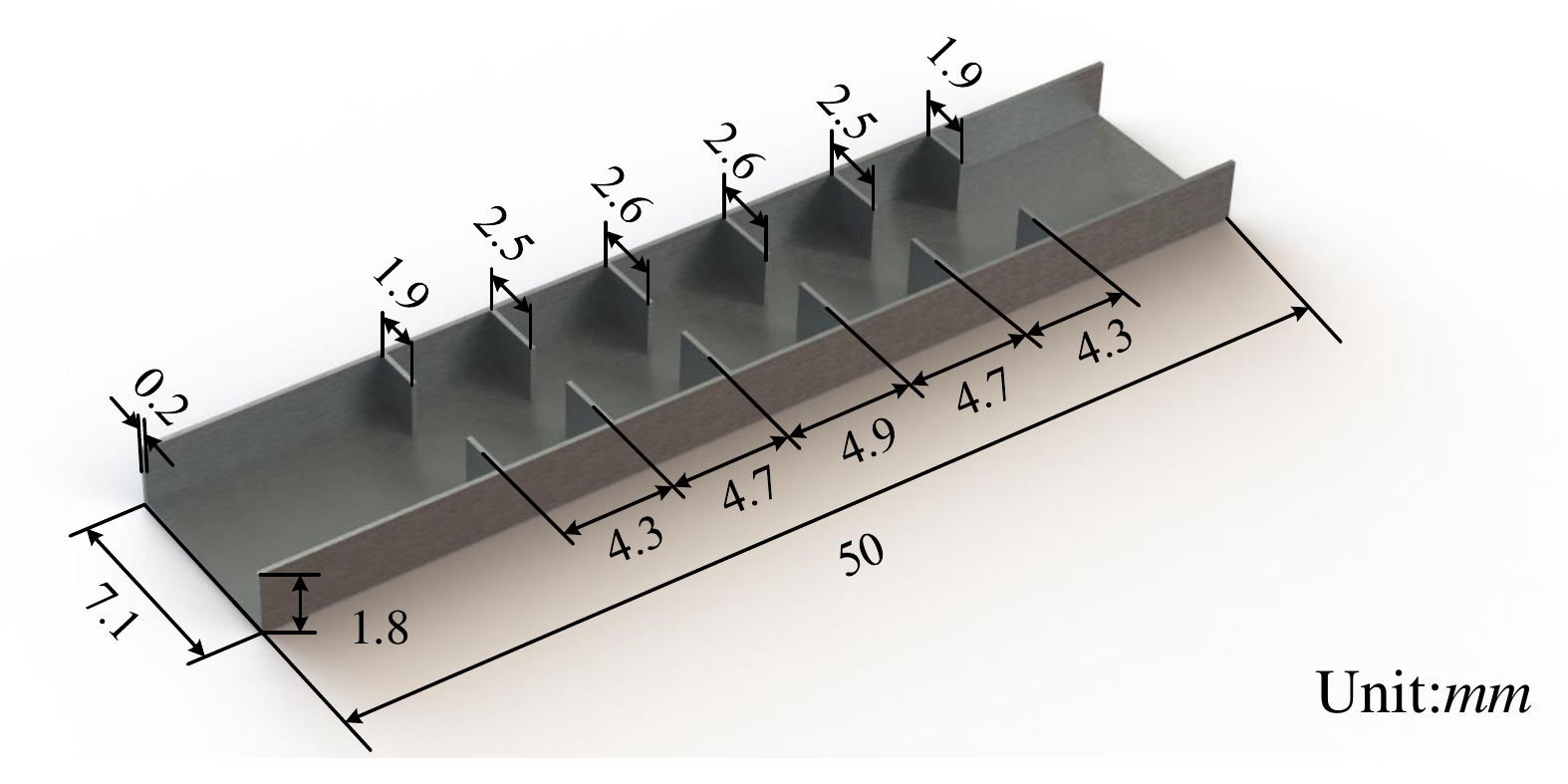}
	}
	\\
	\vspace{-0.1cm}
	\subfigure[]{
		\includegraphics[scale=0.45]{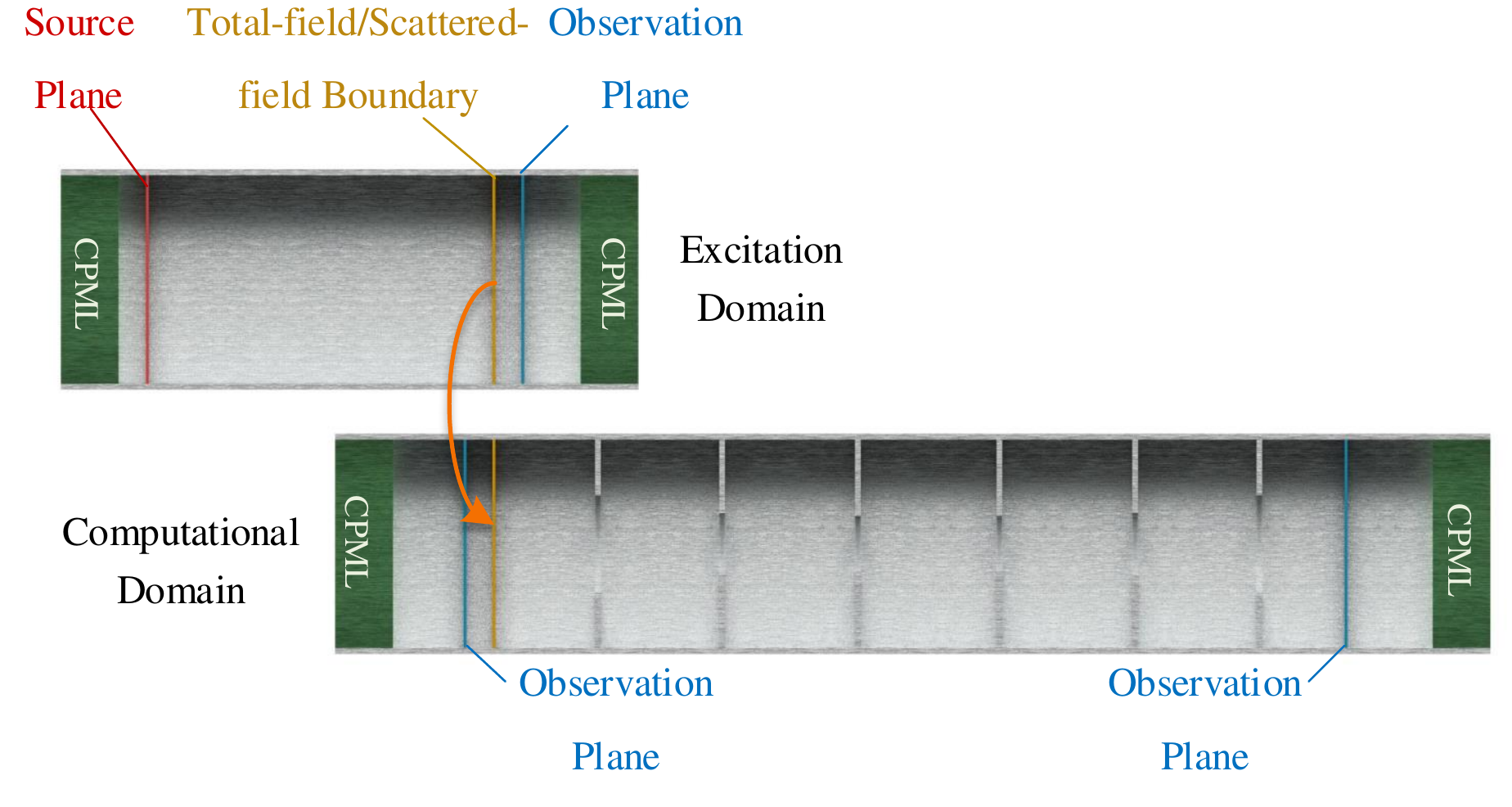}
	}
	\caption{(a) The cross-sectional view in the middle plane of the filter, (b) the configures in the simulations.
	}
	\label{F_FILTER_CROSS}
\end{figure}

To further verify the accuracy and efficiency of the proposed SBP-SAT FDTD method, a 5-pole H-plane iris filter is considered. Fig. \ref{F_FILTER_CROSS}(a) shows the geometrical configurations of the cross-sectional view in the middle of side walls of the filter. Its width and height are $7.1~mm$, $3.6~mm$, respectively. The thicknesses of PEC walls and six irises are $0.2~mm$. The filter is completely symmetric with respect to the plane in the middle of filter along the longitude direction. The length of those irises are $1.9~mm$, $2.5~mm$, $2.6~mm$, $2.6~mm$, $2.5~mm$, and $1.9~mm$, respectively. The distances between two irises are $4.3~mm$, $4.7~mm$, $4.9~mm$, $4.7~mm$, and $4.3~mm$, respectively. The length and width of the computational domain are $50~mm$ and $7.5~mm$, respectively. 

To calculate its S-parameter, a modulated Gaussian pulse $f\left( t \right) = {sin(2{\pi}ft)e^{ - 4\pi {{\left( {t - {t_0}} \right)}^2}/t_w{^2}}}$, where $t_w = 0.11~ns$, $t_0 = 0.85~ns$, $f$ = 36.0 GHz, is applied at source plane to generated the TE$_{10}$ mode. The excitation wave is generated from another computational domain with exactly the same cross section and without irises. Then, it is introduced into simulations through the total-field/scattered-field (TF/SF) boundary conditions \cite{32TFSF}, which is set $15~mm$ away from the boundaries of the computational domain in Fig. \ref{F_FILTER_CROSS}(b). The 10-layer convolutional perfectly matched layers (CPMLs) \cite{22CPML} are used in both the excitation domain and the simulation domain to truncate the computational domain. Three observation planes are used to record fields, as shown in Fig. \ref{F_FILTER_CROSS}(b). One of observation planes is located at $0.5~mm$ from the TF/SF boundary, which is used in the excitation domain. The remaining two observation planes are applied in the computational domain. One is located in the scattered field area to record the reflected wave, and the other is placed at the other end to record the transmitted wave. The total physical time is $4~ns$.

By recording fields in observation planes, the power $P(w)$ in the frequency domain passing through the observation plane is calculated by 
\begin{equation}\label{E25}
	\begin{aligned}
		{P(w)} = \sum\limits_{i = 1}^n {{\cal F}\left( {{{{E}}_i}(t)} \right) \times } {\cal F}{\left( {{{{H}}_i}(t)} \right)^*} \Delta {S_i}
	\end{aligned}
\end{equation}
where ${\cal F}\left(  \cdot  \right)$ is the Fourier transform operator. ${{{E}}_i(t)},{{{H}}_i(t)}$ are electric field and magnetic fields of the $i$th cell in the temporal domain, and $\Delta {S_i}$ is the area of the $i$th cell on the observation plane. The operator ${\left(  \cdot  \right)^*}$ denotes the conjugate of a complex number. Therefore, the S-parameter can be calculated as  
\begin{equation}\label{E26}
	\begin{aligned}
		{S_{11}(w)} = {\left| {\frac{{{{{P}}_r}(w)}}{{{{{P}}_i}(w)}}} \right|},
	\end{aligned}
\end{equation}
\begin{equation}\label{E27}
	\begin{aligned}
		{S_{21}(w)} = {\left| {\frac{{{{{P}}_t}(w)}}{{{{{P}}_i}(w)}}} \right|},
	\end{aligned}
\end{equation}
where ${{{P}} _i}(w)$, ${{{P}} _r}(w)$, ${{{P}} _t}(w)$ are the power of the incident wave, the reflected wave, and the transmitted wave, respectively. Since magnetic and electric fields in the FDTD method and the SBP-SAT FDTD method are not co-located, and magnetic fields are one half time step offset from electric fields, the linear interpolation is used to average magnetic fields in the spatial and temporal domain to get correct values. 

\begin{figure}[h]
	\centering
	\subfigure[]{
		\includegraphics[scale=0.45]{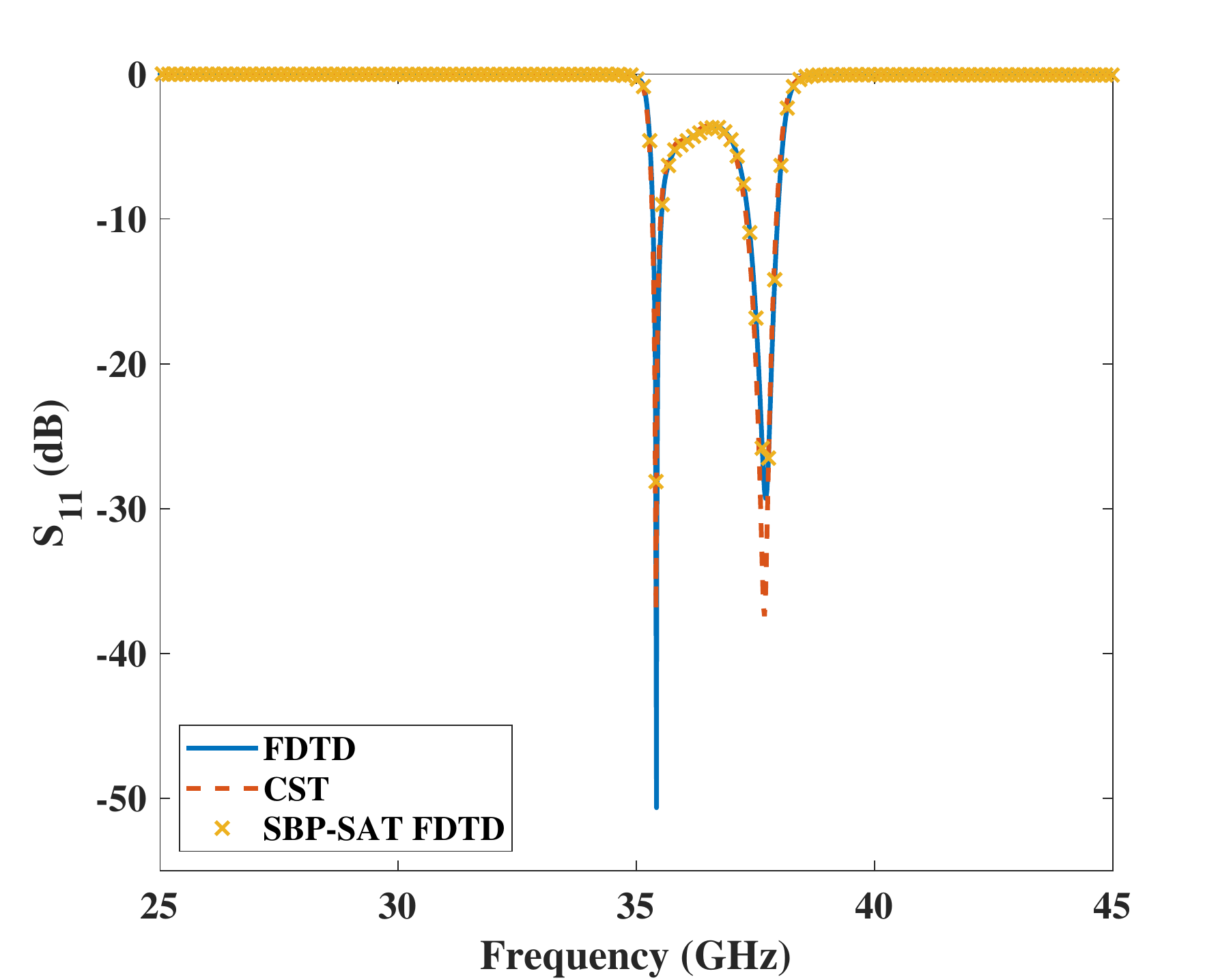}
	}
	\\
	\vspace{-0.1cm}
	\subfigure[]{
		\includegraphics[scale=0.45]{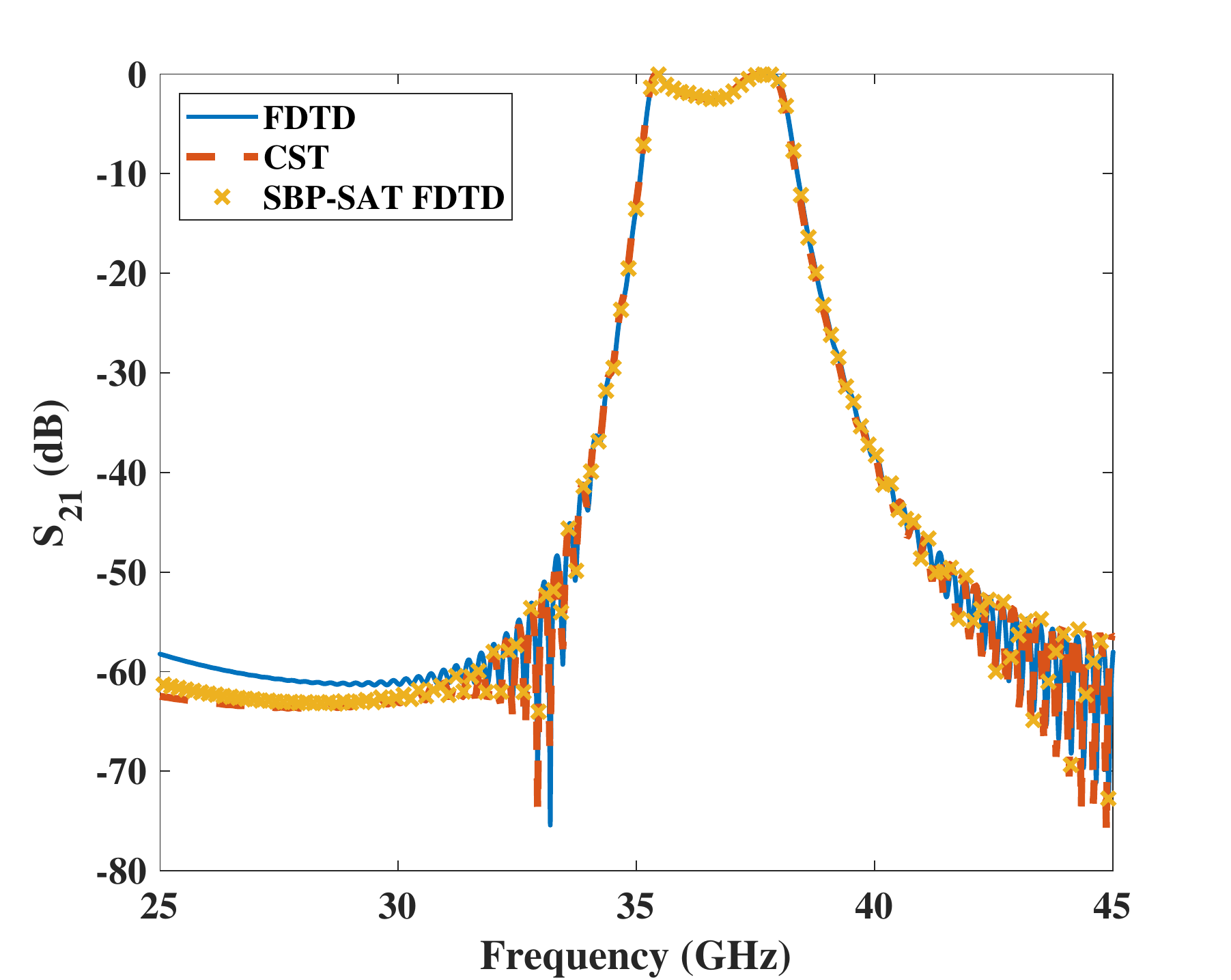}
	}
	\caption{The S-parameter of the iris filter: (a) $S_{11}$ and (b) $S_{21}$ in the frequency range 25.0 GHz to 45.0 GHz.
	}
	\label{F_FILTER_S}
\end{figure}

$S_{11}$ and $S_{21}$ are calculated by the FDTD method and the SBP-SAT FDTD method. For references, it is also simulated by CST \cite{THECST} with two wave ports for reference. $S_{11}$ and $S_{21}$ are shown in Fig. \ref{F_FILTER_S}(a) and (b). It can be found that the pass band is around in frequency range 35.0 GHz to 37.0 GHz. 

$S_{11}$ and $S_{21}$ obtained from three methods show good agreement with each other. $S_{21}$ in the frequency range 25.0 GHz to 32.0 GHz show slight differences between results from CST and the FDTD method. It maybe account for differences in implementations of excitations in two methods. In CST, the characteristic modes are solved through an eigensolver, and then are used in the wave port. In our implementation, the TE${_{10}}$ mode is calculated by another FDTD simulation, as shown in Fig. \ref{F_FILTER_CROSS}(b). It is interesting to note that $S_{21}$ obtained from the SBP-SAT FDTD method show slightly better agreement with that from CST compared with results from the FDTD method.

\begin{table}[h]
	\renewcommand\arraystretch{1.5}
	\centering
	\caption{Comparison of Memory Usage and Runtime of the Filter in the FDTD Method and the Proposed SBP-SAT FDTD Method}
	\label{T4}
	%	\resizebox{9cm}{!}
	%	{
		\setlength{\tabcolsep}{6mm}
		{
			\begin{threeparttable}[b]
				\begin{tabular}{ c|  c|  c }
					\hline
					\hline
					\textbf{Method}	&\textbf{Memory (MB)}	 &\textbf{Time (s)} \\
					\hline
					FDTD &$331.3$  &$6,478.7$  \\
					\hline
					
					Proposed &$338.7$ 	 &$6,588.2$	\\		
					\hline
					\hline
					
				\end{tabular}
				%\tablefootnote{Ratio is defined as the ratio of time cost used in the LOD-FDTD method with fine grid to that in the correspond method.}
			\end{threeparttable}
		}
		%	}
\end{table}

In this simulation, 331.3 MB memory and $6,478.7~s$ are used by the FDTD method. As for the proposed SBP-SAT FDTD method, 338.7 MB memory and $6,588.2~s$ are used as shown in Table \ref{T4}. Therefore, compared with the performance in terms of accuracy, memory consumption, and runtime of the FDTD method, the SBP-SAT FDTD method shows a good performance with a negligible overhead compared with that of the FDTD method.
\begin{figure}[h]
	\centering
	%	\subfigure[CFLN=0.25]{
		\includegraphics[scale=0.045]{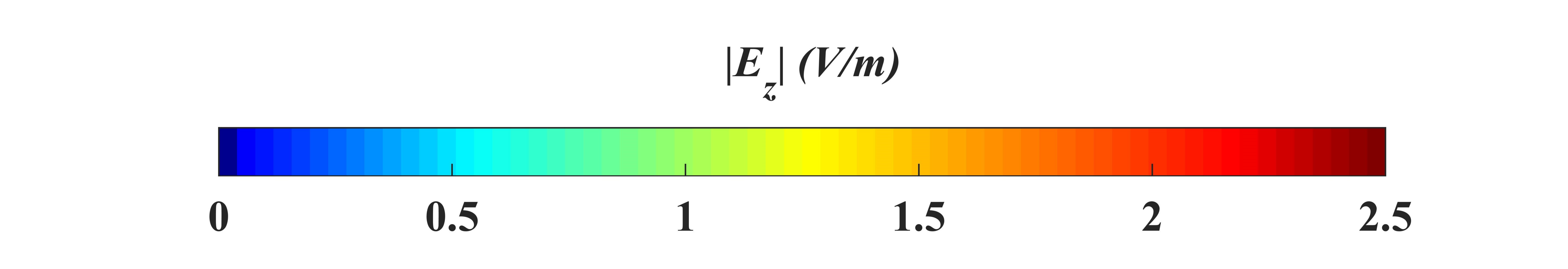}
		\\
		%	}
	\vspace{-0.3cm}
	\subfigure[]{
		\includegraphics[scale=0.045]{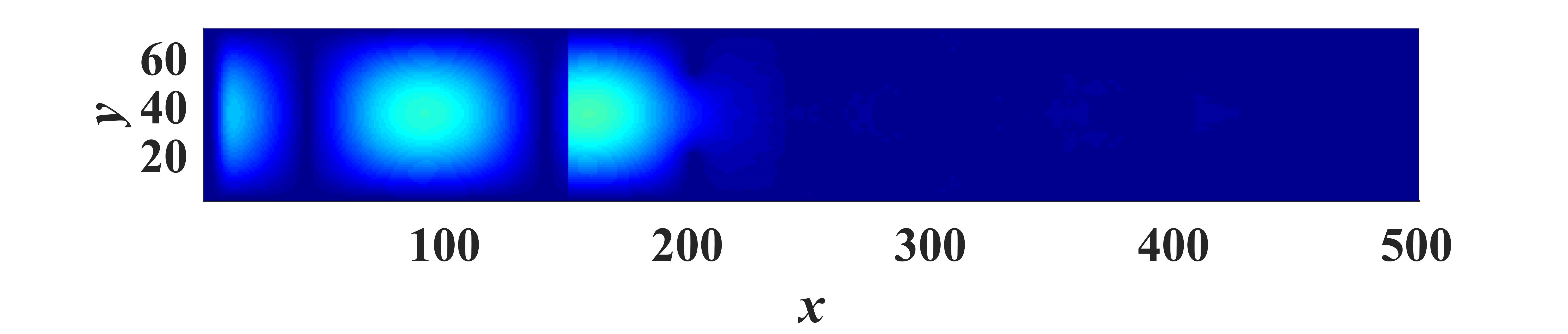}
	}
	\vspace{-0.3cm}
	\\
	\subfigure[]{
		\includegraphics[scale=0.045]{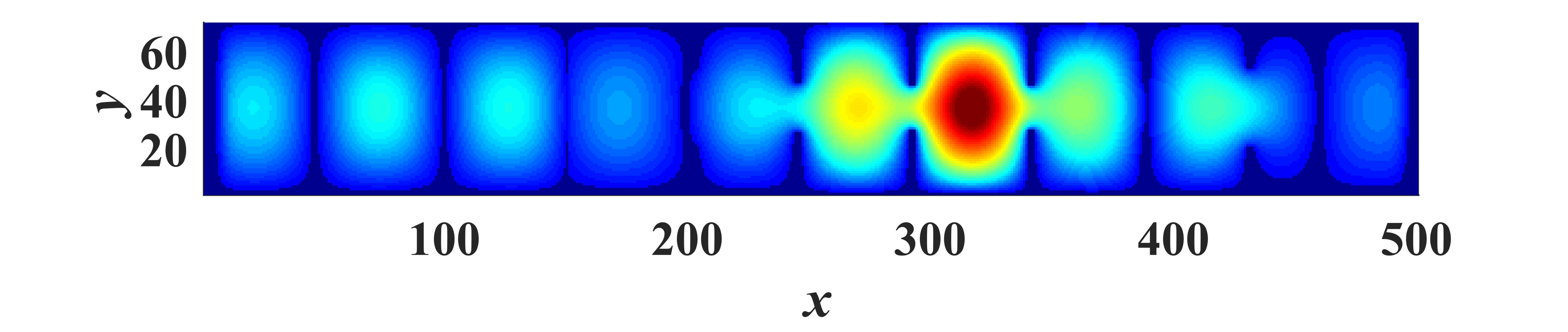}
	}
	\\
	\vspace{-0.3cm}
	\subfigure[]{
		\includegraphics[scale=0.045]{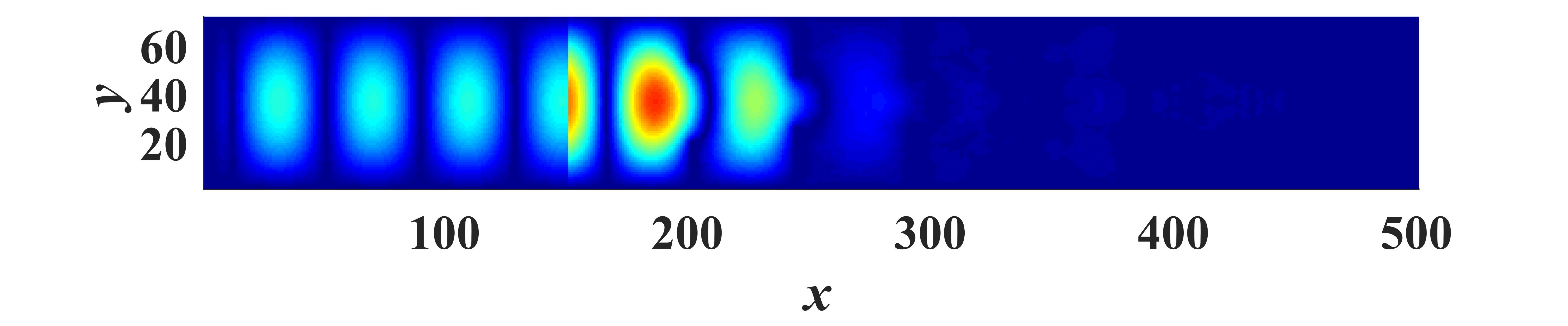}
	}
	\caption{$|E_z|$ of the iris filter at (a) 26.0 GHz, (b) 32.2 GHz and (c) 44.0 GHz.}
	\label{F_10}
\end{figure}

Fig. \ref{F_10} shows $|E_z|$ excited by the incident waves at $f=$26.0 GHz, 32.2 GHz and 44.0 GHz, respectively, which are excited by a sinusoidal current source in the excitation domain. The discontinuity at $x=15~mm$ in Fig. \ref{F_10} is the TF/SF boundary . The reflected fields by the filter are on the left of the TF/SF boundary. It can be found that $E_z$ attenuates and eventually vanishes at $f=26.0$ GHz and $44.0$ GHz since they are in the stop band. However, $E_z$ can pass through the filter at $f$=32.2 GHz in the pass band.

\subsection{The SAR Calculation of A Human Head Model}
To demonstrate the capability of the proposed SAT-SBP FDTD method to solve complex electromagnetic problems, a human head model \cite{23head} illuminated by a plane wave is considered. The plane wave incidents from the $-x$ direction. The human head model with voxels is shown in Fig. 11, which can be decomposed into 117 tissues, and 27 of them are shown for a better visualization including brain stem, white matters, the gray matters, nerve, blood vein, cerebrospinal fluid, eyes, tongue, Ears, Gland, skull, cartilage, spinal, dermis, adipose tissue, and muscle. 

\begin{figure*}[htbp]
%	\begin{minipage}[h]{0.95\linewidth}%多行需修改此处迷你页宽度
	\includegraphics[width=\linewidth]{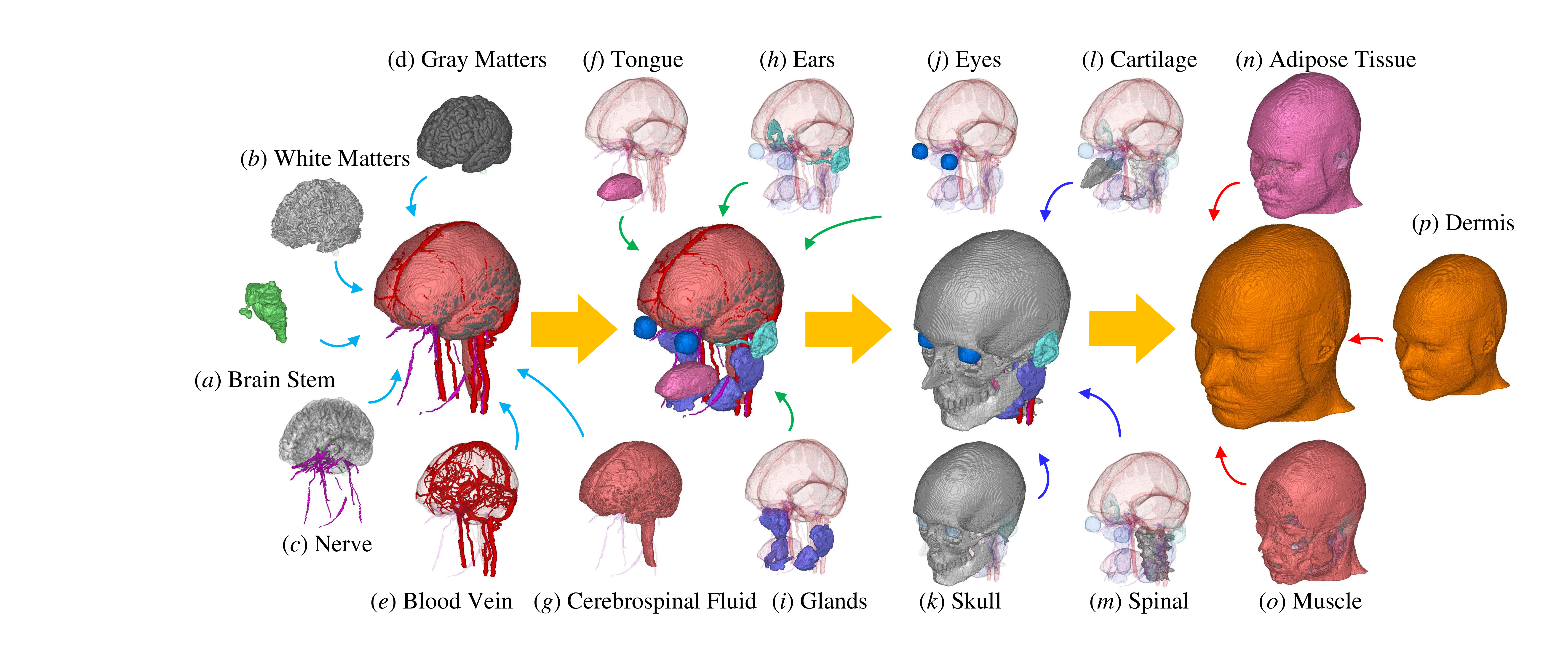}
%	\end{minipage}
	\caption{The human head model: (a) brain stem, (b) the white matters, (c) nerve, (d) the gray matters, (e) blood vein, (f) tongue, (g) cerebrospinal fluid, (h) ears, (i) glands, (j) eyes, (k) skull, (l) cartilage, (m) spinal, (n) adipose tissue (o) muscle, (p) dermis.}
	\label{F_11}
\end{figure*}

\begin{figure}[htbp]
	\centering
	\includegraphics[scale=0.075]{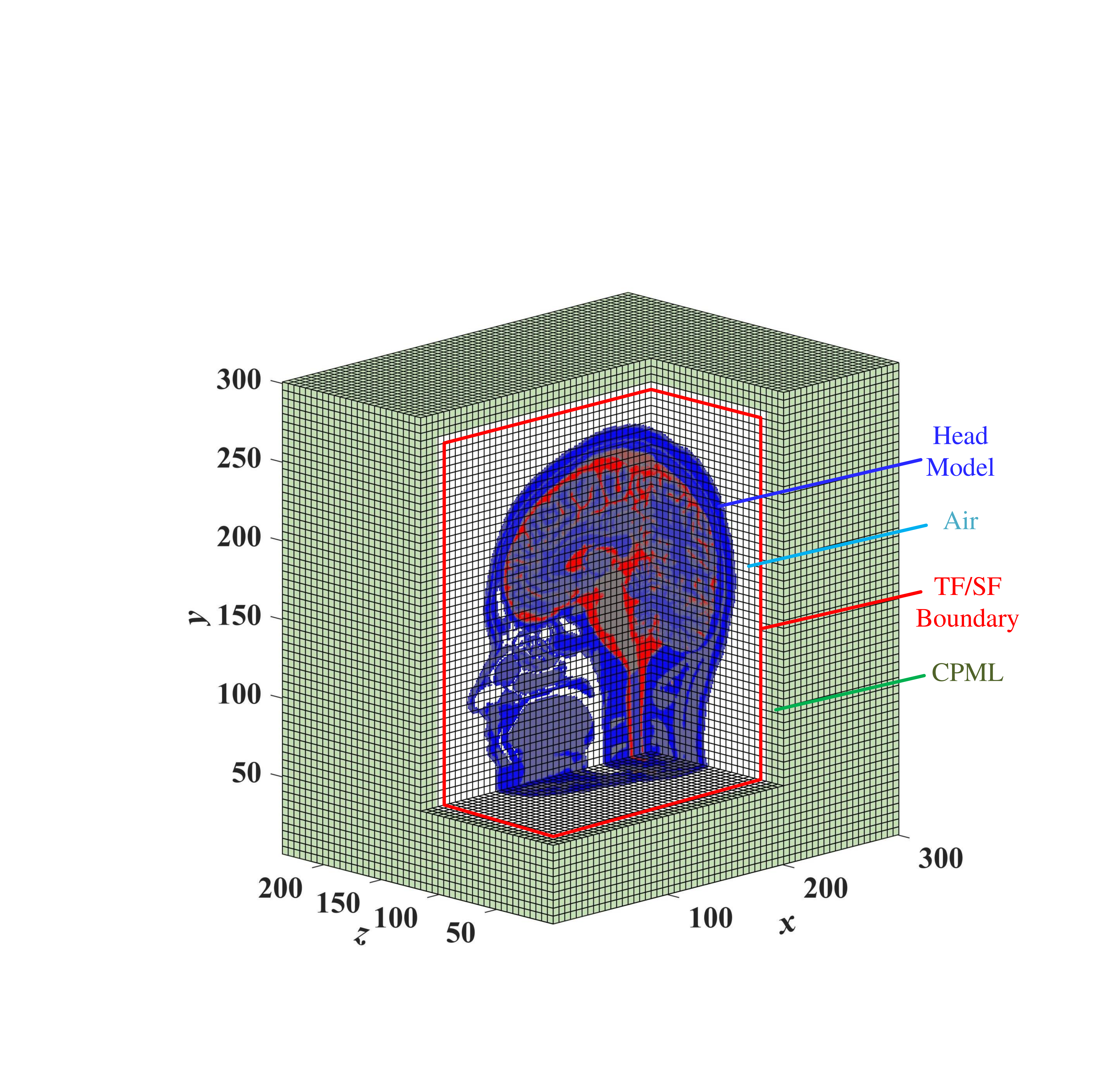}
	\caption{The mesh configuration of the human head model in our simulation.}
	\label{F_12}
\end{figure}

\begin{figure*}[htbp]
	%	\begin{minipage}[h]{0.15\linewidth}%多行需修改此处迷你页宽度
		\subfigure[]{
			\includegraphics[height=36.5mm]{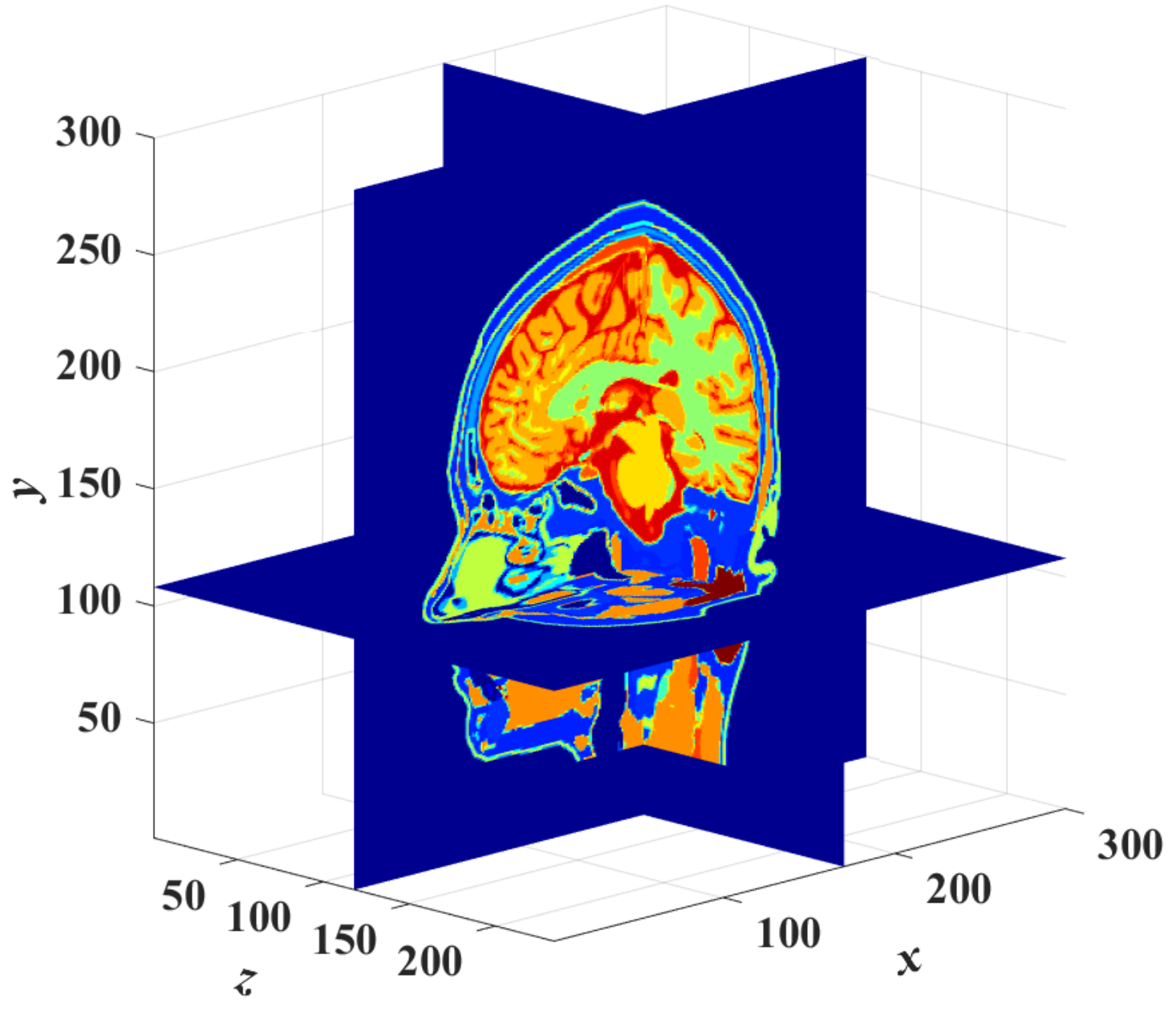}
		}
		%		\centerline{(a)}
		%	\end{minipage}
	%	\hfill
	%	\begin{minipage}[h]{0.25\linewidth}
		\!\!\!\!\!\!\!\!\!
		\subfigure[]
		{
			\includegraphics[height=36.5mm]{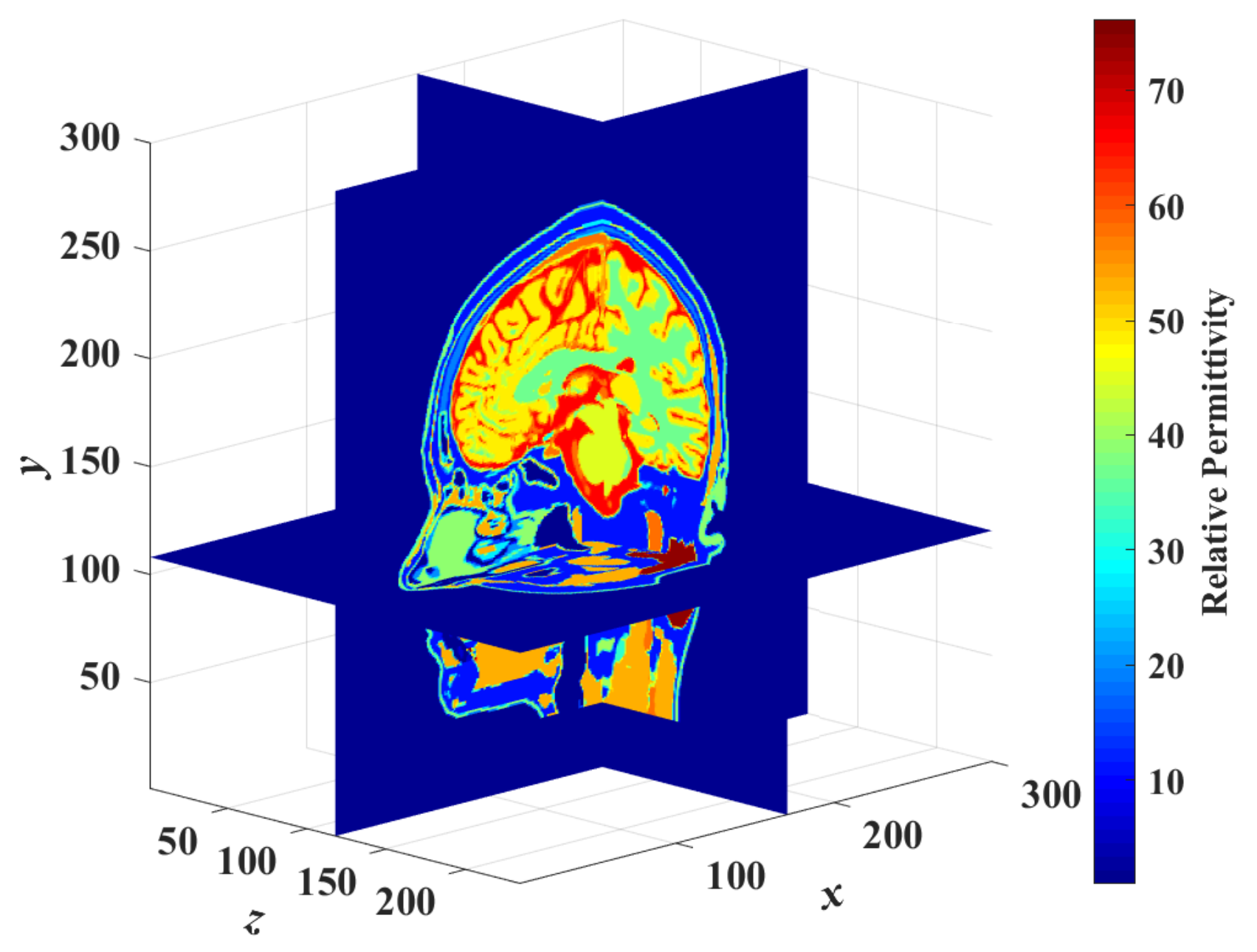}
		}
		%		\centerline{(b)}
		%	\end{minipage}
	%	\hfill
	%	\begin{minipage}[h]{0.20\linewidth}
		\!\!\!\!\!\!\!\!\!
		\subfigure[]{
			\includegraphics[height=36.5mm]{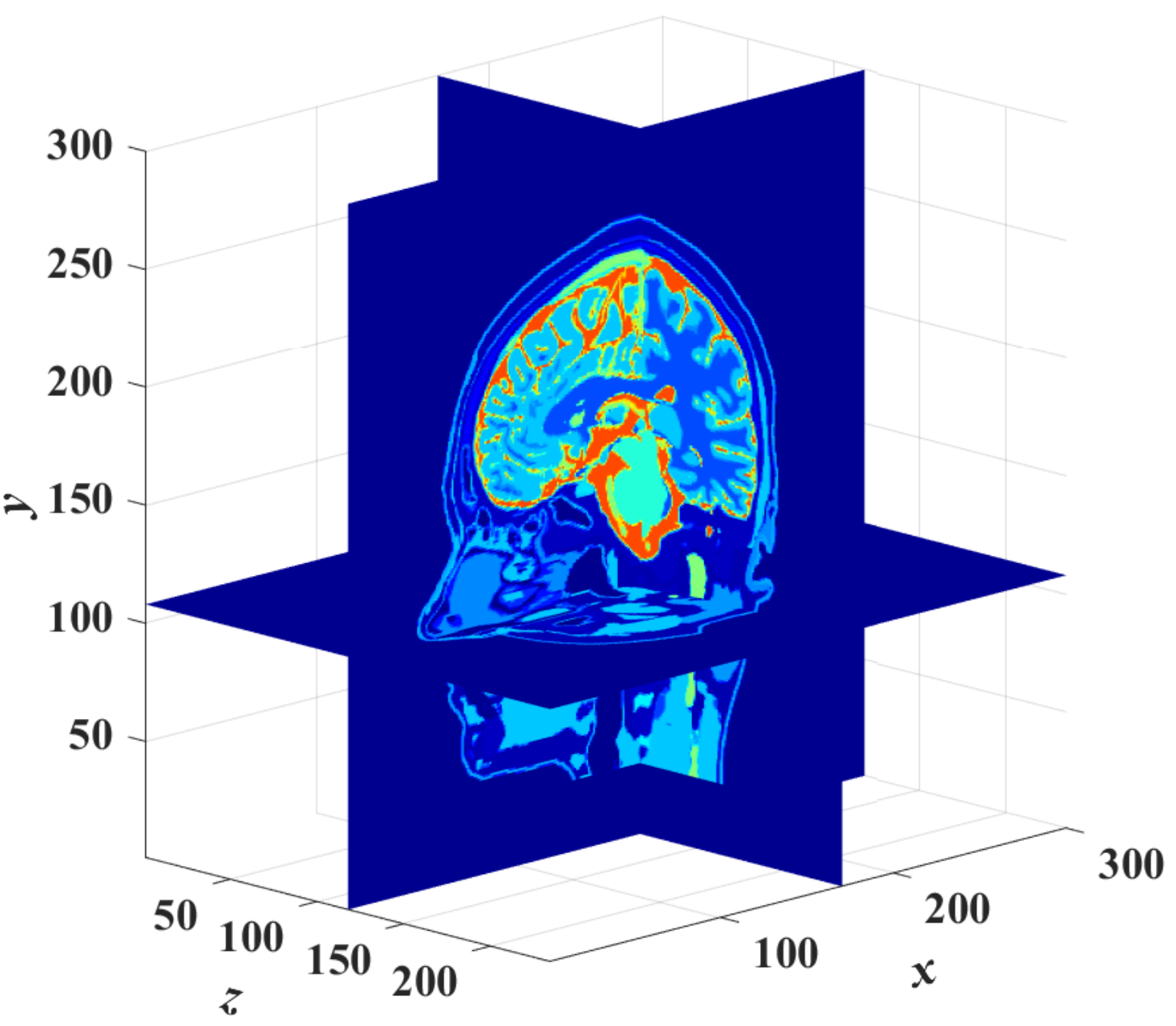}
		}
		%		\centerline{(c)}
		%	\end{minipage}
	%	\hfill
	%	\begin{minipage}[h]{0.20\linewidth}
		\!\!\!\!\!\!\!\!\!
		\subfigure[]{
			\includegraphics[height=36.5mm]{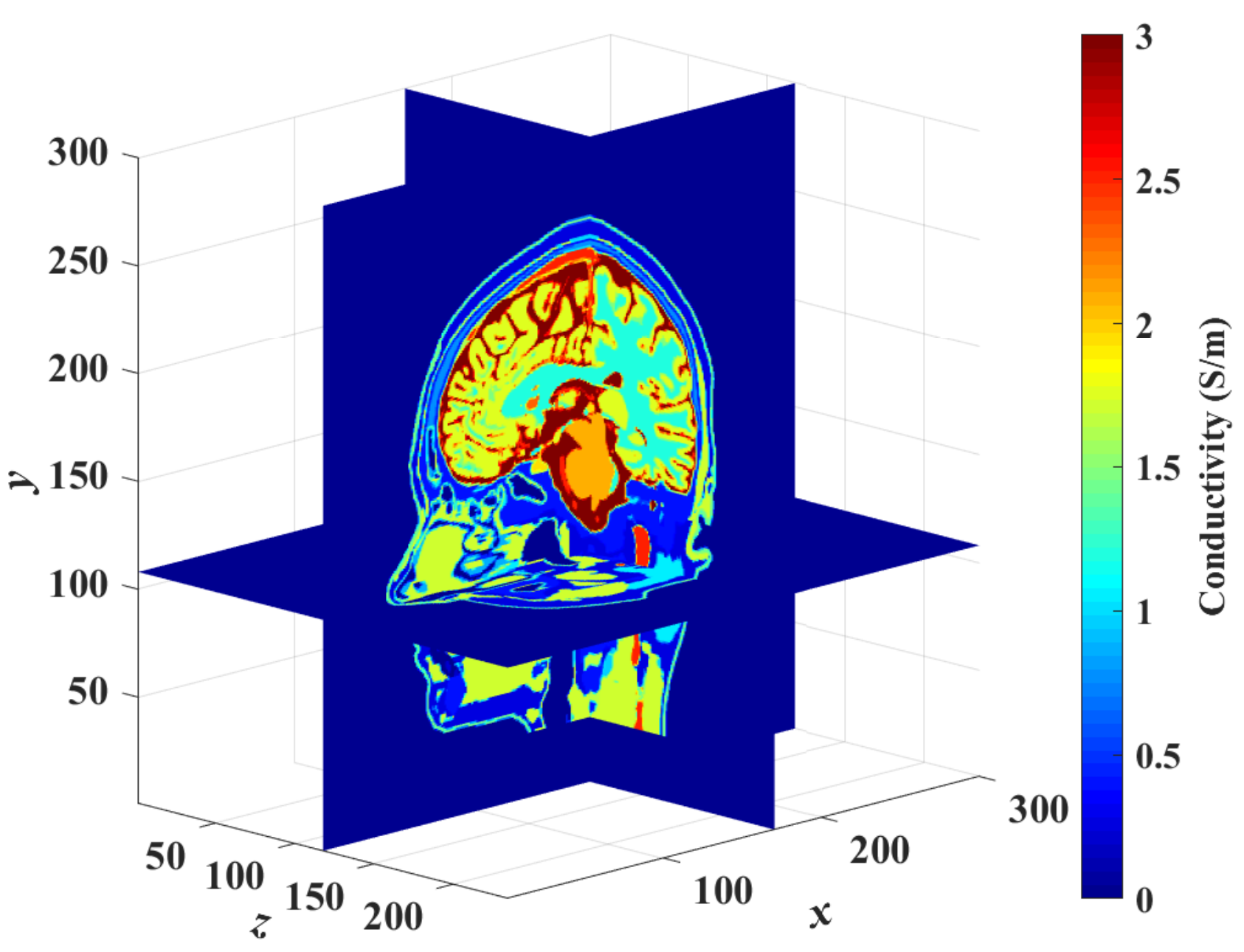}
		}
		%		\centerline{(d)}
		%	\end{minipage}
	\caption{The relative permittivity (a) at 900.0 MHz, (b) at 2.4 GHz, and the conductivity (c) at 900.0 MHz (d) at 2.4 GHz.}
	\label{F_13}
\end{figure*}

\begin{figure*}[h]
	%	\begin{minipage}[h]{0.20\linewidth}%多行需修改此处迷你页宽度
		\subfigure[]{
			\includegraphics[height=38.5mm]{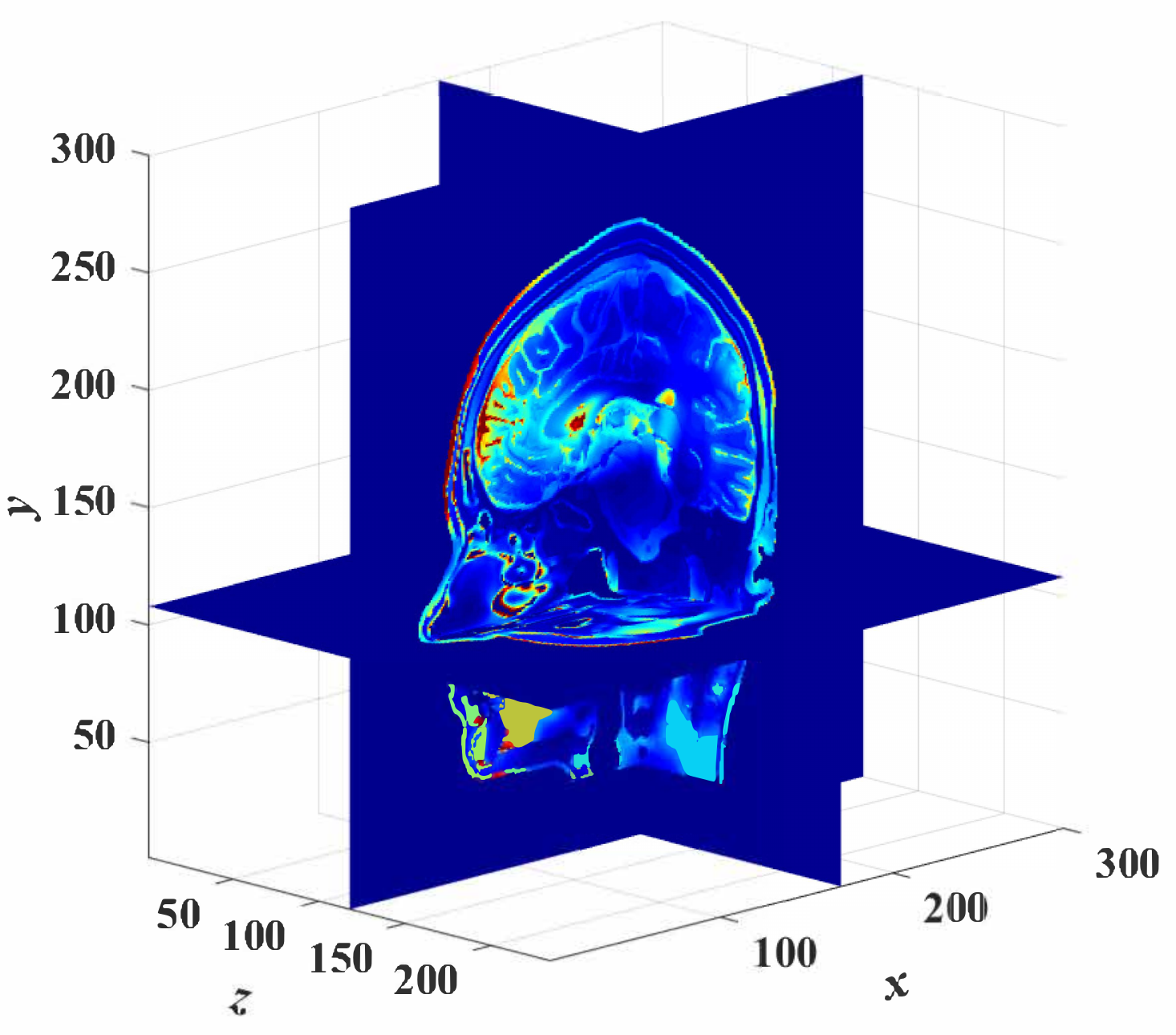}
		}
		%		\centerline{(a)}
		%	\end{minipage}
	%	\hfill
	%	\begin{minipage}[h]{0.20\linewidth}
		\!\!\!\!\!\!\!\!\!
		\subfigure[]{
			\includegraphics[height=38.5mm]{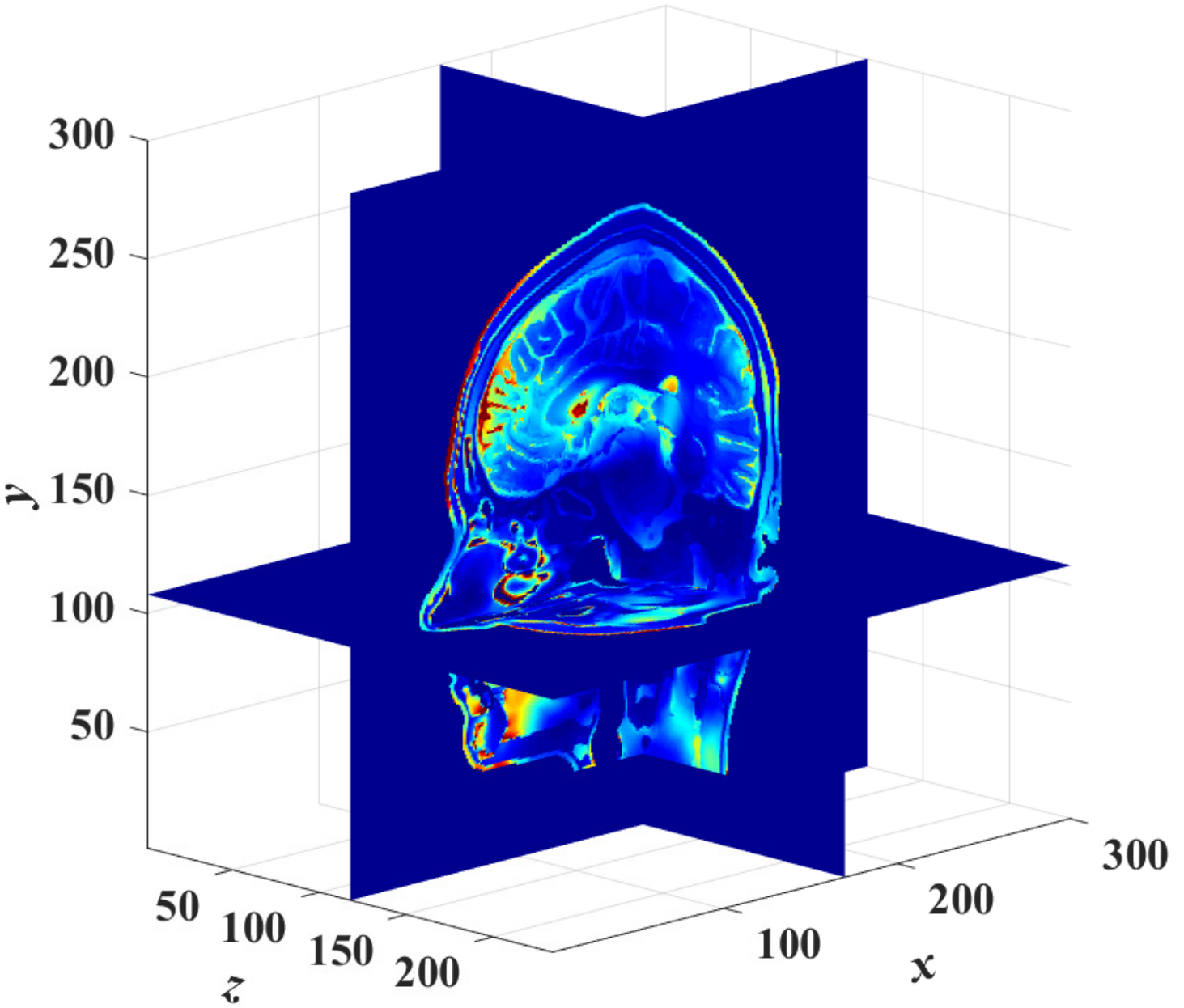}
		}
		%		\centerline{(b)}
		%	\end{minipage}
	%	\hfill
	%	\begin{minipage}[h]{0.20\linewidth}
		\!\!\!\!\!\!\!\!\!
		\subfigure[]{
			\includegraphics[height=38.5mm]{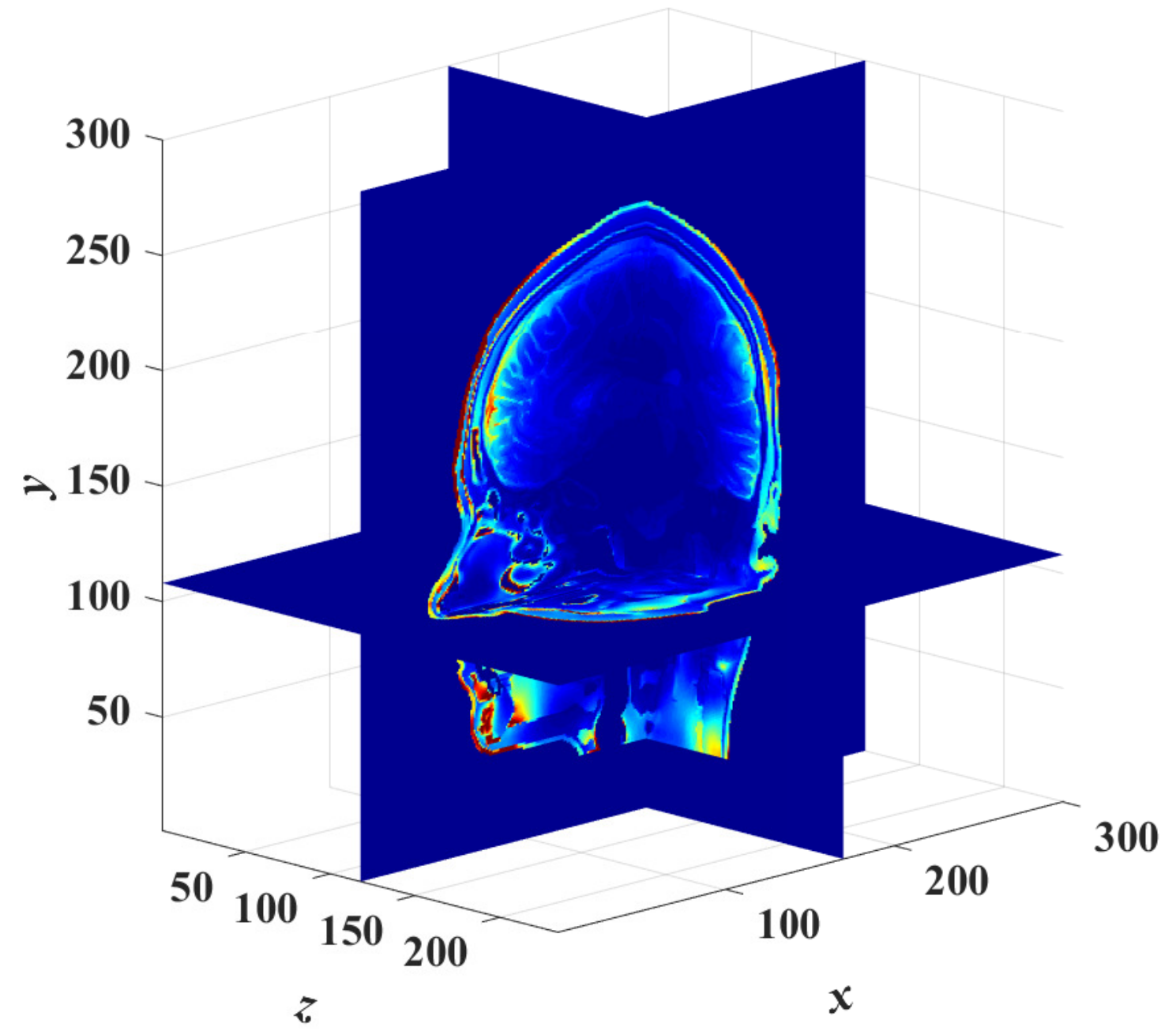}
		}
		%		\centerline{(c)}
		%	\end{minipage}
	%	\hfill
	%	\begin{minipage}[h]{0.20\linewidth}
		\!\!\!\!\!\!\!\!\!
		\subfigure[]{
			\includegraphics[height=40mm]{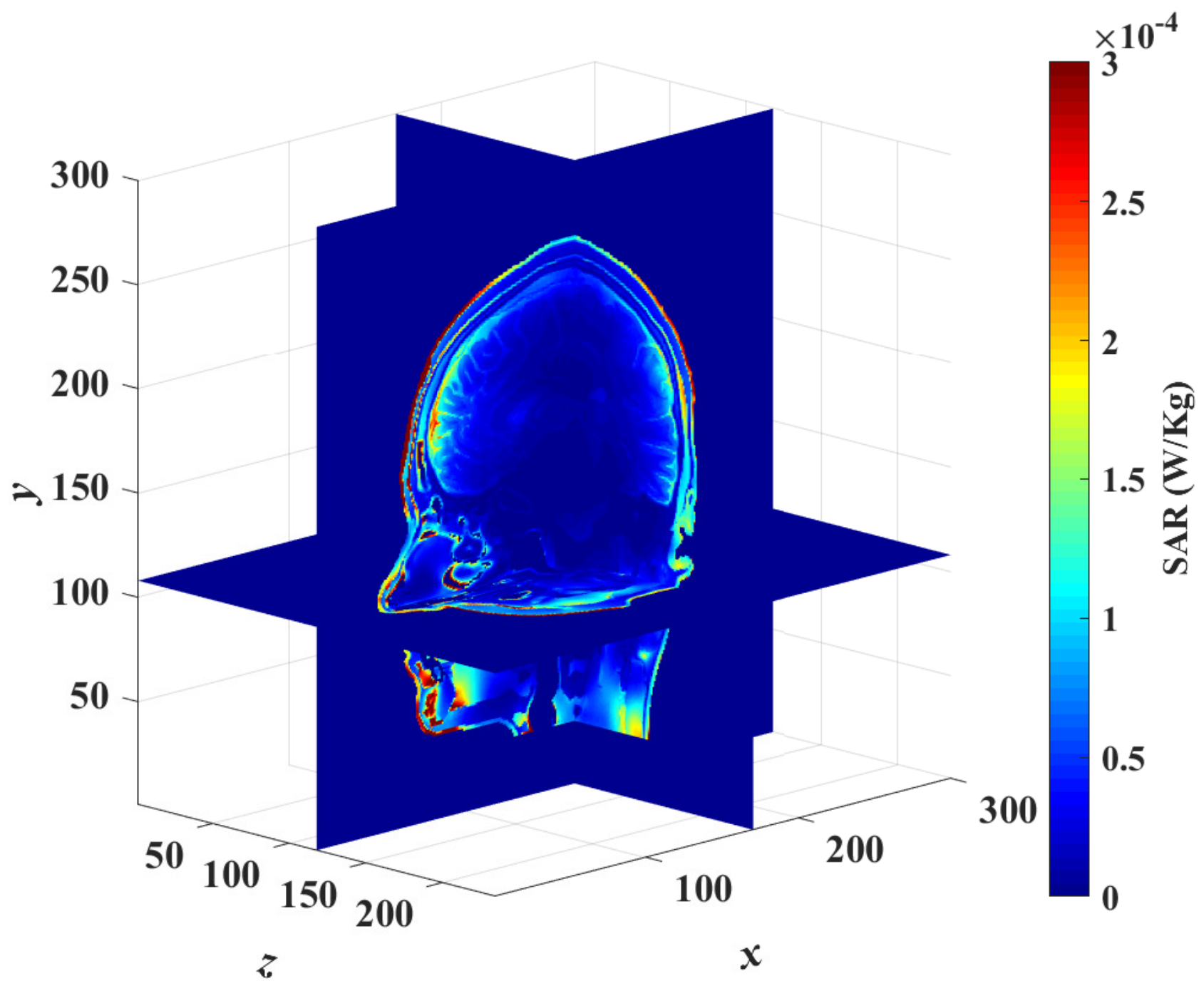}
		}
		%		\centerline{(d)}
		%	\end{minipage}
	\caption{The SAR calculated by (a) the FDTD method at 900.0 MHz, (b) the SBP-SAT FDTD method at 900.0 MHz, (c)the FDTD method at 2.4 GHz, and (d)the SBP-SAT FDTD method at 2.4 GHz.}
	\label{F_14}
\end{figure*}
\begin{figure}[h]
	\centering
	\includegraphics[scale=0.4]{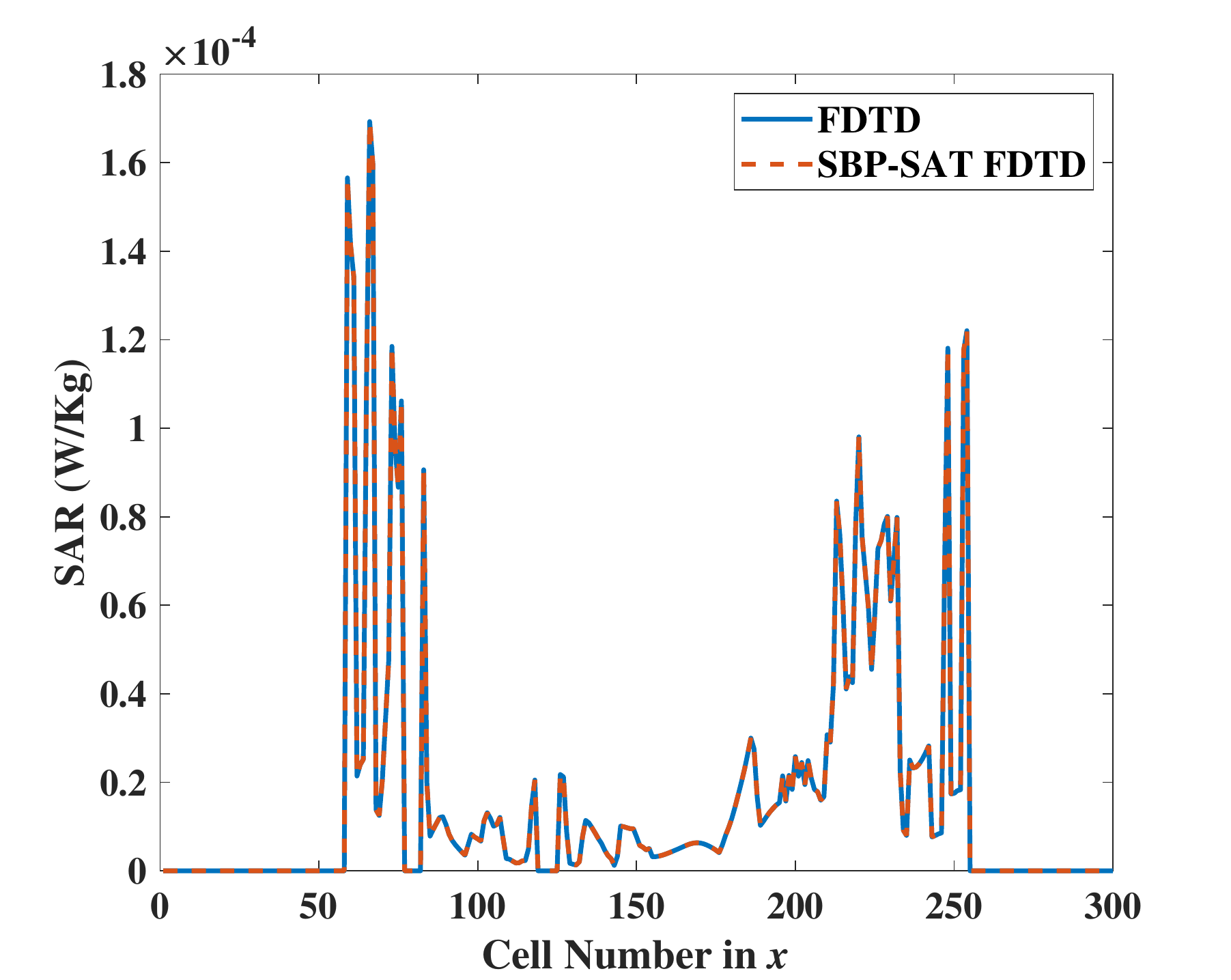}
	\caption{The SAR obtained from the FDTD method and the proposed SBP-SAT FDTD method at 900.0 MHz in the $x$ direction at $y$ = 150 $mm$ and $z$ = 118 $mm$.}
	\label{F_15}
\end{figure}

\begin{figure}[h]
	\centering
%	\subfigure[]{
%		\includegraphics[scale=0.23]{Picture/SAR_x_150_line.eps}
%	}
%	\subfigure[]{
		\includegraphics[scale=0.4]{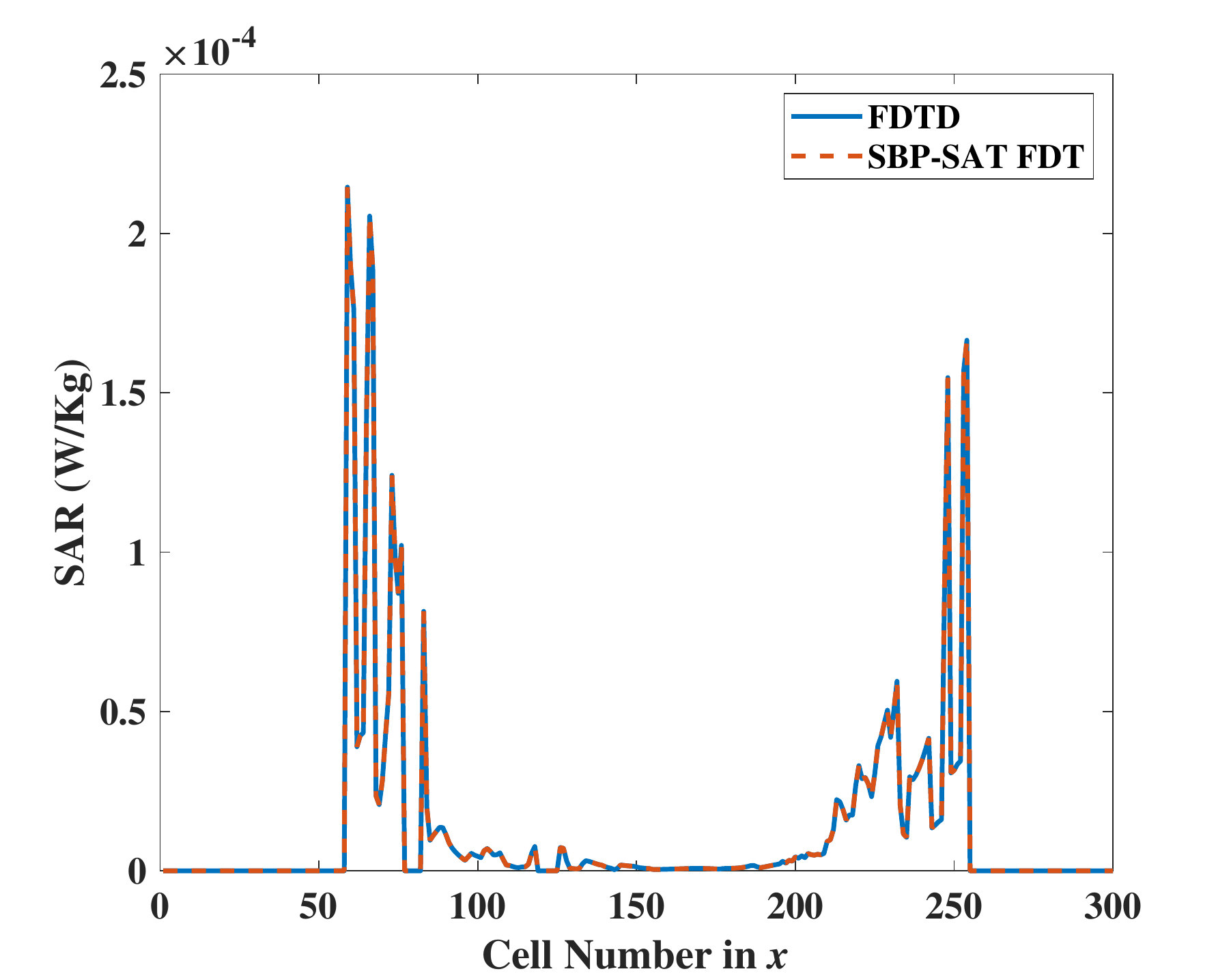}
%	}
	\caption{The SAR obtained from the FDTD method and the proposed SBP-SAT FDTD method at 2.4 GHz in the $x$ direction at $y$ = 150 $mm$ and $z$ = 118 $mm$.}
	\label{F_16}
\end{figure}

The point SAR is calculated by the SAT-SBP FDTD method and the FDTD method, which is given by 
\begin{equation}\label{E28}
	\begin{aligned}
		{\rm{SAR}} = \frac{{\sigma {{\left| E \right|}_{max }^2}}}{{2\rho }},
	\end{aligned}
\end{equation}
where $\sigma$, $\rho$ are the conductivity and density of the corresponding tissues. ${\left| E \right|_{\max }}$ denotes the maximum value of electric fields during the whole simulation. In the Cartesian coordinate system, the point SAR in each cell can be calculated by
\begin{equation}\label{E29}
	\begin{aligned}
		{\rm{SAR}} = \frac{{\sigma {{\left| {E_x^2 + E_y^2 + E_z^2} \right|}_{max }}}}{{2\rho }}.
	\end{aligned}
\end{equation}
In our implementation, $E_x$, $E_y$ and $E_z$ are averaged electric fields value at the center of each cell in the $x$, $y$ and $z$ directions, respectively. 

The human head model is discretized with meshes of cell size $1~mm$. The computational domain is discretized as $300 \times 300 \times 235$ cells in $x$, $y$ and $z$ directions, respectively. 21.15 million cells in total are used in the computational domain. Fig. \ref{F_12} shows the mesh configuration in our simulation. 
For a better visualization of meshes in our simulation, only one line is drawn every five mesh lines, which implies that each cell represents 125 adjacent ones in the computational domain. The CPML and TF/SF boundary are used to truncate the computational domain and excite the plane wave. The TF/SF boundary is 5 cells away from the CPML. The plane wave with $f = 900.0$ MHz and $2.4$ GHz is considered, and it incidents from the $-x$ direction. The total physical time is 10 $ns$.

Fig. \ref{F_13}(a) and (b) show the permittivity and conductivity of the human head model at 900.0 MHz in the cross section of $x= 170~mm$, $y =108~mm$ and $z =118 ~mm$,  and (c), (d) show the permittivity and the conductivity at 2.4 GHz in the same cross section. 

The SAR calculated by the FDTD method and the proposed SBP-SAT FDTD method are shown in Fig. \ref{F_14}. Among these results, Fig. \ref{F_14}(a), (b) show the SAR at 900.0 MHz and (c), (d) show the SAR at 2.4 GHz. It can be found that the SAR calculated by the SBP-SAT FDTD method agrees well with those from the FDTD method in Fig. \ref{F_14}(a) and (b). The patterns of two methods are almost identical to each other. Since the plane wave incidents from the $-x$ direction, the SAR is larger at the front side of the human head than that in other regions, as shown in Fig. \ref{F_14}(a) and (b). In addition, those tissues, such as the cerebrospinal fluid in the brain, have a large electrical conductivity. Therefore, those regions show large SAR values. Another point should be noted that the brainstem of the human head absorbs more electromagnetic energy than the white and gray matter. Compare Fig. \ref{F_14}(a) and (b) with (c) and (d), it can be found that the SAR of the brainstem at $f$ = 2.4 GHz is significantly reduced compared with that at $f$ = 900.0 MHz. The SAR in the surface layer of the human head increases to some extent. It is caused by the pronounced skin effect at high frequencies. Moreover, the liquid parts with the high electrical conductivity, such as the cerebrospinal fluid and blood vessels, also have large SAR values at $f$ = 2.4 GHz.

In order to compare results calculated by these two methods more clearly, the SAR in the $x$ direction at $y$ = 150 $mm$ and $z$ = 0.118 $mm$ in Fig. \ref{F_15} and Fig. \ref{F_16} at 900.0 MHz and 2.4 GHz, respectively. The reason why SAR = 0 is the existence of trachea in the model, and its conductivity is 0 $S/m$, which indicates electromagnetic waves are not absorbed in those regions. In Fig. \ref{F_15} and Fig. \ref{F_16}, it can be found that the SAR is larger near the skin and smaller inside the head at 2.4 GHz due to strong absorption in the high frequency. It can be found that two curves obtained from two methods are completely overlapped in Fig. \ref{F_15} and Fig. \ref{F_16}, which demonstrates that the proposed SBP-SAT FDTD method has the same level of accuracy compared with that of the FDTD method.
%\begin{figure}[h]
%	\centering
%	\subfigure[]{
%		\includegraphics[scale=0.4]{Picture/SAR_x_150_line.eps}
%	}
%	\\
%	\vspace{-0.1cm}
%	\subfigure[]{
%		\includegraphics[scale=0.4]{Picture/SAR_y_150_line_2400MHz.eps}
%	}
%	\caption{The SAR obtained from different methods at 900 MHz and 2.4 GHz: (a) the SAR along $x$ direction at $y$ = 150 $mm$ and $z$ = 118 $mm$ at 900 MHz; (b) the SAR along $y$ direction at $x$ = 150 $mm$ and $z$ = 118 $mm$ at 2.4 GHz.}
%	\label{F_15}
%\end{figure}

The memory usage and runtime of two methods are listed in Table \ref{T3}. It can be found that memory usage of the SBP-SAT FDTD method only increases by 1.09\% compared with that of the FDTD method, and runtime only increases by only 0.13\%. Since only the SATs in (\ref{E13})-(\ref{E16}) are required to be calculated on the boundaries, the negligible overhead in terms of memory usage and runtime is expected. This trend will become even more obvious when large-scale simulations are involved. In general, the proposed SBP-SAT FDTD method shows good accuracy and imposes a very slight overhead in terms of memory usage and runtime. 

\begin{table}[h]
	\renewcommand\arraystretch{1.5}
	\centering
	\caption{Comparison of Memory Usage and Runtime of the Head in the FDTD Method and the Proposed SBP-SAT FDTD Method}
	\label{T3}
%	\resizebox{9cm}{!}
%	{
	\setlength{\tabcolsep}{6mm}
	{
		\begin{threeparttable}[b]
			\begin{tabular}{ c|  c|  c }
				\hline
				\hline
				\textbf{Method}	&\textbf{Memory (MB)}	 &\textbf{Time (s)} \\
				\hline
				FDTD &$973.5$  &$36,138.0$  \\
				\hline
				
				Proposed &$984.1$ 	 &$36,187.0$	\\		
				\hline
				\hline
				
			\end{tabular}
			%\tablefootnote{Ratio is defined as the ratio of time cost used in the LOD-FDTD method with fine grid to that in the correspond method.}
		\end{threeparttable}
	}
%	}
\end{table}

\section{Conclusion}
A three-dimensional SBP-SAT FDTD method is proposed in this article, which has the same level of accuracy compared to that of the FDTD method with a very small overhead. The special emphasis is placed on the fundamental theoretical aspects of the three-dimensional SBP-SAT method and numerical validation. Our theoretical analysis shows that the proposed three-dimensional SBP-SAT FDTD method is long-time stable and have the same level of accuracy as that of the FDTD method. 

Since the boundary conditions are weakly enforced through the SAT techniques, which is similar to the numerical flux in discontinuous galerkin finite element method (DG-FEM) \cite{DGFEMNODE}, the proposed SBP-SAT FDTD method is extremely flexible compared with the FDTD method. It is well-known that the central numerical flux would lead to spurious modes in the DG-FEM \cite{DGFEMSPurious}. In the proposed SBP-SAT FDTD method, the SAT technique used in our implementation  would not suffer from such issues. As our numerical examples, including the simple cavity, the iris filter, and the SAR calculation from a human head model, shown, the SBP-SAT FDTD method only uses slightly 1.09\% more memory and 0.13\% runtime compared with that of the FDTD method.  
   
It provides many possibilities in the FDTD society, such as the theoretically stable subgridding FDTD method, the {\it{hp}}-refinement techniques, and the energy stable hybrid time-domain method. 

Another article upon the subgridding technique based on the proposed three-dimensional SBP-SAT FDTD method and its application to solving the challenging electromagnetic problems, which is the second part of this topic, will be submitted soon.

%\section*{Acknowledgments}
%This should be a simple paragraph before the References to thank those individuals and institutions who have supported your work on this article.
\vspace{-0.3cm}
\section*{Acknowledgments}
The authors wish to thank the IT'IS Foundation in Switzerland for their help with the data accessibility.
%\vspace{-0.2cm}
{\appendix[]}
The entities of ${\mathbb{D}}_{+ }$, ${\mathbb{D}}_{- }$, ${\mathbb{P}}_{+ }$, ${\mathbb{P}}_{-}$, ${\mathbb{Q}}_{+ }$ and ${\mathbb{Q}}_{- }$ are listed as follows.
\begin{flalign}\label{Dplus}
	&{\mathbb{D}}_{+ } = {\frac{1}{\varepsilon}}\left[ 
	{\begin{array}{*{20}{r}}
			{ - 1}&{\frac{1}{2}}&{\frac{1}{2}}&{}&{}&{}&{}\\
			{ - \frac{1}{2}}&{ - \frac{1}{4}}&{\frac{3}{4}}&{}&{}&{}&{}\\
			{}&{}&{ - 1}&1&{}&{}&{}\\
			{}&{}&{}& \ddots & \ddots &{}&{}\\
			{}&{}&{}&{}&{ - 1}&1&{}\\
			{}&{}&{}&{}&{ - \frac{3}{4}}&{\frac{1}{4}}&{\frac{1}{2}}\\
			{}&{}&{}&{}&{ - \frac{1}{2}}&{ - \frac{1}{2}}&1
	\end{array}} \right],   
\end{flalign}

\begin{equation}\label{DMinus}\!\!\!\!\!
	{\mathbb{D}}_{- }= {\frac{1}{\mu}}\left[ 
	{\begin{array}{*{20}{r}}
			{ - 1}&1&{}&{}&{}&{}&{}&{}\\
			{ - 1}&1&{}&{}&{}&{}&{}&{}\\
			{ - \frac{1}{5}}&{ - \frac{3}{5}}&{\frac{4}{5}}&{}&{}&{}&{}&{}\\
			{}&{}&{ - 1}&1&{}&{}&{}&{}\\
			{}&{}&{}& \ddots & \ddots &{}&{}&{}\\
			{}&{}&{}&{}&{ - 1}&1&{}&{}\\
			{}&{}&{}&{}&{}&{ - \frac{4}{5}}&{\frac{3}{5}}&{\frac{1}{5}}\\
			{}&{}&{}&{}&{}&{}&{ - 1}&1\\
			{}&{}&{}&{}&{}&{}&{ - 1}&1 
	\end{array}}\right].  
\end{equation}

${\mathbb{P}}_{+ }$, ${\mathbb{P}}_{-}$, ${\mathbb{Q}}_{+ }$ and ${\mathbb{Q}}_{-}$ are given by
%\begin{align}
	% {\mathbb{P}}_{+ } = diag\left( {\left[ {\frac{1}{2},1,1,...,1,\frac{1}{2}} \right]} \right)h, 
	%\end{align}

\begin{flalign}
	&{\mathbb{P}}_{+ } = diag\left( {\varepsilon}{\left[ {\frac{1}{2},1,1,...,1,\frac{1}{2}} \right]} \right), \quad\quad\quad\quad\quad\quad \\
	&{\mathbb{P}}_{- } = diag\left( {\mu}{\left[ {\frac{1}{2},\frac{1}{4},\frac{5}{4},1,...,1,\frac{5}{4},\frac{1}{4},\frac{1}{2}} \right]} \right), 
\end{flalign}

\begin{flalign}
	&{\mathbb{Q}}_{+ } = \left[ {\begin{array}{*{20}{r}}
			{ - \frac{1}{2}}&{\frac{1}{4}}&{\frac{1}{4}}&{}&{}&{}&{}\\
			{ - \frac{1}{2}}&{ - \frac{1}{4}}&{\frac{3}{4}}&{}&{}&{}&{}\\
			{}&{}&{ - 1}&1&{}&{}&{}\\
			{}&{}&{}& \ddots & \ddots &{}&{}\\
			{}&{}&{}&{}&{ - 1}&1&{}\\
			{}&{}&{}&{}&{ - \frac{3}{4}}&{\frac{1}{4}}&{\frac{1}{2}}\\
			{}&{}&{}&{}&{ - \frac{1}{4}}&{ - \frac{1}{4}}&{\frac{1}{2}}
	\end{array}} \right],  
\end{flalign}

\begin{align}
	&{\mathbb{Q}}_{- } = \left[ {\begin{array}{*{20}{r}}
			{ - \frac{1}{2}}&{\frac{1}{2}}&{}&{}&{}&{}&{}&{}\\
			{ - \frac{1}{4}}&{\frac{1}{4}}&{}&{}&{}&{}&{}&{}\\
			{ - \frac{1}{4}}&{ - \frac{3}{4}}&1&{}&{}&{}&{}&{}\\
			{}&{}&{ - 1}&1&{}&{}&{}&{}\\
			{}&{}&{}& \ddots & \ddots &{}&{}&{}\\
			{}&{}&{}&{}&{ - 1}&1&{}&{}\\
			{}&{}&{}&{}&{}&{ - 1}&{\frac{3}{4}}&{\frac{1}{4}}\\
			{}&{}&{}&{}&{}&{}&{ - \frac{1}{4}}&{\frac{1}{4}}\\
			{}&{}&{}&{}&{}&{}&{ - \frac{1}{2}}&{\frac{1}{2}}
	\end{array}} \right]. 
\end{align}

\end{document}